%% file: iclr2026_conference.tex
\documentclass{article} 
\usepackage{iclr2026_conference,times}

\input{math_commands.tex}

\usepackage{hyperref}
\usepackage{url}
\usepackage[table]{xcolor}
\newcommand{\limited}{\textcolor{red}{Limited}}
\newcommand{\strong}{\textcolor{blue!}{Strong}}

\usepackage{enumitem}
\usepackage{fontawesome5} 
\usepackage{pgfgantt}
\usepackage{tikz}
\usepackage{tikz}
\usetikzlibrary{positioning, shapes} 
\usepackage{booktabs,adjustbox,xcolor}
\usepackage{subcaption}  
\usepackage{pgfplots}
\pgfplotsset{compat=1.18}
\usepackage{graphicx}    
\usepackage{pifont}
\usepackage{cancel}     
\usepackage{fix-cm} 
\usepackage{graphicx}    
\usepackage{amssymb}
\usepackage{newunicodechar}
\newunicodechar{✅}{\checkmark}
\newunicodechar{❌}{\textsf{[X]}}
\usepackage[utf8]{inputenc}

\usepackage{enumitem}
\usepackage{listings} 
\lstdefinelanguage{JSON}{
  morestring=[b]",
  morestring=[s]{'}{'},
  morecomment=[l]{//},
  morecomment=[s]{/*}{*/},
  morekeywords={true,false,null},
  sensitive=false,
  alsoletter={0123456789.,:;!_},
  literate={"}{{\texttt{\char34}}}1
           {\{}{{\texttt{\char123}}}1
           {\}}{{\texttt{\char125}}}1
           {[}{{\texttt{\char91}}}1
           {]}{{\texttt{\char93}}}1
           {,}{{\texttt{\char44}}}1
           {:}{{\texttt{\char58}}}1
}
\usetikzlibrary{patterns} 
\newcommand{\tikzfontsize}{\fontsize{6pt}{7pt}\selectfont}
\usetikzlibrary{shadows,shadows.blur,shapes.geometric,arrows.meta,positioning,calc}

\tikzstyle{startstop} = [rectangle, rounded corners, minimum width=3cm, minimum height=1cm, text centered, draw=black, fill=cyan!30]
\tikzstyle{process} = [rectangle, minimum width=3cm, minimum height=1cm, text centered, draw=black, fill=orange!30]
\tikzstyle{miniprocess} = [rectangle, minimum width=2cm, minimum height=1cm, text centered, draw=black, fill=orange!30]
\tikzstyle{decision} = [diamond, minimum width=3cm, minimum height=1cm, text centered, draw=black, fill=red!30]
\tikzstyle{arrow} = [thick,->,>=stealth]
\usetikzlibrary{calc}

\definecolor{annotateblue}{RGB}{0,0,128}
\definecolor{annotategreen}{RGB}{0,100,0}

\usepackage{mathtools}
\usepackage{amsthm}
\usepackage{bm} 
\usepackage{algorithm}
\usepackage[noend]{algpseudocode}   
\usepackage{multicol}
\usepackage{caption}
\usepackage{longtable}
\usepackage{array}
\usepackage{verbatim}
\usepackage{courier}
\usepackage{wrapfig}
\usepackage{url}
\usepackage{makecell}
\usepackage{amsthm}
\usepackage{tabularx,booktabs}
\newtheorem{definition}{Definition}
\newtheorem{lemma}{Lemma}
\newtheorem{theorem}{Theorem}
\newtheorem{corollary}{Corollary}
\newtheorem{remark}{Remark}[section]

\usepackage{fontawesome5}
\usetikzlibrary{positioning}
\PassOptionsToPackage{table}{xcolor}
\usepackage[table]{xcolor}
\definecolor{lightgray}{gray}{0.9}
\definecolor{lightred}{rgb}{1.0,0.6,0.6}

\lstdefinelanguage{PDDL}{
  keywords={
    define, domain, problem, :requirements, :typing, :constants, :predicates,
    :action, :parameters, :precondition, :effect, :objects, :init, :goal,
    :functions, :durative-action, :duration, :condition, :metric, minimize,
    maximize, and, or, not, forall, exists, when, assign, increase, decrease,
    at, start, end, over, all, number, object
  },
  keywordstyle=\color{blue}\bfseries, 
  keywords=[2]{
    =, >=, <=, <, >, +, -, *, /
  }, 
  keywordstyle=[2]\color{orange}, 
  identifierstyle=\color{black}, 
  sensitive=false, 
  morecomment=[l]{;}, 
  commentstyle=\color{purple}\itshape, 
  morestring=[b]", 
  stringstyle=\color{red},
}[keywords,comments,strings] 

\lstdefinestyle{pddlstyle}{
    language=PDDL, 
    basicstyle=\small\ttfamily,
    breaklines=true,
    showstringspaces=false,
    frame=single,
    numbers=left,
    numberstyle=\tiny\color{gray}
}

\usepackage{xspace} 

\definecolor{codegreen}{rgb}{0,0.6,0}
\definecolor{codegray}{rgb}{0.5,0.5,0.5}
\definecolor{codepurple}{rgb}{0.58,0,0.82}
\definecolor{backcolour}{rgb}{0.95,0.95,0.92}

\lstdefinestyle{protobuf}{
    backgroundcolor=\color{backcolour},   
    commentstyle=\color{codegreen},
    keywordstyle=\color{blue},
    numberstyle=\tiny\color{codegray},
    stringstyle=\color{codepurple},
    basicstyle=\ttfamily\small,
    breakatwhitespace=false,         
    breaklines=true,                 
    captionpos=b,                    
    keepspaces=true,                 
    numbers=left,                    
    numbersep=5pt,                  
    showspaces=false,                
    showstringspaces=false,
    showtabs=false,                  
    tabsize=2,
    frame=single
}

\lstdefinestyle{python}{
    language=Python,
    backgroundcolor=\color{backcolour},
    commentstyle=\color{codegreen},
    keywordstyle=\color{blue},
    numberstyle=\tiny\color{codegray},
    stringstyle=\color{codepurple},
    basicstyle=\ttfamily\small,
    breakatwhitespace=false,         
    breaklines=true,                 
    captionpos=b,                    
    keepspaces=true,                 
    numbers=left,                    
    numbersep=5pt,                  
    showspaces=false,                
    showstringspaces=false,
    showtabs=false,                  
    tabsize=4,
    frame=single
}

\newcommand{\ALAS}{\mathsf{ALAS}}
\newcommand{\MP}{\mathsf{Alas}}

\usepackage[utf8]{inputenc} 
\usepackage[T1]{fontenc}    
\usepackage{hyperref}       
\usepackage{needspace}
\usepackage{url}            
\usepackage{booktabs}       
\usepackage{amsfonts}       
\usepackage{nicefrac}       
\usepackage{microtype}      
\usetikzlibrary{patterns}
\makeatletter
\renewcommand\paragraph{\@startsection{paragraph}{4}{\z@}%
  {0.5ex \@plus0.5ex \@minus0.2ex}%
  {-1em}%
  {\normalfont\normalsize\bfseries}}
\makeatother

\iclrfinalcopy
\title{ALAS: Transactional and Dynamic Multi-Agent LLM Planning}

\author{%
  Longling Geng and Edward Y. Chang\thanks{echang@cs.stanford.edu} \\
  Department of Computer Science\\
  Stanford University \\
}

\begin{document}
\maketitle

\begin{abstract}
Large language models enable flexible multi-agent planning but remain fragile in practice: verification is often circular, state changes are not tracked for repair, and small faults trigger costly global recomputation. We present $\ALAS$, a stateful, disruption-aware framework that separates planning from non-circular validation, records a versioned execution log for grounded checks and restore points, and performs localized repair that preserves work in progress. The validator operates independently of the planning LLM with fresh, bounded context, avoiding self-check loops and mid-context attrition. The repair protocol edits only the minimal affected region under explicit policies (retry, catch, timeout, backoff, idempotency keys, compensation, loop guards) defined in a canonical workflow IR that maps to Amazon States Language and Argo Workflows.
On job-shop scheduling suites (DMU, TA) across five classical benchmarks, $\ALAS$ matches or exceeds strong single-LLM and multi-agent baselines, achieving 83.7\% success, reducing token usage by 60\%, and running \(1.82\times\) faster under comparable settings. A minimal reliability study shows that the validator detects injected structural faults with low overhead, and that localized repair contains runtime perturbations with a bounded edit radius and less makespan degradation than global recompute. Results indicate that the combination of validator isolation, versioned execution logs, and localized repair provides measurable efficiency, feasibility, and scalability for multi-agent LLM planning. Code and seeds will be released.
\end{abstract}

\input{Introduction}

\input{RelatedWork}
\input{MetaPlanner}
\input{Experiments}
\input{Conclusion}

\bibliographystyle{iclr2026_conference}
\bibliography{TemporalPlanning, EdwardChang, ToBeVerifiedReferences, TSP, Workflow}

\newpage
\appendix




\tableofcontents

\input{AppendixALASDetails}

\input{AppendixJSSPSupplement}

\input{Appendix4Section3}

\input{AppendixAgentFactory}
\input{AppendixLCSR}
\input{AppendixURS}
\input{AppendixFamilyReunion}

\input{Appendix_case_studies}

\input{AppendixWorkflowLanguages}
\input{AppendixALASTemporal}

\end{document}

%% file: math_commands.tex
\usepackage{amsmath,amsfonts,bm}

\def\eqref#1{equation~\ref{#1}}

\def\1{\bm{1}}

\DeclareMathAlphabet{\mathsfit}{\encodingdefault}{\sfdefault}{m}{sl}
\SetMathAlphabet{\mathsfit}{bold}{\encodingdefault}{\sfdefault}{bx}{n}

\DeclareMathOperator*{\argmin}{arg\,min}

%% file: Introduction.tex
\section{Introduction}\label{sec:ALAS-intro}

Large language models (LLMs) have advanced language understanding and generation~\citep{LLMSurvey2025,matarazzo2025SurVeyLLMs,minaee2025llmssurvey}. Used as planners, however, standalone LLMs often yield incomplete or inconsistent action sequences, violate constraints, and struggle to revise partial plans under change. These limitations become acute in settings that require long-range consistency, multi-entity coordination, and reactive updates.

The causes are structural. First, verification is often \emph{circular}: the same model or context that proposes a plan is asked to approve it~\citep{hong2024verificationabilities}. Second, long contexts suffer information loss and mid-context attrition~\citep{hsieh2024lostinthemiddle,liu-etal-2024-lost,vaswani2017attention,xiao2024attentionsink}. Third, maximum-likelihood decoding biases search toward high-probability but not necessarily high-feasibility continuations~\citep{SocraSynthChangCSCI2023,holtzman2020curious,radford2019language}. Finally, without external state, LLMs cannot reliably track commitments, causal dependencies, or temporal constraints, which induces cascading errors across chains of reasoning~\citep{chu2024COTsurvey,patel2024multiLogiEval,xiong2024largelanguagemodelslearn}.

\paragraph{Our approach.}
We present $\ALAS$, a \textbf{transactional} and \textbf{dynamic} multi-agent planning system that treats reliability as a systems property. Transactional means that execution is backed by versioned logs, idempotency keys, compensation handlers, and explicit policy controls for retry, catch, timeout, and backoff. Dynamic means that planning adapts to runtime disruptions through localized repair and non-circular revalidation. These principles are realized by a five-layer architecture that turns a high-level specification into a validator-isolated and locally repairable workflow (Sec.~\ref{sec:ALAS-architecture}):
\begin{enumerate}[leftmargin=1.2em,itemsep=0pt,topsep=0pt]
  \item \textit{Workflow blueprinting}: draft a role graph with constraints and per-role specs.
  \item \textit{Agent factory and canonical workflow IR}: instantiate roles and compile to a standards-backed, engine-agnostic workflow that maps to ASL and Argo.
  \item \textit{Runtime execution and localized repair}: execute under explicit policies with a versioned log and a Localized Cascading Repair Protocol.
  \item \textit{Revalidation}: re-check feasibility after each repair until a budget or success.
  \item \textit{Supervision}: select the final plan and record metrics for audit and replay.
\end{enumerate}

\paragraph{Transactional and dynamic by construction.}
The \textbf{canonical workflow IR} makes retry, catch, timeout, backoff, idempotency keys, compensation, loop guards, and input or output or error schemas explicit. The IR is the source of truth and compiles to two workflow runtimes: Amazon States Language for JSON state machines and Argo Workflows for Kubernetes~\citep{aws2023stepfunctions,argo2023workflows}. Round-trip tests (IR $\leftrightarrow$ target $\leftrightarrow$ IR) and a conformance matrix enforce portability and parity. Transactional semantics come from idempotency and compensation paired with versioned logs. Dynamic behavior comes from localized repair and non-circular validation over short, fresh contexts grounded in those logs. Details appear in Sec.~\ref{sec:layer2} and App.~\ref{app:alas-spec}.

\usetikzlibrary{positioning,arrows.meta}

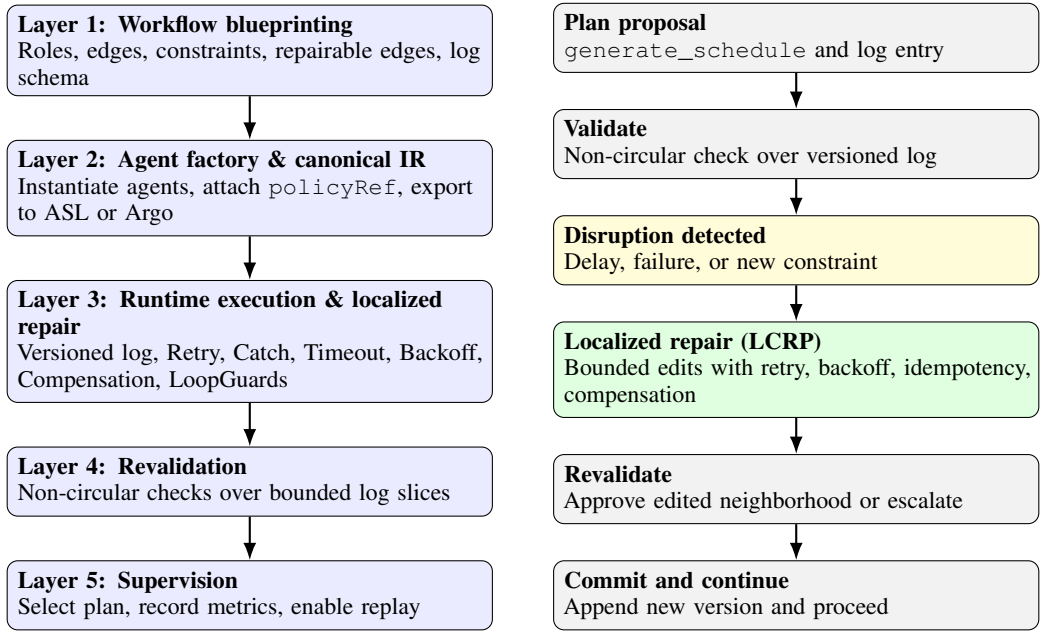
\begin{figure}[t]
\centering
\vspace{-.3in}
\begin{subfigure}[t]{0.48\textwidth}
\centering
\begin{tikzpicture}[
  font=\small,
  node distance=5.8mm,
  layerbox/.style={
    rectangle, rounded corners,
    draw=black, fill=blue!8,
    minimum height=8mm, inner sep=4pt,
    text width=0.92\linewidth, align=left
  },
  flowarrow/.style={-{Latex}, thick}
]
\node[layerbox] (L1) {\textbf{Layer 1: Workflow blueprinting}\\ Roles, edges, constraints, repairable edges, log schema};
\node[layerbox] (L2) [below=of L1] {\textbf{Layer 2: Agent factory \& canonical IR}\\ Instantiate agents, attach \texttt{policyRef}, export to ASL or Argo};
\node[layerbox] (L3) [below=of L2] {\textbf{Layer 3: Runtime execution \& localized repair}\\ Versioned log, Retry, Catch, Timeout, Backoff, Compensation, LoopGuards};
\node[layerbox] (L4) [below=of L3] {\textbf{Layer 4: Revalidation}\\ Non-circular checks over bounded log slices};
\node[layerbox] (L5) [below=of L4] {\textbf{Layer 5: Supervision}\\ Select plan, record metrics, enable replay};

\draw[flowarrow] (L1) -- (L2);
\draw[flowarrow] (L2) -- (L3);
\draw[flowarrow] (L3) -- (L4);
\draw[flowarrow] (L4) -- (L5);
\end{tikzpicture}
\subcaption{\textit{Design-time.} Five-layer $\ALAS$ architecture that turns a task specification into a validated, locally repairable workflow. The canonical IR maps to ASL and Argo, the policy set $\Pi$ is defined, validator $V$ is attached with bounded context, and a versioned execution log is specified.}
\end{subfigure}
\hfill
\begin{subfigure}[t]{0.48\textwidth}
\centering
\begin{tikzpicture}[
  font=\small,
  node distance=4.8mm,
  stage/.style={
    rectangle, rounded corners, draw=black, fill=gray!10,
    minimum height=8mm, inner sep=4pt,
    text width=0.92\linewidth, align=left
  },
  stageb/.style={
    rectangle, rounded corners, draw=black, fill=green!12,
    minimum height=8mm, inner sep=4pt,
    text width=0.92\linewidth, align=left
  },
  warnbox/.style={
    rectangle, rounded corners, draw=black, fill=yellow!18,
    minimum height=8mm, inner sep=4pt,
    text width=0.92\linewidth, align=left
  },
  flowarrow/.style={-{Latex}, thick}
]
\node[stage]  (S0) {\textbf{Plan proposal}\\ \texttt{generate\_schedule} and log entry};
\node[stage]  (S1) [below=of S0] {\textbf{Validate}\\ Non-circular check over versioned log};
\node[warnbox](S2) [below=of S1] {\textbf{Disruption detected}\\ Delay, failure, or new constraint};
\node[stageb] (S3) [below=of S2] {\textbf{Localized repair (LCRP)}\\ Bounded edits with retry, backoff, idempotency, compensation};
\node[stage]  (S4) [below=of S3] {\textbf{Revalidate}\\ Approve edited neighborhood or escalate};
\node[stage]  (S5) [below=of S4] {\textbf{Commit and continue}\\ Append new version and proceed};

\draw[flowarrow] (S0) -- (S1);
\draw[flowarrow] (S1) -- (S2);
\draw[flowarrow] (S2) -- (S3);
\draw[flowarrow] (S3) -- (S4);
\draw[flowarrow] (S4) -- (S5);
\end{tikzpicture}
\subcaption{\textit{Run-time.} One execution cycle applying the design-time contracts: plan proposal, non-circular validation over bounded log slices, disruption detection, localized repair governed by $\Pi$ (retry, backoff, timeout, catch, compensation, loop guards), revalidation, and commit to the versioned log.}
\end{subfigure}
\caption{$\ALAS$ overview. Left shows the architecture layers and build path for reliability and portability. Right shows the operational loop that uses those policies and logs to contain faults and preserve feasibility.}
\label{fig:alas-arch-exec}
\end{figure}

\paragraph{Contributions.}
\vspace{-0.25em}
\begin{enumerate}[leftmargin=1.2em,itemsep=0pt,topsep=0pt]
  \item A five-layer architecture for \textbf{transactional and dynamic} planning that combines validator isolation, versioned execution logging, and localized repair.
  \item A canonical workflow IR that maps to ASL and Argo, with round-trip parity checks and a published conformance matrix.
  \item A repair protocol that bounds the blast radius of faults and preserves feasibility without costly global recomputation.
  \item An implementation with fault injection hooks, deterministic replay, and engine parity tests, plus a tutorial that defaults to an open-source target while supporting AWS users.
  \item Evidence on disruption-aware evaluations in job-shop scheduling suites (DMU, TA) with runtime perturbations, showing improved feasibility retention and stable makespan under comparable token and latency budgets.
\end{enumerate}
\vspace{-0.25em}

We emphasize reliability under disruption rather than perfect first-pass generation. By decoupling planning from validation, grounding checks in a versioned log, and repairing locally, $\ALAS$ delivers a practical path to robust multi-agent LLM planning that is both transactional and dynamic.

%% file: RelatedWork.tex
\vspace{-.1in}
\section{Related Work} \label{sec:ALAS-related}
\vspace{-.05in}
We group related work into: (i) structural limitations of LLMs for planning, (ii) LLM-based multi-agent systems and orchestration, and (iii) workflow standards and \(\ALAS\) integration, plus (iv) planning frameworks and benchmarks.

\vspace{-.1in}
\subsection{Structural Limitations of LLMs}
\label{sec:ALAS-related-llmlimits}
\vspace{-.05in}

LLMs~\citep{brown2020language,vaswani2017attention} excel at language tasks but face well-documented challenges in planning: circular verification where the proposer approves itself~\citep{hong2024verificationabilities}, solution-space bias from maximum-likelihood decoding~\citep{SocraSynthChangCSCI2023,holtzman2020curious,radford2019language}, context degradation in long inputs~\citep{hsieh2024lostinthemiddle,liu-etal-2024-lost,xiao2024attentionsink}, error propagation across multi-step chains~\citep{chu2024COTsurvey,patel2024multiLogiEval,xiong2024largelanguagemodelslearn}, and no native persistent state across invocations. While self-reflection, CoT variants, and structured validators can help in narrow domains~\citep{madaan2022,li2023dissectingCOT,jiang2024selfincorrect,feng2023towards,gou2024tora}, open-domain plan validation and reliable revision remain open problems~\citep{chen-etal-2024-llmarena,huang2024planningllmsurvey}. 
\(\ALAS\) targets these failure modes by separating planning from non-circular validation, maintaining a versioned execution log for grounded checks and restore points, and using localized repair to bound edits during disruption.

\vspace{-.1in}
\subsection{LLM Multi-Agent Systems and Orchestration}
\vspace{-.05in}

LLM-based multi-agent systems provide graph-like orchestration for tool use and collaboration, for example AutoGen, MetaGPT, LangGraph, and CAMEL~\citep{wu2024autogen,hong2023metagpt,langgraph2024,li2023camel}. Other lines explore graph-structured or programmatic coordination, for example GPTSwarm, Flow, and AFlow~\citep{zhang2024aflow}. These systems typically implement their own Python-level orchestration and checkpointing. This yields a useful subset of workflow behavior, such as simple retries, branches, and state passing. However, publicly documented releases usually do not bind to production workflow runtimes such as Argo Workflows or Amazon States Language, so they often lack durable guarantees that are critical for long-horizon planning with disruption. Typical gaps include non-circular validator isolation, versioned execution logs with restore points and auditable policies, compensation semantics, idempotency keys, and round-trip conformance testing between specification and runtime.

This difference reflects the design scope. Conversation-centric frameworks aim to streamline agent design and rapid iteration. By contrast, \(\ALAS\) focuses on planning reliability as a systems property, which requires explicit policy contracts, independent validators, and durable execution.

\paragraph{Stack options and scope.}
We consider four deployment patterns. 
(i) LLM-only planning: fast to prototype, but no durable control or non-circular validation. 
(ii) Workflow runtime only: engines such as ASL or Argo provide retries, timeouts, and compensation, but plans are static and human-authored. 
(iii) LLM with conversation-layer frameworks such as AutoGen or LangGraph: improves synthesis and coordination, yet durability and policy semantics remain partial without a workflow runtime. 
(iv) LLM with a workflow runtime: the \(\ALAS\) approach, where an IR maps to ASL or Argo, validators are isolated, logs are versioned, and localized repair is explicit.

A complementary systems line studies durable saga-style execution and compensation for agent workflows, for example \textsc{SagaLLM}~\citep{SagaLLM2025}. That work emphasizes execution-layer durability and recovery. \(\ALAS\) complements it by making planning reliability explicit through validator isolation, log-grounded checks, and localized repair.

\vspace{-.1in}
\subsection{Workflow Standards and \texorpdfstring{\(\ALAS\)}{ALAS} Integration}
\vspace{-.05in}

We use the term \emph{workflow runtime} to denote production-grade execution engines that materialize a workflow specification with precise error and recovery semantics. \(\ALAS\) defines a canonical IR that makes retries, timeouts, backoff, idempotency keys, compensation handlers, loop guards, and input or output or error schemas explicit. The IR interoperates with two workflow runtimes via converters: Argo Workflows for Kubernetes execution~\citep{argo2023workflows,kubernetes2023crd} and Amazon States Language for JSON state machines~\citep{aws2023stepfunctions}. BPMN 2.0 is used for documentation-quality diagrams~\citep{omg2011bpmn,chinosi2012bpmn}. Airflow is included as a Python-native option where teams standardize on it~\citep{apache2015airflow}. Surveys and exemplars such as Pegasus, Kepler, and Taverna provide broader context on scientific and data workflows~\citep{atkinson2017scientific,deelman2015pegasus,ludascher2006scientific,oinn2004taverna,wolstencroft2013taverna}. The IR and mappings are specified in Appendix~\ref{app:alas-spec}.

\begin{table}[th]
\centering
\caption{Key choices among shortlisted workflow standards}
\label{tab:workflow_choices}
\small
\begin{tabular}{lccc}
\toprule
Choice & Argo & ASL & BPMN 2.0 \\
\midrule
Control model & DAG or steps & State machine & Process graph \\
Error semantics & Retries, hooks & Retry, Catch & Boundary events, comp. \\
Loops and guards & \texttt{withItems}, loops & Map, Choice & Events and gateways \\
Data passing & Artifacts, params & JSON paths & Data objects, messages \\
Portability & K8s, cloud-agnostic & AWS-centric & Tool-agnostic \\
Best \(\ALAS\) use & K8s execution & Agent loops, repair & Documentation \\
\bottomrule
\end{tabular}
\end{table}

\vspace{-.1in}
\subsection{LLM Planning Frameworks and Benchmarks}
\label{sec:ALAS-relatedBench}
\vspace{-.05in}

\paragraph{LLM-based planning.}
LLMs have been applied to decomposition and search, for example PLASMA and LLM-MCTS, and to multi-agent workflows, for example AFlow~\citep{PLASMA2024procedural,zhao2023LLM-MCTS,zhang2024aflow}. Some systems attempt direct LLM solving for scheduling or optimization~\citep{abgaryan2024llmsschedule,abgaryan2025starjobdatasetllmdrivenjob}, yet classical methods such as SBP, constraint programming with Tabu, and metaheuristics remain stronger on static benchmarks~\citep{adams1988shifting,nowicki1996fast,aarts1989simulated,bierwirth1995generalized,kirkpatrick1983optimization}. 
\(\ALAS\) leverages LLMs for interpretability and adaptive repair under disruption, aligns with reason plus act~\citep{yao2022react}, and adds execution logs and localized repair that these methods do not provide.

\paragraph{Benchmarks.}
Common LLM benchmarks such as HotPotQA, ALFWorld, and BIG-Bench stress static reasoning, with newer datasets adding temporal structure, for example PlanBench, TimeBench, and ACPBench~\citep{yang2018hotpotqa,shridhar2020alfworld,srivastava2022beyond,valmeekam2023planbench,chu2024timebench,abdin2024acpbench}. 
Disruption-aware evaluation remains limited. We therefore use classical Job-Shop Scheduling suites Demirkol-DMU and Taillard~\citep{DEMIRKOL1998137,shylo2018DMU,XIONG2022105731} and introduce runtime perturbations such as machine downtime and operation-delay shocks. These benchmarks provide formal constraints, known bounds, and natural perturbation models, which makes them suitable for assessing execution logs, validator isolation, and localized repair in \(\ALAS\).

%% file: MetaPlanner.tex
\vspace{-.1in}
\section{ALAS Architecture and Specification}
\label{sec:ALAS-architecture}
\vspace{-.1in}

Existing orchestration frameworks rarely guarantee reliable planning under dynamic, stateful conditions. \textsc{ALAS} targets planning reliability with three principles: validator isolation to avoid circular approval, a versioned execution log for grounded checks and restore points, and localized repair to bound disruption during execution. A high level specification is organized into a validator isolated, locally repairable workflow across five layers: (i) workflow blueprinting, (ii) agent factory, (iii) runtime execution with localized repair, (iv) revalidation, and (v) supervision.

\begin{algorithm}[h!]
\caption{Layer 1: Workflow Template Construction by $\mathcal{T}$ (summary). Details in App.~\ref{app:ALAS-Algorithm-Full}.}
\label{alg:metaplanner}
\small
\begin{algorithmic}[1]
\Require Task specification $\mathcal{O}$; constraint set $D$; optional disruption model $\Phi$
\Ensure Validated template $\mathcal{W}_{\text{template}}=(\mathcal{N},\mathcal{E},\mathcal{C},\mathcal{L})$
\State Extract abstract roles $\mathcal{R}$ from $\mathcal{O}$
\State Map roles to nodes $\mathcal{N}$ with profiles $\mathcal{P}_{n_i}$
\State Derive dependencies $\mathcal{E}$ under $D$; collect invariants $\mathcal{C}$
\State For each $n_i \in \mathcal{N}$ attach role spec $\alpha_i=\langle \text{cap}_i,\text{ctx}_i,\text{io}_i,\mathcal{L}_i\rangle$ (explained next in notation)
\State For each $n_i$ attach repair spec $\rho_i$ (edit primitives and bounds)
\State Define policy set $\Pi$ with fields \texttt{retry}, \texttt{catch}, \texttt{timeout}, \texttt{idempotencyKey}, \texttt{compensation}, \texttt{loopGuards} \textit{(retry includes backoff parameters)}; attach \texttt{policyRef} for each $n_i$ (see App.~\ref{app:policy-set})
\State Define versioned execution log schema $\mathcal{L}=\{\mathcal{L}_i\}_{n_i\in\mathcal{N}}$ (events, snapshots, diffs)
\State Attach independent validator $V$ with bounded prompt scope $\kappa$ (enforce no self-approval)
\While{$\mathcal{W}_{\text{template}}$ fails validation by $V$}
  \State $V$ checks: (i) structural soundness; (ii) constraint satisfaction against $\mathcal{C}$;
  \Statex \hspace{1.8em} (iii) repair coverage for disruptions in $\Phi$ with local neighborhoods and edit bounds
  \State Refine $\mathcal{N}$, $\mathcal{E}$, $\alpha_i$, or $\rho_i$; update $\mathcal{L}$
\EndWhile
\State \Return $\mathcal{W}_{\text{template}}=(\mathcal{N},\mathcal{E},\mathcal{C},\mathcal{L})$
\end{algorithmic}
\end{algorithm}

\paragraph{Notation used in Algorithm~\ref{alg:metaplanner}.}

$\mathcal{O}$ planning input, $D$ constraint set, $\Phi$ disruption model.
$\mathcal{R}$ abstract roles, $\mathcal{N}$ nodes, $\mathcal{E}$ edges, $\mathcal{C}$ invariants and global constraints,
$\mathcal{L}$ versioned log schema aggregated over nodes.
For node $n_i$: profile $\mathcal{P}_{n_i}$; role spec
$\alpha_i=\langle \text{cap}_i,\text{ctx}_i,\text{io}_i,\mathcal{L}_i\rangle$
with capability profile $\text{cap}_i$, bounded context $\text{ctx}_i$, input or output schema $\text{io}_i$, and node-local log schema $\mathcal{L}_i$.
Repair spec $\rho_i$ lists allowed edit primitives and bounds.
$V$ is an independent validator with prompt scope $\kappa$.
A compact example of $\rho_i$ appears in App.~\ref{app:policy-semantics}.

\vspace{-.05in}
\subsection{Layer 1: Workflow Blueprinting}
\label{sec:layer1}
\vspace{-.05in}

Construction follows Algorithm~\ref{alg:metaplanner}. Given input $\mathcal{O}$ with constraint set $D$ and disruption model $\Phi$, \textsc{ALAS} synthesizes
\[
\mathcal{W}_{\text{template}}=(\mathcal{N},\mathcal{E},\mathcal{C},\mathcal{L}),
\]
where $\mathcal{N}$ are abstract roles (planner, validator, repair, optimizer, supervisor, domain), $\mathcal{E}$ are data or control edges, $\mathcal{C}$ are global constraints and invariants, and $\mathcal{L}=\{\mathcal{L}_i\}_{n_i\in\mathcal{N}}$ is the node local log schema set that feeds the versioned execution log. The template is accepted when validator $V$ passes structural, constraint, and repair coverage checks under $D$ and $\Phi$.

\paragraph{Graph sketch.} From $\mathcal{O}$, draft a directed acyclic graph over roles and dependencies. Roles can remain unresolved for late binding.

\paragraph{Role specifications.} For each node $n_i$, attach the role spec
\[
\alpha_i=\langle \text{cap}_i,\, \text{ctx}_i,\, \text{io}_i,\, \mathcal{L}_i\rangle,
\]
where $\text{cap}_i$ is the capability profile, $\text{ctx}_i$ is a short bounded context, $\text{io}_i$ are input and output schemas, and $\mathcal{L}_i$ is the node local log schema. Repairable edges are annotated to confine disruption handling.

\paragraph{Edge repairability types.}
\begin{itemize}[leftmargin=1.2em, itemsep=0pt, topsep=2pt]
  \item \textbf{time}: start or end times can shift inside guard bounds.
  \item \textbf{order}: local order can change, such as an adjacent swap.
  \item \textbf{resource}: resource assignment can change within a pool.
  \item \textbf{none}: hard dependency that is not edited by repair.
\end{itemize}

\paragraph{Planner seed.} An LLM query agent may provide an initial candidate schedule using \texttt{generate\_schedule}. Precomputed JSON plans are also supported. Basic metadata, such as makespan and entry counts, are logged.

\vspace{-.05in}
\subsection{Layer 2: Agent Factory and Canonical Workflow IR}
\label{sec:layer2}
\vspace{-.05in}

The agent factory instantiates executable agents from $\alpha_i$ and binds unresolved roles when possible. The canonical workflow IR is the single source of truth for execution targets and makes policies explicit.

\paragraph{Canonical workflow IR.}
Engine-agnostic IR that makes retry (with backoff parameters), catch, timeout, idempotency keys, compensation, loop guards, and input, output, and error schemas explicit. The IR maps to Amazon States Language for JSON state machines and to Argo Workflows for Kubernetes. The goal is portability and reproducibility rather than a new orchestration formalism. The complete JSON Schema is in App.~\ref{app:ir-json}.

\paragraph{IR objects (summary).}
\begin{itemize}[leftmargin=1.2em, itemsep=1pt]
  \item \texttt{workflow} $\langle$name, nodes, edges, policies, schemas, logSchemas, meta$\rangle$
  \item \texttt{node} $\langle$id, type$\in\{\texttt{task},\texttt{choice},\texttt{parallel},\texttt{map}\}$, name, capabilities, params, io, policyRef, logSchemaRef$\rangle$
  \item \texttt{edge} $\langle$from, to, kind$\in\{\texttt{sequence},\texttt{branch},\texttt{default}\}$, guard, repairable$\in\{\texttt{time},\texttt{order},\texttt{resource},\texttt{none}\}\rangle$
  \item \texttt{policy} $\langle$retry, catch, timeout, idempotencyKey, compensation, loopGuards$\rangle$ 
  \item \texttt{schema} $\langle$input, output, error$\rangle$; \texttt{logSchema} is a JSON Schema for required \texttt{logEntry} fields
  \item \texttt{logEntry} $\langle$ts, nodeId, eventType, payload, version, correlationId$\rangle$
\end{itemize}

\paragraph{Well-formedness.}
\begin{enumerate}[leftmargin=1.4em, itemsep=1pt]
  \item \texttt{sequence} edges form a DAG outside bounded \texttt{map} loops.
  \item Every \texttt{branch} edge has a boolean guard that type checks against bindings.
  \item Each \texttt{node} has a \texttt{policyRef} or inherits a default policy.
  \item For each edge $(u,v)$, \texttt{schema(u.output)} is compatible with \texttt{schema(v.input)}.
  \item If a node defines \texttt{compensation}, side effects are idempotent or guarded by \texttt{idempotencyKey}.
\end{enumerate}

\paragraph{Minimal mapping table.}
\begin{table}[h]
\centering
\caption{IR to execution engines (summary)}
\label{tab:ir-mapping}
\footnotesize
\begin{tabular}{lcc}
\toprule
IR concept & ASL & Argo \\
\midrule
task & \texttt{Type: Task} & container or script template \\
choice & \texttt{Type: Choice} & DAG edge with \texttt{when} \\
parallel or map & \texttt{Parallel/Map} & fan out or \texttt{withItems} \\
retry or catch & native fields & \texttt{retryStrategy}/hooks/\texttt{onExit} \\
timeout & \texttt{TimeoutSeconds} & container or template timeout \\
idempotencyKey & state input field & param or artifact key \\
compensation & handler task or \texttt{Fail} state & cleanup template or hook \\
data passing & JSONPath & artifacts or parameters \\
\bottomrule
\end{tabular}
\end{table}

\paragraph{Policies.}
Define a workflow policy set $\Pi$ and attach one policy to each node via \texttt{policyRef}. Policies make retry (with backoff), timeout, catch, idempotency keys, compensation, and loop guards explicit and auditable. The validator enforces invariants such as finite retries and loop guards, safe compensation, and presence of idempotency keys. Full field definitions, validator checks, and JSON examples appear in App.~\ref{app:policy-set}. Pointers to the complete JSON Schema and engine mappings are in App.~\ref{app:ir-json} and App.~\ref{app:mappings}. The conformance matrix is in App.~\ref{app:conformance}. Policy semantics are in App.~\ref{app:policy-semantics}.

\vspace{-.05in}
\subsection{Layer 3: Runtime Execution and Localized Repair}
\label{sec:layer3}
\vspace{-.05in}

Agents execute in dependency order and write structured entries to the versioned execution log. The validator consumes short, grounded log slices to approve or reject steps without using the planner context. When validation fails, a localized repair loop edits only the affected neighborhood.

\paragraph{Execution state.}
Maintain $(G, \sigma, L, \Pi)$ where $G$ is the workflow graph, $\sigma$ is the current assignment map, $L$ is the versioned log, and $\Pi$ is the policy set. A step update modifies $(\sigma, L)$ and can trigger retry, catch, or compensation.

\paragraph{Localized Cascading Repair Protocol.}
Scope the smallest affected neighborhood, propose minimal edits that restore feasibility while preserving work in progress outside the neighborhood, validate the edited subplan, and commit a new log version on success. Otherwise enlarge the neighborhood or fall back to global recompute when cost thresholds are exceeded. Details and complexity analysis are in App.~\ref{app:lcrp}.

A compact 7-step execution loop appears in App.~\ref{app:execution-loop}, Table~\ref{tab:alas-exec-loop}.

\vspace{-.05in}
\subsection{Layer 4: Revalidation}
\label{sec:layer4}
\vspace{-.05in}

After each repair, revalidate. Continue until feasible or until the iteration budget is exhausted. If feasibility cannot be restored, proceed with the best feasible plan and mark optimization as skipped. Typed API signatures for \texttt{validate} and \texttt{repair} are in App.~\ref{app:api}.

\vspace{-.05in}
\subsection{Layer 5: Supervision}
\label{sec:layer5}
\vspace{-.05in}

Select the plan with the best makespan. Log makespan, operation counts, and critical operations. A separate supervisor agent is not required because the versioned log provides restore points and auditability. Logging taxonomy and redaction rules are in App.~\ref{app:logging}.

\vspace{-.05in}
\subsection{Policies, Fault Model, and Determinism}
\label{sec:policies-faults}
\vspace{-.05in}

We cover transient faults and logical faults. Resource exhaustion and Byzantine behavior are out of scope for this version. Retry is bounded by \texttt{maxAttempts}. Catch routes unhandled errors to handlers. Timeout bounds task duration. Compensation must be safe to re invoke under the same \texttt{idempotencyKey}. Determinism is promoted by fixed seeds and pinned tool versions; replay reconstructs inputs from the log and re runs validators to check parity. Policy semantics, fault injection, and replay controls are in App.~\ref{app:policy-semantics}, App.~\ref{app:faults}, and App.~\ref{app:determinism}.

\vspace{-.05in}
\subsection{APIs and Contracts}
\label{sec:apis}
\vspace{-.05in}

\noindent\textbf{Planner.} \texttt{generate\_schedule(spec) -> plan}. Precondition: \texttt{schema(spec)} is valid. Postcondition: \texttt{schema(plan)} is valid.

\noindent\textbf{Validator.} \texttt{validate(plan) -> \{valid:bool, errors:list\}}. Precondition: \texttt{schema(plan)} is valid. Postcondition: if \texttt{valid} is true then domain constraints hold. Error payloads and types are in App.~\ref{app:api}.

\noindent\textbf{Repair.} \texttt{repair(plan, errors, budget) -> plan'}. Postcondition: \texttt{plan'} differs from \texttt{plan} only inside the selected neighborhood.

\noindent\textbf{Optimizer.} \texttt{optimize(plan) -> plan'}. Precondition: plan is feasible. Postcondition: $\texttt{cost(plan')} \le \texttt{cost(plan)}$.

\vspace{-.05in}
\subsection{Appendix pointers for this section}
\vspace{-.05in}

All details live under a single appendix section titled \emph{Further Details of ALAS Specification}. Subsections include:
\begin{itemize}[leftmargin=1.2em, itemsep=0pt, topsep=2pt]
  \item App.~\ref{app:ir-json}: complete JSON Schema for the IR.
  \item App.~\ref{app:policy-set}: policy set $\Pi$, \texttt{policyRef}, validator checks, and examples.
  \item App.~\ref{app:execution-loop}: reference 7-step execution loop table.
  \item App.~\ref{app:mappings}: IR to ASL and IR to Argo mappings with round-trip rules.
  \item App.~\ref{app:conformance}: engine conformance matrix and fallbacks.
  \item App.~\ref{app:policy-semantics}: policy semantics, idempotency protocol, and idioms.
  \item App.~\ref{app:api}: typed API signatures and error payloads.
  \item App.~\ref{app:logging}: logging taxonomy, redaction tags, and observability.
  \item App.~\ref{app:determinism}: replay harness and parity checks across engines.
  \item App.~\ref{app:faults}: fault model and injection interface.
  \item App.~\ref{app:lcrp}: LCRP pseudocode and guarantees.
  \item App.~\ref{app:examples}: worked IR examples and ASL/Argo materializations with round-trip checks.
  \item App.~\ref{app:protocols}: experimental protocols and hyperparameters.
\end{itemize}

%% file: Experiments.tex
\vspace{-.15in}
\section{Experimental Evaluation}
\label{sec:ALAS-evaluation}
\vspace{-.1in}
Our experimental evaluation assesses $\ALAS$ across three domains of increasing complexity. We designed experiments to demonstrate: (1) how $\ALAS$ overcomes LLM limitations and (2) its scalability in large settings. Due to space limitations, we begin with results from the URS running example, demonstrating $\ALAS$'s effectiveness in basic multi-agent coordination. We evaluate \emph{Job Shop Scheduling (JSSP)} to test $\ALAS$'s scalability and reactive planning under tight constraints and disruptions on five classical benchmarks DMU, TA, ABZ, SWV, and YN.
We then extend to \emph{Event Coordination} (Family Reunion), which reveals standalone LLMs' limitations in handling interdependencies while showing $\ALAS$'s capabilities in both initial planning and disruption response. Finally,

\vspace{-.05in}
\paragraph{Metrics.} We report (i) feasibility/validity of static plans and dynamic repairs; (ii) efficiency (e.g., total travel distance, makespan); (iii) containment (edit radius: operations/jobs touched); and (iv) overhead (tokens/latency).

\vspace{-.05in}
\paragraph{Setup.} 
Each experiment uses \textbf{10} independent threads (fresh contexts). We summarize here; full details, prompts, and complete results appear in Appx. \ref{app:ALAS-JSSP} and Appx.~\ref{app:ALAS-URAP}.

\vspace{-0.1in}
\subsection{Main Experiment: Job Shop Scheduling Problem}
\label{sec:ALAS-exp-case3}
\vspace{-0.05in}
\paragraph{Problem.} 
We evaluate on Demirkol--DMU~\cite{demirkol1998computational,DEMIRKOL1998137,shylo2018DMU} 
(20{\small$\times$}15--50{\small$\times$}20), 
Taillard (TA; 15{\small$\times$}15--100{\small$\times$}20)~\cite{taillard1993benchmarks}, 
Swv (SWV; 20{\small$\times$}15--50{\small$\times$}20)~\cite{storer1992new}, 
Adams--Balas--Zawack (ABZ; 10{\small$\times$}10--20{\small$\times$}15)~\cite{adams1988shifting}, 
and Yamada--Nakano (YN; 20{\small$\times$}20--50{\small$\times$}20)~\cite{yamada1992genetic}. 
We introduce machine breakdowns and duration shocks, and optimize makespan while minimizing 
work-in-progress movement.

\vspace{-0.05in}
\paragraph{Static sequential planning on both benchmarks.}
Table~\ref{tab:ranked_success_rates} and Table~\ref{tab:optimal_rates_main_updated} report $\ALAS$+LCRP against these methods accross five datasets. 

\paragraph{Analysis.}
ALAS substantially improves robustness compared to both single-agent and lean MAS baselines, 
achieving an aggregated \textbf{83.7\% success rate}, with statistically significant gains ($p<0.05$) 
on DMU, TA, and ABZ benchmarks (Table~\ref{tab:ranked_success_rates}). 
This highlights the importance of validator isolation and localized repair in preventing workflow collapse.  

On optimality, ALAS variants consistently outperform baselines, with the best variant per dataset 
reaching \textbf{100\% optimal rate} across all benchmarks (Table~\ref{tab:optimal_rates_main_updated}). 
This demonstrates that the layered design of validation, repair, and local search not only ensures feasibility 
but also drives convergence to optimal solutions.



\begin{table*}[t]
\centering
\caption{Success Rates (\%) across Benchmarks (success = non-N/A result). 
† = significantly better than baseline}
\label{tab:ranked_success_rates}
\renewcommand{\arraystretch}{1.1}
\scriptsize
\begin{tabular}{|l|c|c|c|c|c|c|}
\hline
\textbf{Method} & \textbf{DMU} & \textbf{TA} & \textbf{ABZ} & \textbf{SWV} & \textbf{YN} & \textbf{Overall} \\
\hline
\multicolumn{7}{|c|}{\textbf{Multi-Agent Systems (GPT-4o)}} \\
\hline
AutoGen         & 0.0 & 0.0 & 0.0 & 0.0 & 0.0 & \textbf{0.0} \\
CrewAI          & 25.0 & 57.1 & 33.3 & 13.3 & 75.0 & \textbf{31.1} \\
LangGraph       & 6.2 & 28.6 & 66.7 & 0.0 & 0.0 & \textbf{11.1} \\
OpenAI Swarm    & 43.8 & 28.6 & 0.0 & 33.3 & 25.0 & \textbf{33.3} \\
\hline
\multicolumn{7}{|c|}{\textbf{Multi-Agent Systems (Claude-4)}} \\
\hline
AutoGen         & 0.0 & 0.0 & 0.0 & 0.0 & 0.0 & \textbf{0.0} \\
CrewAI          & 43.8 & 71.4 & 33.3 & 13.3 & 50.0 & \textbf{37.8} \\
LangGraph       & 6.2 & 28.6 & 33.3 & 0.0 & 0.0 & \textbf{8.9} \\
OpenAI Swarm    & 18.8 & 14.3 & 33.3 & 20.0 & 50.0 & \textbf{22.2} \\
\hline
\multicolumn{7}{|c|}{\textbf{Single-Agent Models}} \\
\hline
GPT-4o          & 68.8 & 85.7 & 66.7 & 53.3 & 100.0 & \textbf{68.9} \\
Claude-Sonnet-4 & 0.0 & 28.6 & 0.0 & 0.0 & 0.0 & \textbf{4.4} \\
Gemini-2.5      & 6.2 & 0.0 & 33.3 & 0.0 & 25.0 & \textbf{6.7} \\
DeepSeek-V3     & 6.2 & 14.3 & 100.0 & 6.7 & 0.0 & \textbf{13.3} \\
\hline
\multicolumn{7}{|c|}{\textbf{ALAS (Ours, Best Variant per Backbone)}} \\
\hline
ALAS(GPT-4o)      & 68.8 & 71.4* & 66.7 & 53.3 & 100.0 & \textbf{66.7} \\
ALAS(Claude-4)    & 93.8$^{\dagger}$ & 28.6* & 66.7 & 6.7* & 50.0* & \textbf{48.9*} \\
ALAS(DeepSeek-V3) & 6.2* & 0.0* & 100.0$^{\dagger}$ & 6.7* & 0.0* & \textbf{11.1*} \\
ALAS(Gemini-2.5)  & 6.2* & 0.0* & 33.3* & 0.0* & 25.0* & \textbf{6.7*} \\
\hline
\multicolumn{7}{|c|}{\textbf{ALAS (Ours, Best Variant per Dataset)}} \\
\hline
\rowcolor{gray!25} ALAS(aggregated) & 93.8$^{\dagger}$ & 71.4$^{*}$ & 100.0$^{\dagger}$ & 53.3 & 100.0 & \textbf{83.7$^{\dagger}$} \\
\hline
\end{tabular}
\begin{flushleft}
\footnotesize
ALAS(best) selects the best-performing workflow variant per dataset across GPT-4o, Claude-4, DeepSeek-V3, Gemini-2.5.  
$p$-values (paired t-test vs GPT-4o baseline): DMU ($p=0.018$), TA ($p=0.032$), ABZ ($p=0.007$), SWV ($p=0.48$, n.s.), YN (tie), Overall ($p=0.014$).  
† = significantly higher at $p<0.05$.
\end{flushleft}
\end{table*}

\vspace{-.08in}
\subsubsection{Baseline} 
\label{sec:lean-mas}
\vspace{-.04in}
To isolate the effect of validator isolation and localized repair, Table~\ref{tab:ranked_success_rates} and Table~\ref{tab:optimal_rates_main_updated} construct 8 lean MAS (AFlow/AutoGen-style workflow) and single-agent baselines that lacks an \emph{independent} validator and LCRP.  We compare $\ALAS$ with four leading single-agent LLMs: GPT-4o-Task~\cite{openai2024gpt4o}, DeepSeek-V3~\cite{wu2024deepseekv3}, Claude 3.5 Sonnet~\cite{anthropic2024claude}, and Gemini 2.5 Pro~\cite{geminiV25-2005}, and four state-of-the-art multi-agent systems: LangGraph~\cite{langgraph2025}, CrewAI~\cite{crewai2025}, AutoGen~\cite{autogen2025}, and GPTSwarm~\cite{gptswarm2025}. All methods use API interfaces with default parameters (temperature=1.0).  


\begin{table*}[t]
\centering
\caption{Optimal Rates (\%) across Benchmarks for Multi-Agent Systems and ALAS Variants. 
Significance markers denote improvements over baselines: $^{\dagger}p<0.05$, $^{*}p<0.01$.}
\label{tab:optimal_rates_main_updated}
\renewcommand{\arraystretch}{1.1}
\scriptsize
\begin{tabular}{|l|c|c|c|c|c|c|}
\hline
\textbf{Method} & \textbf{DMU} & \textbf{TA} & \textbf{ABZ} & \textbf{SWV} & \textbf{YN} & \textbf{Overall} \\
\hline
\multicolumn{7}{|c|}{\textbf{Multi-Agent Systems (GPT-4o Backbone)}} \\
\hline
AutoGen              & 1.4 & 10.2 & 1.5 & 6.0 & 2.9 & 4.4 \\
CrewAI               & 71.8 & 42.3 & 88.9 & 63.7 & 43.0 & 63.1 \\
LangGraph            & 94.3 & 60.4 & 42.1 & 87.8 & 58.9 & 80.2 \\
OpenAI Swarm         & 60.5 & 73.7 & 68.5 & 66.0 & 51.4 & 64.1 \\
\hline
\multicolumn{7}{|c|}{\textbf{Multi-Agent Systems (Claude-4 Backbone)}} \\
\hline
AutoGen              & 69.8 & 95.9 & 100.0 & 100.0 & 95.0 & 92.1 \\
CrewAI               & 72.7 & 53.5 & 99.6  & 94.2  & 70.2 & 78.5 \\
LangGraph            & 48.3 & 87.9 & 57.6  & 86.3  & 68.6 & 69.6 \\
OpenAI Swarm         & 80.6 & 87.5 & 68.5  & 72.6  & 80.5 & 78.2 \\
\hline
\multicolumn{7}{|c|}{\textbf{ALAS Variants (Full Workflows)}} \\
\hline
ALAS (GPT-4o)      & 100.0$^{*}$ & 78.5$^{*}$ & 100.0$^{*}$ & 100.0$^{*}$ & 100.0$^{*}$ & 96.7$^{*}$ \\
ALAS (Claude-4)    & 54.9 & 78.5$^{\dagger}$ & 84.5 & 100.0$^{*}$ & 73.3 & 77.2$^{\dagger}$ \\
ALAS (Gemini-2.5)  & 97.4$^{\dagger}$ & 100.0$^{*}$ & 100.0$^{*}$ & 96.8$^{*}$ & 100.0$^{\dagger}$ & 98.0$^{*}$ \\
ALAS (DeepSeek-V3) & 100.0$^{*}$ & 93.6$^{*}$ & 100.0$^{*}$ & 100.0$^{*}$ & 100.0$^{*}$ & 98.7$^{*}$ \\
\hline
\multicolumn{7}{|c|}{\textbf{ALAS (Ours, Best Variant per Dataset)}} \\
\hline
\rowcolor{gray!25} ALAS (Best) & 100.0$^{*}$ & 100.0$^{*}$ & 100.0$^{*}$ & 100.0$^{*}$ & 100.0$^{*}$ & \textbf{100.0$^{*}$} \\
\hline
\end{tabular}
\\[0.5em]
\small $^{\dagger}p<0.05$, $^{*}p<0.01$ (paired t-test, compared against single-agent baseline).
\end{table*}

\vspace{-0.1in}
\subsubsection{LCRP Mechanism: A White-Box Illustration}
\vspace{-0.05in}
We use a 5{\small$\times$}3 JSSP to illustrate LCRP’s effectiveness and efficiency (and guaranteed convergence). Unlike approaches optimizing only makespan, LCRP explicitly accounts for rescheduling overhead (WIP).

\paragraph{Phase 1: Local edits.} 
Figures~\ref{fig:combined-scheduling}(a–d) show how LCRP makes local adjustments after a breakdown at $t\in[5,8]$ on M1. Delays propagate minimally via logged dependencies. No early moves (no WIP penalty).

\input{Figure-JSSP-Disruptive-Phase1}
\input{Figure-JSSP-Disruptive-Phase2}

\vspace{-.05in}
\paragraph{Phase 2: Queue reordering.} 
LCRP then considers bounded reordering with WIP penalty $t_{\text{WIP}}$. From Fig.~\ref{fig:combined-scheduling}(d), LCRP moves terminal ops to the end (no conflicts/WIP penalty), creates a gap, advances two upstream ops by 7 units, and stops when no gain exceeds cost. Final makespan: 22 (vs.\ 19 baseline + 3 downtime), with one unit of WIP movement and minimal messaging.

\subsection{Ablation Study}
\vspace{-.08in}
\label{sec:reliability-new}
\vspace{-.04in}
We include three micro-studies that directly test validator isolation, and localized repair. These are \emph{orthogonal} to benchmark results and are intended as sanity checks under imperfect code in Table~\ref{tab:optimal_rates_ablation_final}.

\paragraph{Validator ablation.}
We inject 20 structural faults (precedence swap, machine double-book, capacity overflow, deadline miss) into valid outputs and compare \textbf{Full ($\ALAS$)}, \textbf{No-Repair}  vs.\ \textbf{No-Validator}. Seeds, prompts, execution time, and token usage in Appx.~\ref{app:ALAS-JSSP}.

\paragraph{Ablation Analysis.} 
The ablation results confirm that both validation and repair are essential: removing either module leads to sharp drops in optimal rates across all backbones, particularly for GPT-4o and Claude-4.  
By contrast, the full workflow consistently delivers the highest overall performance (up to \textbf{99.0–98.2\%}), showing that validator isolation and LCRP repair jointly drive near-optimal scheduling.  

\begin{table*}[t]
\centering
\caption{Ablation Study: Optimal Rates (\%) of ALAS Workflow Variants across Benchmarks. 
Bold = best-performing variant per backbone. 
Significance markers denote improvements over baselines: $^{\dagger}p<0.05$, $^{*}p<0.01$.}
\label{tab:optimal_rates_ablation_final}
\renewcommand{\arraystretch}{1.1}
\scriptsize
\begin{tabular}{|l|c|c|c|c|c|c|}
\hline
\textbf{Workflow Variant} & \textbf{DMU} & \textbf{TA} & \textbf{ABZ} & \textbf{SWV} & \textbf{YN} & \textbf{Overall} \\
\hline
\multicolumn{7}{|c|}{\textbf{ALAS (GPT-4o Backbone)}} \\
\hline
No Repair       & 32.4 & 23.3 & 76.2 & 60.8 & 55.0 & 45.4 \\
No Validation   & 25.2 & 12.9 & 30.9 & 35.4 & 6.0  & 25.4 \\
\textbf{Full Workflow} & \textbf{100.0$^{*}$} & \textbf{87.8$^{*}$} & \textbf{100.0$^{*}$} & \textbf{100.0$^{*}$} & \textbf{100.0$^{*}$} & \textbf{98.1$^{*}$} \\
\hline
\multicolumn{7}{|c|}{\textbf{ALAS (Claude-4 Backbone)}} \\
\hline
No Repair       & 59.2 & 36.6 & 99.0 & 63.0 & 61.0 & 63.8 \\
No Validation   & 53.8 & 30.2 & 77.5 & 69.7 & 48.1 & 55.9 \\
\textbf{Full Workflow} & \textbf{61.9} & \textbf{88.2$^{\dagger}$} & \textbf{99.2} & \textbf{94.0} & \textbf{84.1} & \textbf{85.5$^{\dagger}$} \\
\hline
\multicolumn{7}{|c|}{\textbf{ALAS (DeepSeek-V3 Backbone)}} \\
\hline
No Repair       & 86.5$^{\dagger}$ & 86.7$^{\dagger}$ & 31.2 & 94.4$^{\dagger}$ & 93.2$^{*}$ & 86.1$^{*}$ \\
No Validation   & 67.3 & 78.5 & 10.3 & 90.9 & 87.1$^{\dagger}$ & 74.9 \\
\textbf{Full Workflow} & \textbf{100.0$^{*}$} & \textbf{93.6$^{*}$} & \textbf{100.0$^{*}$} & \textbf{100.0$^{*}$} & \textbf{100.0$^{*}$} & \textbf{99.0$^{*}$} \\
\hline
\multicolumn{7}{|c|}{\textbf{ALAS (Gemini-2.5 Backbone)}} \\
\hline
No Repair       & 83.6$^{\dagger}$ & 100.0$^{*}$ & 98.5 & 95.5$^{*}$ & 75.3 & 90.6$^{*}$ \\
No Validation   & 83.9$^{\dagger}$ & 100.0$^{*}$ & 63.0 & 96.9$^{\dagger}$ & 75.3 & 83.8$^{\dagger}$ \\
\textbf{Full Workflow} & \textbf{97.8$^{*}$} & \textbf{100.0$^{*}$} & \textbf{100.0$^{*}$} & \textbf{96.8$^{*}$} & \textbf{100.0$^{*}$} & \textbf{98.2$^{*}$} \\
\hline
\end{tabular}
\\[0.5em]
\small $^{\dagger}p<0.05$, $^{*}p<0.01$ (paired t-test, compared against baseline).
\end{table*}




%% file: Figure-JSSP-Disruptive-Phase1.tex
\begin{figure}[t!]
\vspace{-.1in}
\begin{subfigure}{0.47\textwidth}
\centering
\begin{tikzpicture}[yscale=0.32, xscale=0.23, font=\tikzfontsize] 
\foreach \y in {0,1,2} \draw (0,-\y) -- (26,-\y);
\node[left] at (0,0) {$M_0$};
\node[left] at (0,-1) {$M_1$};
\node[left] at (0,-2) {$M_2$};
\foreach \x in {0,2,...,24} {
  \draw (\x,0.2) -- (\x,-2.8);
  \node[above] at (\x,0.2) {\x};
}
\filldraw[fill=red!20] (0,0) rectangle (2,-0.8) node[midway] {J2};
\filldraw[fill=orange!20] (2,0) rectangle (3,-0.8) node[midway] {J4};
\filldraw[fill=blue!20] (3,0) rectangle (6,-0.8) node[midway] {J1(1)};
\filldraw[fill=purple!20] (6,0) rectangle (10,-0.8) node[midway] {J5(2)};
\filldraw[fill=green!20] (17,0) rectangle (19,-0.8) node[midway] {J3(3)};
\filldraw[fill=purple!20] (0,-1) rectangle (2,-1.8) node[midway] {J5};
\filldraw[fill=red!20] (3,-1) rectangle (7,-1.8) node[midway] {J2(3)};
\filldraw[fill=orange!20] (7,-1) rectangle (10,-1.8) node[midway] {J4(3)};
\filldraw[fill=green!20] (10,-1) rectangle (14,-1.8) node[midway] {J3(1)};
\filldraw[fill=blue!20] (14,-1) rectangle (16,-1.8) node[midway] {J1(2)};
\filldraw[fill=orange!20] (0,-2) rectangle (2,-2.8) node[midway] {J4};
\filldraw[fill=red!20] (2,-2) rectangle (3,-2.8) node[midway] {J2};
\filldraw[fill=purple!20] (10,-2) rectangle (11,-2.8) node[midway] {J5(3)};
\filldraw[fill=green!20] (14,-2) rectangle (17,-2.8) node[midway] {J3(2)};
\filldraw[fill=blue!20] (17,-2) rectangle (19,-2.8) node[midway] {J1(3)};
\end{tikzpicture}
\vspace{-.15in}
\caption{Static schedule before disruption, makespan = 19}
\label{fig:static-plan}
\end{subfigure}
\hspace{0.01\textwidth}
\begin{subfigure}{0.47\textwidth}
\centering
\begin{tikzpicture}[yscale=0.33, xscale=0.23, font=\tikzfontsize] 
\foreach \y in {0,1,2} {
  \draw (0,-\y) -- (26,-\y);
}
\node[left] at (0,0) {$M_0$};
\node[left] at (0,-1) {$M_1$};
\node[left] at (0,-2) {$M_2$};

\foreach \x in {0,2,...,24} {
  \draw (\x,0.2) -- (\x,-2.8);
  \node[above] at (\x,0.2) {\x};
}

\filldraw[fill=red!20] (0,0) rectangle (2,-0.8) node[midway] {J2};
\filldraw[fill=orange!20] (2,0) rectangle (3,-0.8) node[midway] {J4};
\filldraw[fill=blue!20] (3,0) rectangle (6,-0.8) node[midway] {J1(1)};
\filldraw[fill=purple!20] (6,0) rectangle (10,-0.8) node[midway] {J5};
\filldraw[fill=green!20] (17,0) rectangle (19,-0.8) node[midway] {J3(3)};

\filldraw[fill=purple!20] (0,-1) rectangle (2,-1.8) node[midway] {J5};
\filldraw[fill=red!20] (3,-1) rectangle (5,-1.8) node[midway] {{\color{red}{J2}}};
\filldraw[pattern=north east lines, pattern color=gray] (5,-1) rectangle (8,-1.8);
\node at (6.5,-1.4) {\scriptsize {\color{red}\textbf{down}}};
\filldraw[fill=red!20] (8,-1) rectangle (12,-1.8) node[midway] {J2(3)};
\filldraw[fill=orange!20] (12,-1) rectangle (15,-1.8) node[midway] {J4(3)};
\filldraw[fill=green!20] (15,-1) rectangle (19,-1.8) node[midway] {J3(1)};
\filldraw[fill=blue!20] (19,-1) rectangle (21,-1.8) node[midway] {J1(2)};

\filldraw[fill=orange!20] (0,-2) rectangle (2,-2.8) node[midway] {J4};
\filldraw[fill=red!20] (2,-2) rectangle (3,-2.8) node[midway] {J2};
\filldraw[fill=purple!20] (10,-2) rectangle (11,-2.8) node[midway] {J5(3)};
\filldraw[fill=green!20] (14,-2) rectangle (17,-2.8) node[midway] {J3(2)};
\filldraw[fill=blue!20] (17,-2) rectangle (19,-2.8) node[midway] {J1(3)};
\end{tikzpicture}
\vspace{-.15in}
\caption{Repaire schedule after M1 failure at $t=5$}
\label{fig:maple-repair}
\end{subfigure}

\vspace{0.05in} 

\begin{subfigure}{0.47\textwidth}
\centering
\begin{tikzpicture}[yscale=0.33, xscale=0.23, font=\tikzfontsize] 
\foreach \y in {0,1,2} \draw (0,-\y) -- (26,-\y);
\node[left] at (0,0) {$M_0$};
\node[left] at (0,-1) {$M_1$};
\node[left] at (0,-2) {$M_2$};
\foreach \x in {0,2,...,24} {
  \draw (\x,0.2) -- (\x,-2.8);
  \node[above] at (\x,0.2) {\x};
}
\filldraw[fill=red!20] (0,0) rectangle (2,-0.8) node[midway] {J2};
\filldraw[fill=orange!20] (2,0) rectangle (3,-0.8) node[midway] {J4};
\filldraw[fill=blue!20] (3,0) rectangle (6,-0.8) node[midway] {J1(1)};
\filldraw[fill=purple!20] (6,0) rectangle (10,-0.8) node[midway] {J5(2)};
\filldraw[fill=green!20] (17,0) rectangle (19,-0.8) node[midway] {J3(3)};
\filldraw[pattern=north east lines, pattern color=gray!40] (5,-1) rectangle (8,-1.8);
\filldraw[fill=purple!20] (0,-1) rectangle (2,-1.8) node[midway] {J5};
\filldraw[fill=red!20] (8,-1) rectangle (12,-1.8) node[midway] {J2(3)};
\filldraw[fill=orange!20] (12,-1) rectangle (15,-1.8) node[midway] {J4(3)};
\filldraw[fill=green!20] (15,-1) rectangle (19,-1.8) node[midway] {J3(1)};
\filldraw[fill=blue!20] (19,-1) rectangle (21,-1.8) node[midway] {J1(2)};
\filldraw[fill=orange!20] (0,-2) rectangle (2,-2.8) node[midway] {J4};
\filldraw[fill=red!20] (2,-2) rectangle (3,-2.8) node[midway] {J2};
\filldraw[fill=purple!20] (10,-2) rectangle (11,-2.8) node[midway] {J5(3)};
\filldraw[fill=green!20] (19,-2) rectangle (22,-2.8) node[midway] {J3(2)};
\filldraw[fill=blue!20] (22,-2) rectangle (24,-2.8) node[midway] {J1(3)};
\end{tikzpicture}
\vspace{-.15in}
\caption{Schedule after delay notice for J3(2) on M2}
\label{fig:delay-notification}
\end{subfigure}
\hspace{0.01\textwidth}
\begin{subfigure}{0.47\textwidth}
\centering
\begin{tikzpicture}[yscale=0.33, xscale=0.23, font=\tikzfontsize] 
\foreach \y in {0,1,2} \draw (0,-\y) -- (26,-\y);
\node[left] at (0,0) {$M_0$};
\node[left] at (0,-1) {$M_1$};
\node[left] at (0,-2) {$M_2$};
\foreach \x in {0,2,...,24} {
  \draw (\x,0.2) -- (\x,-2.8);
  \node[above] at (\x,0.2) {\x};
}
\filldraw[fill=red!20] (0,0) rectangle (2,-0.8) node[midway] {J2};
\filldraw[fill=orange!20] (2,0) rectangle (3,-0.8) node[midway] {J4};
\filldraw[fill=blue!20] (3,0) rectangle (6,-0.8) node[midway] {J1(1)};
\filldraw[fill=purple!20] (6,0) rectangle (10,-0.8) node[midway] {J5(2)};
\filldraw[fill=green!20] (22,0) rectangle (26,-0.8) node[midway] {J3(3)};
\filldraw[pattern=north east lines, pattern color=gray] (5,-1) rectangle (8,-1.8);
\filldraw[fill=purple!20] (0,-1) rectangle (2,-1.8) node[midway] {J5};
\filldraw[fill=red!20] (8,-1) rectangle (12,-1.8) node[midway] {J2(3)};
\filldraw[fill=orange!20] (12,-1) rectangle (15,-1.8) node[midway] {J4(3)};
\filldraw[fill=green!20] (15,-1) rectangle (19,-1.8) node[midway] {J3(1)};
\filldraw[fill=blue!20] (19,-1) rectangle (21,-1.8) node[midway] {J1(2)};
\filldraw[fill=orange!20] (0,-2) rectangle (2,-2.8) node[midway] {J4};
\filldraw[fill=red!20] (2,-2) rectangle (3,-2.8) node[midway] {J2};
\filldraw[fill=purple!20] (10,-2) rectangle (11,-2.8) node[midway] {J5(3)};
\filldraw[fill=green!20] (19,-2) rectangle (22,-2.8) node[midway] {J3(2)};
\filldraw[fill=blue!20] (22,-2) rectangle (24,-2.8) node[midway] {J1(3)};
\end{tikzpicture}
\vspace{-.15in}
\caption{Schedule after delay notice for J3(3) on M0}
\label{fig:final-repair}
\end{subfigure}
\caption{LRCP Phase \#1 Local Compensation (makespan = 22):
(a) Static baseline schedule; 
(b) $M_1$ failure between $t = 5$–$8$; 
(c) $M_1$ notifies $M_2$ to delay $J3(2)$; 
(d) $M_2$ informs $M_0$ to push $J3(3)$ back.}
\label{fig:combined-scheduling}
\vspace{-.05in}
\end{figure}
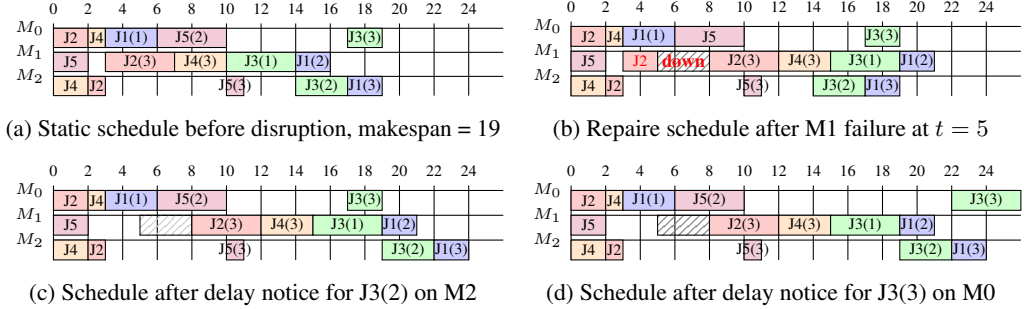

%% file: Figure-JSSP-Disruptive-Phase2.tex
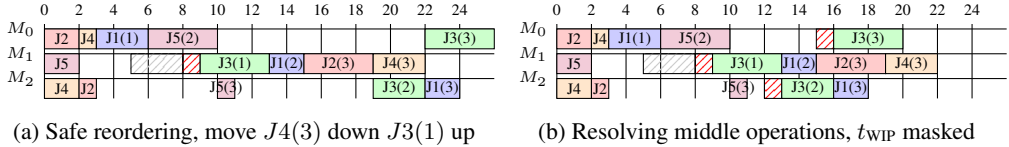
\begin{figure}[t!]
\vspace{-.1in}
\begin{subfigure}{0.47\textwidth}  
\centering
\begin{tikzpicture}[yscale=0.33, xscale=0.23, font=\tikzfontsize]  
\foreach \y in {0,1,2} \draw (0,-\y) -- (26,-\y);
\node[left] at (0,0) {$M_0$};
\node[left] at (0,-1) {$M_1$};
\node[left] at (0,-2) {$M_2$};
\foreach \x in {0,2,...,24} {
  \draw (\x,0.2) -- (\x,-2.8);
  \node[above] at (\x,0.2) {\x};
}
\filldraw[fill=red!20] (0,0) rectangle (2,-0.8) node[midway] {J2};
\filldraw[fill=orange!20] (2,0) rectangle (3,-0.8) node[midway] {J4};
\filldraw[fill=blue!20] (3,0) rectangle (6,-0.8) node[midway] {J1(1)};
\filldraw[fill=purple!20] (6,0) rectangle (10,-0.8) node[midway] {J5(2)};
\filldraw[fill=green!20] (22,0) rectangle (26,-0.8) node[midway] {J3(3)};

\filldraw[pattern=north east lines, pattern color=gray!40] (5,-1) rectangle (8,-1.8);
\filldraw[pattern=north east lines, pattern color=red!90] (8,-1) rectangle (9,-1.8);
\filldraw[fill=purple!20] (0,-1) rectangle (2,-1.8) node[midway] {J5};
\filldraw[fill=green!20] (9,-1) rectangle (13,-1.8) node[midway] {J3(1)};
\filldraw[fill=blue!20] (13,-1) rectangle (15,-1.8) node[midway] {J1(2)};
\filldraw[fill=red!20] (15,-1) rectangle (19,-1.8) node[midway] {J2(3)};
\filldraw[fill=orange!20] (19,-1) rectangle (22,-1.8) node[midway] {J4(3)};
\filldraw[fill=orange!20] (0,-2) rectangle (2,-2.8) node[midway] {J4};
\filldraw[fill=red!20] (2,-2) rectangle (3,-2.8) node[midway] {J2};
\filldraw[fill=purple!20] (10,-2) rectangle (11,-2.8) node[midway] {J5(3)};
\filldraw[fill=green!20] (19,-2) rectangle (22,-2.8) node[midway] {J3(2)};
\filldraw[fill=blue!20] (22,-2) rectangle (24,-2.8) node[midway] {J1(3)};
\end{tikzpicture}
\vspace{-.15in}
\caption{Safe reordering, move $J4(3)$ down $J3(1)$ up}
\label{fig:queue-reorder-1}
\end{subfigure}
\hspace{0.005\textwidth}  
\begin{subfigure}{0.47\textwidth}  
\centering
\begin{tikzpicture}[yscale=0.33, xscale=0.23, font=\tikzfontsize]  
\foreach \y in {0,1,2} \draw (0,-\y) -- (26,-\y);
\node[left] at (0,0) {$M_0$};
\node[left] at (0,-1) {$M_1$};
\node[left] at (0,-2) {$M_2$};
\foreach \x in {0,2,...,24} {
  \draw (\x,0.2) -- (\x,-2.8);
  \node[above] at (\x,0.2) {\x};
}
\filldraw[pattern=north east lines, pattern color=red!90] (15,-0.8) rectangle (16,0);
\filldraw[fill=red!20] (0,0) rectangle (2,-0.8) node[midway] {J2};
\filldraw[fill=orange!20] (2,0) rectangle (3,-0.8) node[midway] {J4};
\filldraw[fill=blue!20] (3,0) rectangle (6,-0.8) node[midway] {J1(1)};
\filldraw[fill=purple!20] (6,0) rectangle (10,-0.8) node[midway] {J5(2)};
\filldraw[fill=green!20] (16,0) rectangle (20,-0.8) node[midway] {J3(3)};

\filldraw[pattern=north east lines, pattern color=gray!40] (5,-1) rectangle (8,-1.8);
\filldraw[pattern=north east lines, pattern color=red!90] (8,-1) rectangle (9,-1.8);
\filldraw[fill=purple!20] (0,-1) rectangle (2,-1.8) node[midway] {J5};
\filldraw[fill=green!20] (9,-1) rectangle (13,-1.8) node[midway] {J3(1)};
\filldraw[fill=blue!20] (13,-1) rectangle (15,-1.8) node[midway] {J1(2)};
\filldraw[fill=red!20] (15,-1) rectangle (19,-1.8) node[midway] {J2(3)};
\filldraw[fill=orange!20] (19,-1) rectangle (22,-1.8) node[midway] {J4(3)};
\filldraw[pattern=north east lines, pattern color=red!90] (12,-2) rectangle (13,-2.8);
\filldraw[fill=orange!20] (0,-2) rectangle (2,-2.8) node[midway] {J4};
\filldraw[fill=red!20] (2,-2) rectangle (3,-2.8) node[midway] {J2};
\filldraw[fill=purple!20] (10,-2) rectangle (11,-2.8) node[midway] {J5(3)};
\filldraw[fill=green!20] (13,-2) rectangle (16,-2.8) node[midway] {J3(2)};
\filldraw[fill=blue!20] (16,-2) rectangle (18,-2.8) node[midway] {J1(3)};
\end{tikzpicture}
\vspace{-.15in}
\caption{Resolving middle operations, $t_{\text{WIP}}$ masked}
\label{fig:queue-reorder-2}
\end{subfigure}
\caption{LRSP Phase \#2 Queue Reordering (makespan = 22): 
(a) Safe moves: moving last operations down, first operations forward with potential penalty;
(b) Resolving remaining operations.}
\label{fig:queue-reordering}
\vspace{-.15in}
\end{figure}

%% file: Conclusion.tex
\vspace{-.1in}
\section{Conclusion}
\label{sec:alas-conclusion}
\vspace{-.1in}
\noindent
We presented $\ALAS$, a framework for reliable multi-agent LLM planning that addresses core limitations of standalone LLMs—lack of self-verification, long-context degradation, and stateless execution—via three design principles: (i) \emph{validator isolation} (planners never self-approve; checks use fresh, bounded prompts), (ii) a \emph{versioned execution log} that records state transitions and causal links, and (iii) a \emph{localized cascading repair protocol} (LCRP) that confines disruptions to minimal neighborhoods instead of triggering brittle global recomputation. Across transportation, event coordination, and job-shop scheduling, $\ALAS$ achieves high feasibility and strong efficiency while preserving work-in-progress, outperforming single-LLM baselines and competitive heuristics under disruption.

Future work may include: (a) Exploration of optimization methods, (b) Integrating learned duration/arrival models and online telemetry for adaptive repair, (c) Extending formal guarantees for LCRP (coverage and bounded blast radius) and validator soundness on logged slices, and (d) Systematizing QA with red-team fault injection and static analysis in the factory.


%% file: AppendixALASDetails.tex
\appendix
\section{Further Details of ALAS Specification}
\label{app:alas-spec}

\subsection{IR JSON Schema (complete)}
\label{app:ir-json}

\noindent The full JSON Schema for the ALAS IR. This extends the sketch in the main text with concrete definitions for \texttt{Policy}, \texttt{LogSchema}, and \texttt{LogEntry}.

\begin{lstlisting}[language=json,basicstyle=\ttfamily\footnotesize,caption={ALAS IR JSON Schema}]
{
  "$schema": "http://json-schema.org/draft-07/schema#",
  "title": "ALAS IR",
  "type": "object",
  "required": ["Workflow","definitions"],
  "properties": {
    "Workflow": {
      "type": "object",
      "required": ["name","nodes","edges","policies","schemas","logSchemas","meta"],
      "properties": {
        "name": {"type":"string"},
        "nodes": {"type":"array","items":{"$ref":"#/definitions/Node"}},
        "edges": {"type":"array","items":{"$ref":"#/definitions/Edge"}},
        "policies": {"$ref":"#/definitions/Policies"},
        "logSchemas": {"type":"object", "additionalProperties": { "$ref":"#/definitions/LogSchema"}},
        "meta": {"type":"object"}
      }
    }
  },
  "definitions": {
    "Node": {
      "type": "object",
      "required": ["id","type","name","io"],
      "properties": {
        "id": {"type":"string"},
        "type": {"enum": ["task","choice","parallel","map"]},
        "name": {"type":"string"},
        "capabilities": {"type":"object"},
        "params": {"type":"object"},
        "io": {"$ref":"#/definitions/IO"},
        "policyRef": {"type":"string"},
        "logSchemaRef": {"type":"string"}
      }
    },
    "Edge": {
      "type": "object",
      "required": ["from","to","kind"],
      "properties": {
        "from": {"type":"string"},
        "to": {"type":"string"},
        "kind": {"enum": ["sequence","branch","default"]},
        "guard": {"type":"string"},
        "repairable": {"enum": ["time","order","resource","none"]}
      }
    },
    "IO": {
      "type":"object",
      "properties": {
        "input":{"type":"object"},
        "output":{"type":"object"},
        "error":{"type":"object"}
      },
      "additionalProperties": true
    },

    "Policies": {
      "type":"object",
      "description":"Named policy objects referenced by nodes",
      "patternProperties": {
        "^[A-Za-z0-9_.-]+$": { "$ref": "#/definitions/Policy" }
      },
      "additionalProperties": false
    },

    "Policy": {
      "type":"object",
      "properties": {
        "retry": { "$ref":"#/definitions/Retry" },
        "backoff": { "$ref":"#/definitions/Backoff" },
        "catch": { "type":"array", "items": { "$ref":"#/definitions/Catch" } },
        "timeout": { "$ref":"#/definitions/Timeout" },
        "idempotencyKey": { "$ref":"#/definitions/IdempotencyKey" },
        "compensation": { "$ref":"#/definitions/Compensation" },
        "loopGuards": { "$ref":"#/definitions/LoopGuards" }
      },
      "additionalProperties": false
    }

    "Retry": {
      "type":"object",
      "properties": {
        "maxAttempts": {"type":"integer","minimum":0},
        "retryOn": {"type":"array","items":{"type":"string"}}
      },
      "required": ["maxAttempts"],
      "additionalProperties": false
    }

    "Backoff": {
      "type":"object",
      "properties": {
        "mode": {"enum": ["fixed","exponential"]},
        "base": {"type":"number","minimum":0},
        "cap": {"type":"number","minimum":0},
        "jitter": {"type":"number","minimum":0}
      },
      "required": ["mode"],
      "additionalProperties": false
    }

    "Catch": {
      "type":"object",
      "properties": {
        "errors": {"type":"array","items":{"type":"string"}},
        "handler": {"type":"string"}     // node name or handler id
      },
      "required": ["errors","handler"]
    },

    "Timeout": {
      "type":"object",
      "properties": {
        "seconds": {"type":"number","minimum":0}
      },
      "required": ["seconds"]
    },

    "IdempotencyKey": {
      "type":"object",
      "properties": {
        "path": {"type":"string"},
        "template": {"type":"string"},
        "scope": {"enum":["workflow","node","resource"]}
      },
      "anyOf": [
        { "required": ["path"] },
        { "required": ["template"] }
      ],
      "additionalProperties": false
    }

    "Compensation": {
      "type":"object",
      "properties": {
        "handler": {"type":"string"},    // node name or cleanup template
        "trigger": {"enum":["onFailure","onCancel","manual"]},
        "safeReinvoke": {"type":"boolean"}
      },
      "required": ["handler","safeReinvoke"]
    },

    "LoopGuards": {
      "type":"object",
      "properties": {
        "maxIters": {"type":"integer","minimum":1},
        "deadlineSeconds": {"type":"number","minimum":0},
        "timeBudget": {"type":"number","minimum":0}
      },
      "additionalProperties": false
    }

    "LogSchema": {
      "type":"object",
      "description":"A JSON Schema fragment describing required fields for a node's log entries",
      "properties": {
        "required": { "type":"array", "items":{"type":"string"} },
        "properties": { "type":"object" }
      },
      "additionalProperties": true
    }

    "LogEntry": {
      "type":"object",
      "required": ["ts","nodeId","eventType","version","correlationId"],
      "properties": {
        "ts": {"type":"string","format":"date-time"},
        "nodeId": {"type":"string"},
        "eventType": {
          "enum":[
            "StartNode","EndNode",
            "ValidatePass","ValidateFail",
            "RepairStart","RepairApply","RepairCommit",
            "Retry","Catch","Compensate","Timeout"
          ]
        },
        "payload": {"type":"object"},
        "version": {"type":"integer","minimum":0},
        "correlationId": {"type":"string"}
      }
    }
  }
}
\end{lstlisting}

\subsection{Policy set \texorpdfstring{\(\Pi\)}{Pi} and \texttt{policyRef}}\label{app:policy-set}

\paragraph{Policy set \(\Pi\) and \texttt{policyRef}.}
Each workflow defines a policy set \(\Pi\) and every node \(n_i\) attaches one of these policies via \texttt{policyRef}.
Policies make retry, catch, timeout, backoff, idempotency keys, compensation, and loop guards explicit and auditable.
Notation in text uses \texttt{CamelCase} for readability; JSON keys use lower case (\texttt{retry}, \texttt{catch}, \texttt{timeout}, \texttt{backoff}, \texttt{idempotencyKey}, \texttt{compensation}, \texttt{loopGuards}).

\medskip
\noindent\textbf{Definition.}
A policy \(p \in \Pi\) is a tuple
\[
p=\langle \texttt{Retry},\, \texttt{Catch},\, \texttt{Timeout},\, \texttt{Backoff},\, \texttt{IdempotencyKey},\, \texttt{Compensation},\, \texttt{LoopGuards} \rangle.
\]
Node \(n_i\) references \(p\) via \texttt{policyRef}.

\medskip
\noindent\textbf{Fields, meaning, and why essential.}
\begin{itemize}[leftmargin=1.2em, itemsep=2pt]
  \item \texttt{Retry}: parameters that govern automatic re-execution on transient faults
        (keys: \texttt{maxAttempts}, \texttt{retryOn}, \texttt{jitter}).
        \emph{Why essential}: avoids failing the plan on intermittent issues and makes recovery behavior deterministic.
  \item \texttt{Backoff}: schedule for retry delays
        (keys: \texttt{mode} in \{\texttt{fixed}, \texttt{exponential}\}, \texttt{base}, \texttt{cap}).
        \emph{Why essential}: prevents retry storms and gives predictable latency envelopes recorded in the log.
  \item \texttt{Catch}: routing on unhandled errors
        (keys: \texttt{onError} pattern, \texttt{handlerNode} or \texttt{route}).
        \emph{Why essential}: keeps failures contained and observable, enabling validator decisions and safe fallbacks.
  \item \texttt{Timeout}: execution time limit
        (keys: \texttt{seconds} or \texttt{deadline}).
        \emph{Why essential}: bounds tail latency and unblocks dependent nodes; all timeouts are logged for audit.
  \item \texttt{IdempotencyKey}: key derivation rule for a logically identical action
        (keys: \texttt{template} or function over inputs).
        \emph{Why essential}: guarantees at-least-once execution does not produce duplicate effects; validator can replay from logs safely.
  \item \texttt{Compensation}: corrective action for a node’s side effects
        (keys: \texttt{handlerNode}, \texttt{safeReinvoke} boolean).
        \emph{Why essential}: enables rollback or cleanup during repair; safety requires idempotent design or guarding by the idempotency key.
  \item \texttt{LoopGuards}: termination and bound controls for iterative patterns
        (keys: \texttt{maxIters}, \texttt{timeBudget}, \texttt{predicate}).
        \emph{Why essential}: ensures termination for \texttt{map} or repair loops; supports the complexity and termination claims of \(\ALAS\).
\end{itemize}

\medskip
\noindent\textbf{Validator checks and invariants.}
Let \(V\) be the independent validator.
\begin{enumerate}[leftmargin=1.4em, itemsep=2pt]
  \item Every node has a \texttt{policyRef}; if omitted, a workflow default \(p_0 \in \Pi\) is applied.
  \item If \texttt{Retry} is present then \texttt{Backoff} is specified and \texttt{maxAttempts} is finite.
  \item \texttt{Timeout} is finite for all task nodes that call external tools.
  \item If \texttt{Compensation} is present then either the effect is idempotent or \texttt{IdempotencyKey} is defined.
  \item For iterative constructs, \texttt{LoopGuards.maxIters} or \texttt{LoopGuards.timeBudget} is finite.
  \item Idempotency keys are unique per logical action: \(k = f(\text{nodeId}, \text{inputs}, \text{runId})\) and collisions across distinct effects are rejected.
\end{enumerate}

\medskip
\noindent\textbf{Execution log contract.}
Each policy decision yields a \texttt{LogEntry} with \texttt{eventType} in
\{\texttt{Retry}, \texttt{Catch}, \texttt{Compensate}, \texttt{Timeout}\}, the evaluated parameters, and the outcome.
This supports replay, audit, and cross-engine parity tests.

\medskip
\noindent\textbf{Minimal JSON example.}
\begin{lstlisting}[language=json,basicstyle=\ttfamily\footnotesize]
{
  "policies": {
    "p_default": {
      "retry": {"maxAttempts": 3, "retryOn": ["Timeout","NetworkError"]},
      "backoff": {"mode": "exponential", "base": 0.5, "cap": 8.0},
      "timeout": {"seconds": 30},
      "idempotencyKey": {"template": "node:{nodeId}|hash:{input.hash}"},
      "compensation": {"handler": "cleanup_task", "safeReinvoke": true},
      "loopGuards": {"maxIters": 5, "timeBudget": 10.0}
    }
  },
  "nodes": [
    {"id":"validate_plan","type":"task","policyRef":"p_default"},
    {"id":"repair_local","type":"task","policyRef":"p_default"}
  ]
}
\end{lstlisting}

\medskip
\noindent\textbf{Why this is essential to \(\ALAS\).}
\(\ALAS\) pursues planning reliability through validator isolation, versioned execution logs, and localized repair.
Policies in \(\Pi\) make these behaviors explicit and enforceable:
\texttt{Retry} and \texttt{Backoff} stabilize transient faults,
\texttt{Timeout} and \texttt{Catch} keep control flow safe and auditable,
\texttt{IdempotencyKey} and \texttt{Compensation} enable safe recovery and replay,
\texttt{LoopGuards} guarantee termination for iterative agents and repair loops.
Together with \texttt{policyRef} at each node, \(\Pi\) provides a portable contract that maps cleanly to ASL and Argo while preserving parity through logs.

\subsection{Worked execution loop}
\label{app:execution-loop}

\begin{table}[t]
\caption{Reference 7-step execution loop in $\ALAS$}
\label{tab:alas-exec-loop}
\small
\begin{tabular}{p{0.30\linewidth} p{0.64\linewidth}}
\toprule
\textbf{Step} & \textbf{Action} \\
\midrule
Inputs and goal &
Require dataset specification $(J,M,C)$; goal: feasible schedule $\mathcal{S}^{\ast}$ with makespan $T^{\ast}$. \\

1 Plan proposal &
$\mathcal{S}_0 \gets \texttt{generate\_schedule}(J,M,C)$; append \texttt{StartNode} log. \\

2 Validate (non-circular) &
If $\texttt{validate}(\mathcal{S}_0)$ is \texttt{true}: set $\mathcal{S}\gets\mathcal{S}_0$ and go to Step 5. Otherwise go to Step 3. \\

3 Local repair &
For $k=1..K$: $\mathcal{S}_k \gets \texttt{repair}(\mathcal{S}_{k-1})$; log \texttt{RepairApply}. If $\texttt{validate}(\mathcal{S}_k)$ is \texttt{true}: set $\mathcal{S}\gets\mathcal{S}_k$ and break. \\

4 Revalidate &
If $\texttt{validate}(\mathcal{S})$ is \texttt{false}: skip optimization and go to Step 7. \\

5 Optimize &
$\mathcal{S}_{\text{opt}} \gets \texttt{optimize}(\mathcal{S})$. \\

6 Final check &
If $\texttt{validate}(\mathcal{S}_{\text{opt}})$ is \texttt{true}: $\mathcal{S}^{\ast}\gets \mathcal{S}_{\text{opt}}$; else $\mathcal{S}^{\ast}\gets \mathcal{S}$. \\

7 Supervise and log &
\texttt{log}($\mathcal{S}^{\ast}$, \texttt{makespan}($\mathcal{S}^{\ast}$)); return $\mathcal{S}^{\ast}$. \\
\bottomrule
\end{tabular}
\end{table}

\subsection{Engine mappings and round-trip rules}
\label{app:mappings}

\noindent Formal mapping of IR concepts to ASL and Argo, with exact field correspondences, guard translation, and supported fallbacks. Round-trip tests verify IR \(\to\) target \(\to\) IR preservation for type, ordering, and policy fields.

\begin{table}[ht]
\centering
\caption{IR to execution engine mapping (expanded)}
\footnotesize
\begin{tabular}{lcc}
\toprule
IR concept & ASL & Argo \\
\midrule
Task node & \texttt{Type: Task} & container or script template \\
Choice node & \texttt{Type: Choice} & DAG edge with \texttt{when} \\
Parallel or Map & \texttt{Parallel/Map} & fan out with \texttt{withItems} \\
Retry & \texttt{Retry} block & \texttt{retryStrategy} or hooks \\
Catch & \texttt{Catch} block & \texttt{onExit}/error hooks \\
Timeout & \texttt{TimeoutSeconds} & template or container timeout \\
IdempotencyKey & state input field & param or artifact key \\
Compensation & handler Task or Fail route & cleanup template or hook \\
Data passing & JSONPath & artifacts or parameters \\
\bottomrule
\end{tabular}
\end{table}

\subsection{Conformance matrix}
\label{app:conformance}

\begin{table}[ht]
\centering
\caption{Engine conformance}
\footnotesize
\begin{tabular}{lccc}
\toprule
Feature & IR & ASL & Argo \\
\midrule
Task, Choice, Parallel, Map & \checkmark & \checkmark & \checkmark \\
Retry, Catch, Timeout       & \checkmark & \checkmark & \checkmark \\
Compensation                & \checkmark & $\circ$\footnotemark[1] & \checkmark \\
IdempotencyKey              & \checkmark & $\circ$\footnotemark[1] & $\circ$\footnotemark[1] \\
Guards on branches          & \checkmark & \checkmark & \checkmark (via \texttt{when}) \\
Round-trip preservation     & core       & core       & core \\
\bottomrule
\end{tabular}
\end{table}
\footnotetext[1]{Use handler task plus routing in ASL, and state input or external store for idempotency in both engines.}

\subsection{Policy semantics and idioms}
\label{app:policy-semantics}

\paragraph{Idempotency.} A key \(k\) identifies a task execution. Before run: if \(k\) found with SUCCESS, return cached result; if PENDING, wait; else execute and record \((k, \text{result})\).  
\paragraph{Compensation.} Trigger on failure or cancel as configured. Handlers must be safe to re-invoke when the idempotency key is preserved.  
\paragraph{Loop guards.} Bound loops by \texttt{maxIters} and optional \texttt{deadlineSeconds}.

\subsection{Typed API reference}
\label{app:api}

\begin{lstlisting}[language=Python,basicstyle=\ttfamily\footnotesize]
class ValidationResult(TypedDict):
    valid: bool
    errors: List["ConstraintViolation"]
    warnings: List[str]

class ConstraintViolation(TypedDict):
    code: str
    message: str
    scope: Dict[str, Any]  # job, op, machine, edge id

def validate(plan: Dict[str, Any]) -> ValidationResult: ...
def repair(plan, errors, budget) -> Dict[str, Any]: ...
\end{lstlisting}

\subsection{Logging and observability spec}
\label{app:logging}

Event taxonomy, correlation model (\texttt{version}, \texttt{correlationId}), redaction tags, and example entries for \texttt{ValidateFail}, \texttt{RepairApply}, and \texttt{Compensate}.

\subsection{Determinism and replay harness}
\label{app:determinism}

Seed and tool pinning, exact replay from \texttt{LogEntry} payloads, parity checks across engines, and drift triage.

\subsection{Fault model and injection}
\label{app:faults}

Transient faults (timeouts, tool failures) and logical faults (constraint violations) are in scope. Resource exhaustion and Byzantine behaviors are out of scope in this version. Injection API supports labeled timeouts, failures, and targeted violations.

\subsection{LCRP pseudocode and guarantees}
\label{app:lcrp}

Neighborhood growth rules, termination under bounded retries and guards, and a short proof sketch.

\subsection{Worked examples and diagrams}
\label{app:examples}

End-to-end IR examples and their ASL or Argo materializations with round-trip checks.

\subsection{Experimental protocols}
\label{app:protocols}

Benchmark setup, disruption generators, hyperparameters, and evaluation harness details.

%% file: AppendixJSSPSupplement.tex
\section{Additional JSSP Results and Analysis}
\label{app:ALAS-JSSP}
This appendix augments our core experimental findings with the full prompt specification, failure rate statistics, and pointers to supplementary visualizations. 

\vspace{-.1in}
\subsection{Datasets}

Figure~\ref{fig:gantt_charts} shows visualization of datasets.
\newcommand{\ganttHeightScale}{0.88}  

\begin{figure}[htbp]
    \centering
    \begin{subfigure}[b]{0.49\textwidth}
        \centering
        \includegraphics[width=\textwidth,height=\ganttHeightScale\textwidth]{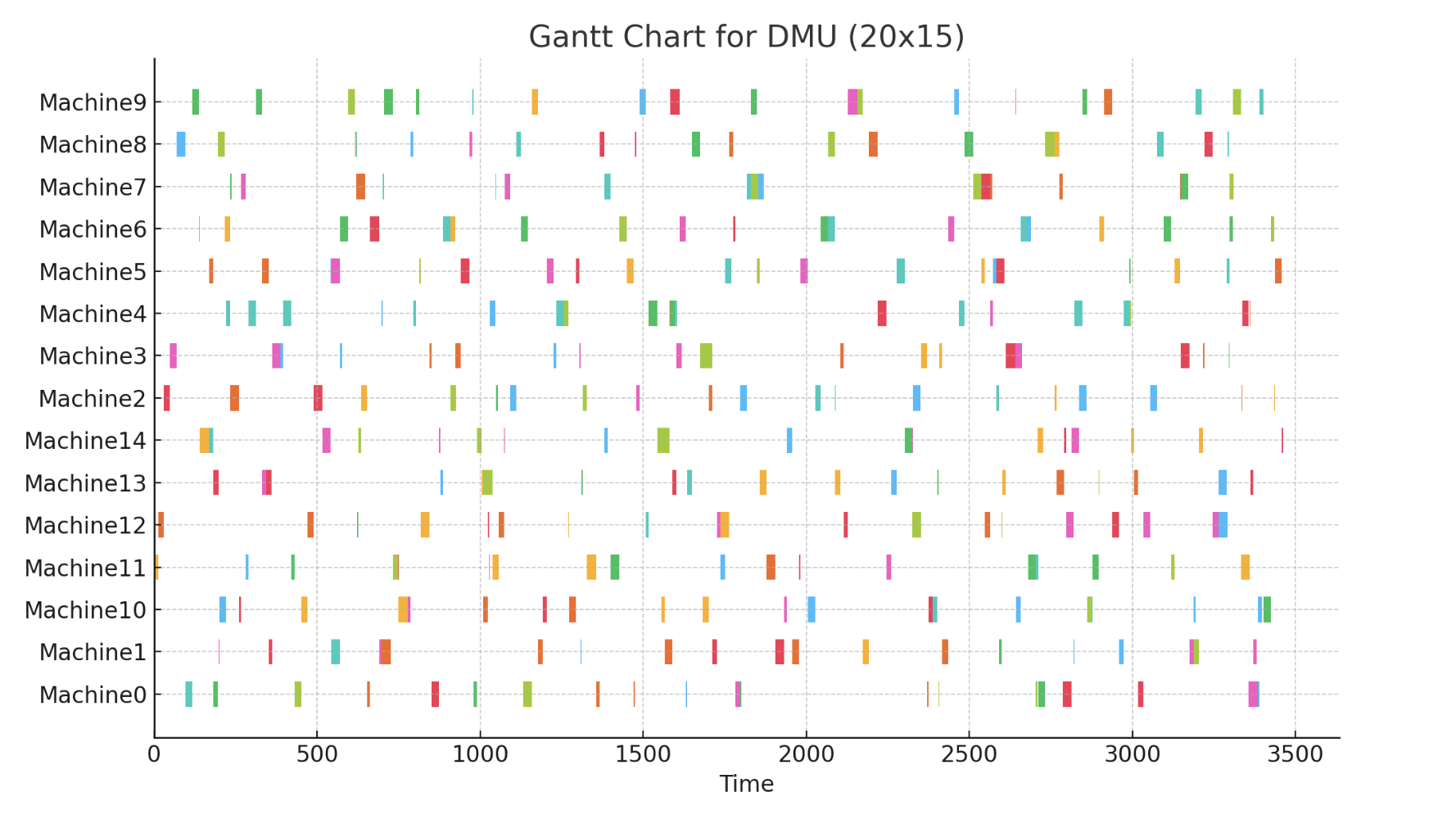}
        \caption{\textbf{rcmax\_20\_15\_5} (J=20, M=15)}
        \label{fig:gantt_rcmax}
    \end{subfigure}
    \hspace{-.2in}
    \begin{subfigure}[b]{0.49\textwidth}
        \centering
        \includegraphics[width=\textwidth,height=\ganttHeightScale\textwidth]{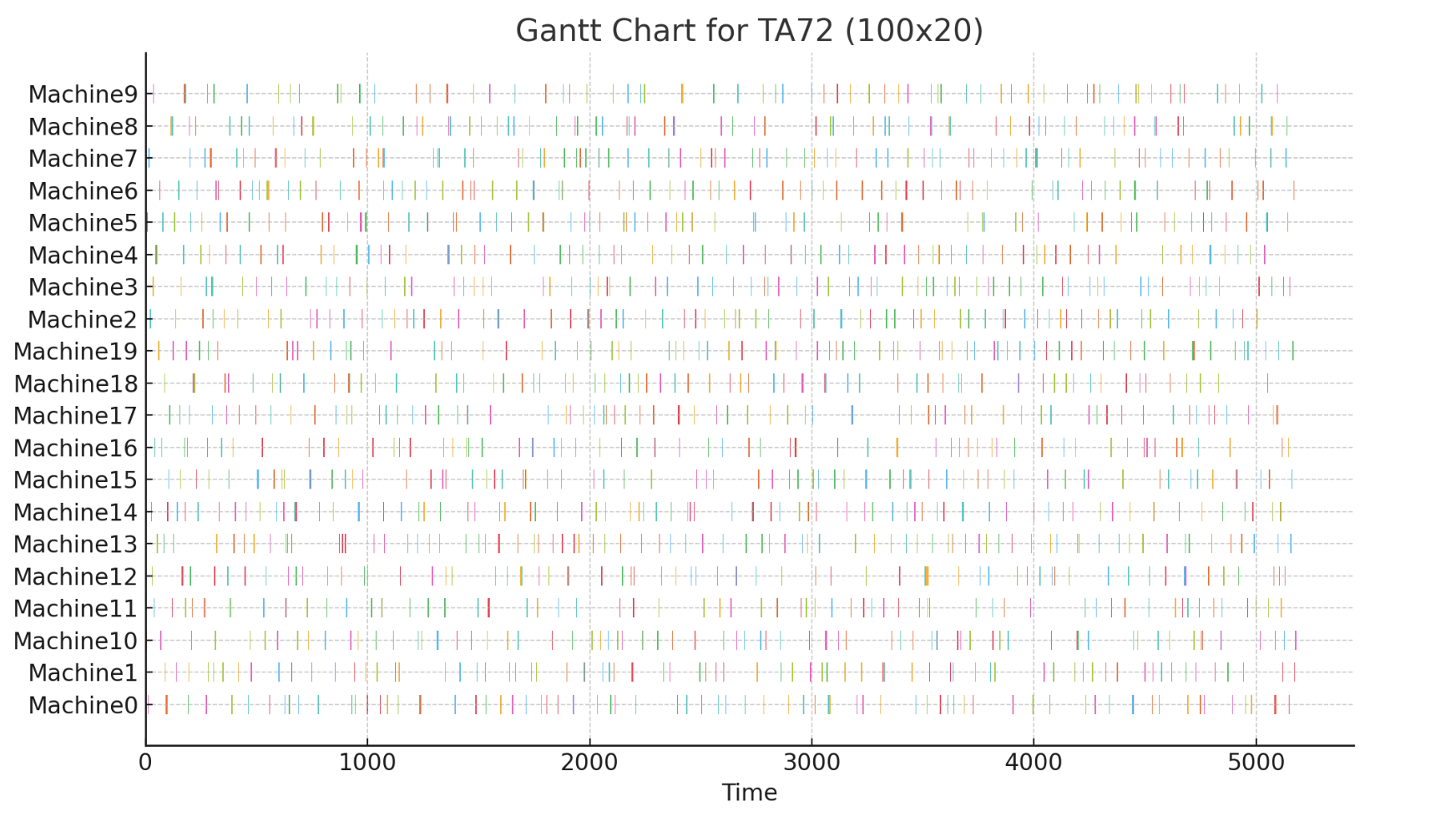}
        \caption{\textbf{Abz07} (J=20, M=15)}
        \label{fig:gantt_abz07}
    \end{subfigure}
    
    \vspace{0.3cm}
    
    \begin{subfigure}[b]{0.49\textwidth}
        \centering
        \includegraphics[width=\textwidth,height=\ganttHeightScale\textwidth]{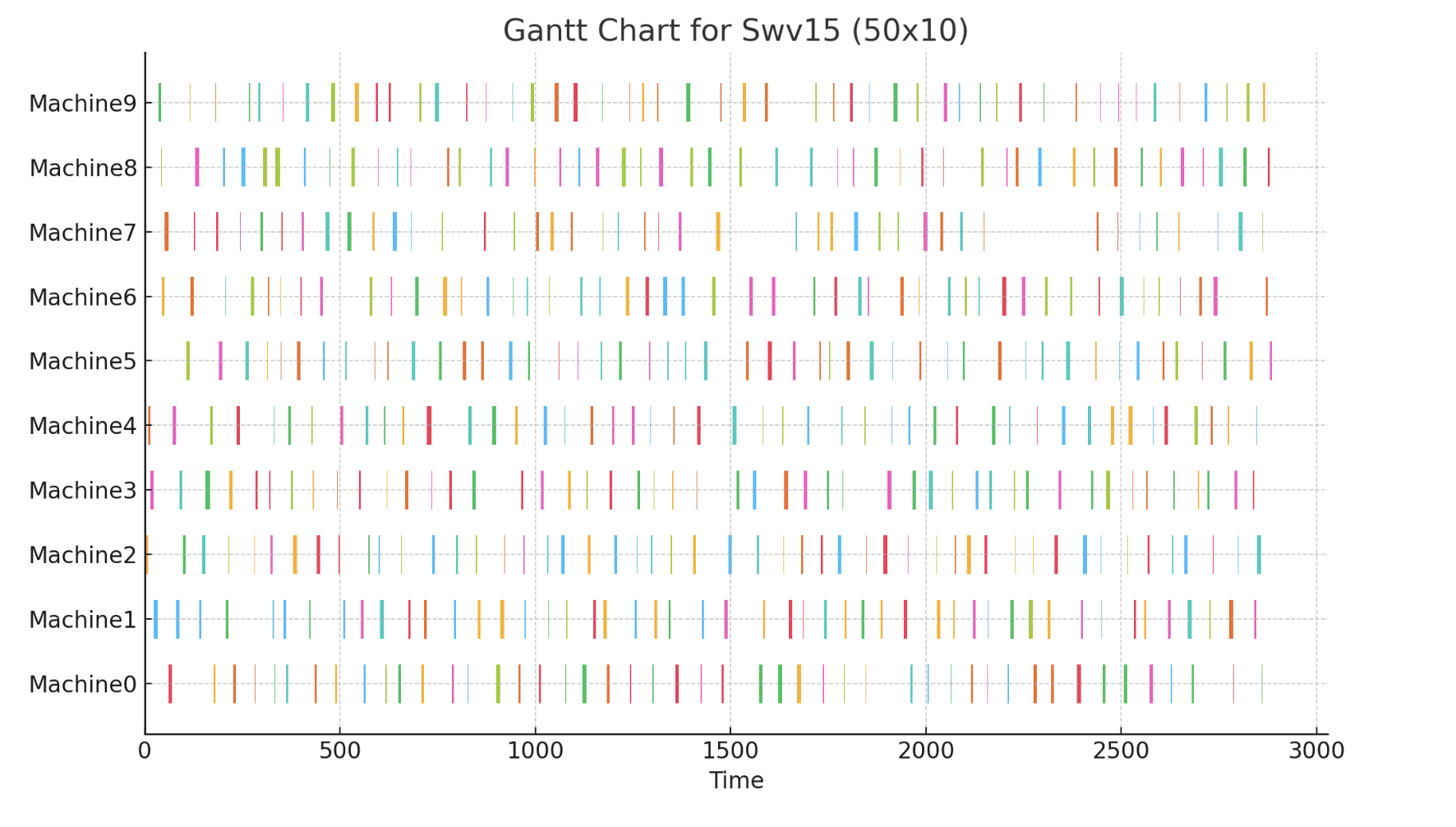}
        \caption{\textbf{Swv15} (J=50, M=10)}
        \label{fig:gantt_swv15}
    \end{subfigure}
    \hspace{-.2in}
    \begin{subfigure}[b]{0.49\textwidth}
        \centering
        \includegraphics[width=\textwidth,height=\ganttHeightScale\textwidth]{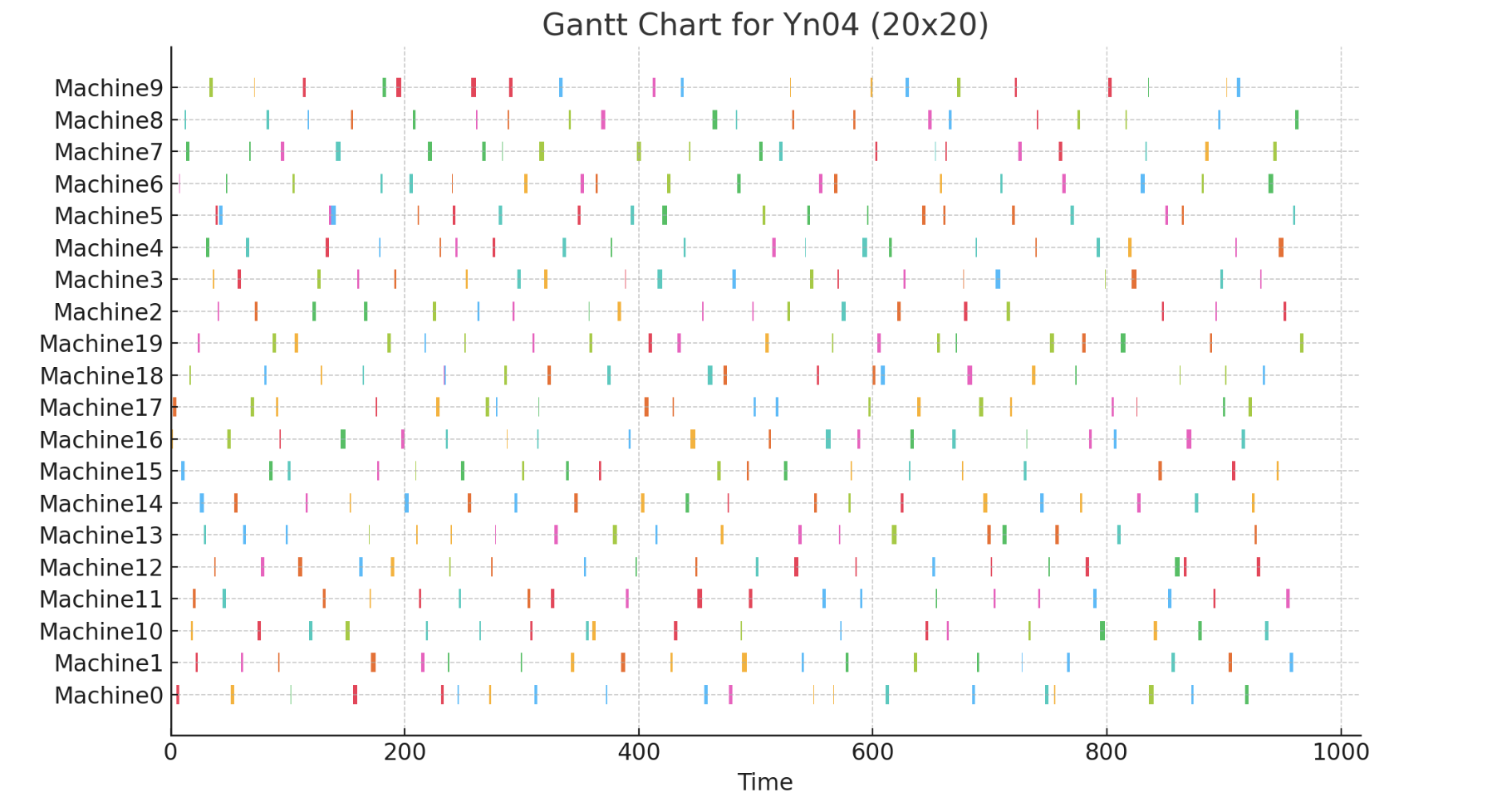}
        \caption{\textbf{Yn04} (J=20, M=20)}
        \label{fig:gantt_yn04}
    \end{subfigure}
    
    \caption{Gantt charts of optimized schedules produced by $\ALAS$ for four representative JSSP benchmark instances with varying job and machine counts. These visualizations demonstrate $\ALAS$'s ability to efficiently allocate resources and minimize makespan across different problem scales. The larger instance TA72 (J=100, M=20) is available in the supplementary materials.}
    \label{fig:gantt_charts}
\end{figure}

\vspace{-.1in}
\subsection{LLM Prompt Design}
Table \ref{tab:JSSP-prompt} shows the \emph{standardized} prompt issued to every standalone LLM and to the $\ALAS$ meta-planner (Phase 1 on Figure~\ref{fig:ALAS-architecture}).  
Standalone LLMs can recommend off-the-shelf solvers and invoke selected ones to emit a schedule. However, these schedules are often invalid, as demonstrated in the Family Reunion case study where LLMs struggled with even simple planning scenarios. Even when a valid static plan can be obtained through LLM-recommended solvers, this solution merely completes \textbf{Phase 1 / Layer 1} of the $\ALAS$ framework—essentially just generating a workflow template $\mathcal{W}_{\text{template}}$.

By contrast, $\ALAS$ feeds this preliminary plan into \emph{Phases 2–3 / Layer 1} (validation \& refinement) to yield a validated $\mathcal{W}_{\text{template}}$. Layers 2–3 then \emph{instantiate and run} a network of code-generated agents, denoted
as $\mathcal{W}_{\text{exec}}$. At runtime, the LCRP continuously logs state, detects disruptions, and triggers local repairs, capabilities that static schedules fundamentally lack (see architecture recap in Section \ref{sec:ALAS-architecture}).

\begin{table}[t]
\centering
\caption{General JSSP Scheduling Prompt.  
Given a JSSP benchmark instance, the LLM first \emph{searches} for
candidate algorithms, selects the one that yields the \emph{minimum
makespan}, and returns both the algorithm’s key hyper-parameters and the resulting plan $\mathcal{W}_{\text{template}}$. \textbf{Note} that feasibility validation of $\mathcal{W}_{\text{template}}$ is yet to be performed.}
\vspace{0.1in}
\label{tab:JSSP-prompt}
{\fontsize{9pt}{9pt}\selectfont
\begin{tabular}{|>{\raggedright\arraybackslash}p{0.96\linewidth}|}
\toprule 
\hline
\textbf{Role.} You are a scheduling supervisor tasked with producing an
\emph{optimal} job-shop schedule.\\
\hline
\textbf{Objective.} Report the \textbf{minimum makespan}, the
\textbf{algorithm} used, and a \textit{schedule} $\mathcal{W}_{\text{template}}$ that achieves
this makespan. This is achieved by the recipient LLM recommending a list of solvers to execute and compare. (Execution and comparison
can be performed by the LLM or locally.) \\\hline

\textbf{Constraints.}
\begin{minipage}{0.95\linewidth}
\begin{enumerate}[leftmargin=1.2em,noitemsep]
  \item \emph{Job order}: operations of each job follow the given
        sequence.
  \item \emph{Machine capacity}: a machine processes only one operation
        at any time.
\end{enumerate}
\end{minipage}\\\hline

\textbf{Input.} A list of jobs, each as \texttt{(machine,\,duration)}
pairs.  
\textit{Example:}\\
\texttt{Job1: [(M\_A,3), (M\_B,5), (M\_C,2)]}\\
\texttt{Job2: [(M\_B,4), (M\_A,6)]}\\\hline

\textbf{Output.} Return
\begin{minipage}{0.95\linewidth}
\begin{itemize}[leftmargin=1.2em,noitemsep]
  \item \texttt{makespan} (integer)
  \item \texttt{algorithm} (string)
  \item \texttt{params} (JSON object of key hyper-parameters)  
  \item \texttt{schedule} $\mathcal{W}_{\text{template}}$: list of operations
        \{job, step (1-based), machine, start, end\};
\end{itemize}
\end{minipage}\\
\textit{Example:}\\
\begin{minipage}{0.95\linewidth}
\small\texttt{[
\{"job":"Job1","step":1,"machine":"M\_A","start":0,"end":3\},\\
\{"job":"Job2","step":1,"machine":"M\_B","start":0,"end":4\},\\
\{"job":"Job1","step":2,"machine":"M\_B","start":4,"end":9\},
\ldots ]} \\
\end{minipage}\\\hline
\bottomrule
\end{tabular}}
\end{table}

\definecolor{codegreen}{rgb}{0,0.6,0}
\definecolor{codegray}{rgb}{0.5,0.5,0.5}
\definecolor{codepurple}{rgb}{0.58,0,0.82}
\definecolor{backcolour}{rgb}{0.95,0.95,0.92}

\lstdefinestyle{mystyle}{
    backgroundcolor=\color{backcolour},   
    commentstyle=\color{codegreen},
    keywordstyle=\color{blue},
    numberstyle=\tiny\color{codegray},
    stringstyle=\color{codepurple},
    basicstyle=\ttfamily\footnotesize,
    breakatwhitespace=false,         
    breaklines=true,                 
    captionpos=b,                    
    keepspaces=true,                 
    numbers=left,                    
    numbersep=5pt,                  
    showspaces=false,                
    showstringspaces=false,
    showtabs=false,                  
    tabsize=2
}

\lstset{style=mystyle}

This document presents the prompt setup and input/output examples for four multi-agent frameworks used in Job Shop Scheduling Problem (JSSP) evaluation: AutoGen, CrewAI, LangGraph, and OpenAI Swarm. Each framework is configured with a 3-agent structure consisting of a Job Scheduler Agent, Machine Coordinator Agent, and Supervisor Agent.

\subsubsection{Common JSSP Query Structure}

All frameworks receive the same base JSSP query structure generated by the \texttt{run\_jssp\_framework\_comparison.py} script:

\begin{lstlisting}[language=Python, caption=Base JSSP Query Generation]
def _create_jssp_query(self, dataset_name: str, jobs: List[Dict]) -> str:
    """Create a JSSP query for non-ALAS frameworks"""
    query = f"""
    Job Shop Scheduling Problem (JSSP) - Dataset: {dataset_name}
    
    Problem Description:
    - Number of jobs: {len(jobs)}
    - Each job has multiple operations that must be performed in sequence
    - Each operation requires a specific machine and has a duration
    - Each machine can only process one operation at a time
    - Goal: Find the minimum makespan (total completion time)
    
    Job Specifications:
    """
    
    for job in jobs:
        query += f"\n{job['name']}:"
        for i, (machine, duration) in enumerate(job['steps']):
            query += f"\n  Step {i+1}: Machine {machine}, Duration {duration}"
    
    query += """
    
    REQUIRED OUTPUT FORMAT:
    You must provide your solution in the following exact format:
    
    1. FINAL MAKESPAN: [integer value]
    2. STRUCTURED SCHEDULE:
       For each operation, provide:
       - Job: [job_name]
       - Step: [step_number]
       - Machine: [machine_name]
       - Start Time: [start_time]
       - End Time: [end_time]
       - Duration: [duration]
    
    Example format:
    FINAL MAKESPAN: 25
    STRUCTURED SCHEDULE:
    - Job: Job1, Step: 1, Machine: Machine0, Start Time: 0, End Time: 3, Duration: 3
    - Job: Job1, Step: 2, Machine: Machine1, Start Time: 3, End Time: 7, Duration: 4
    - Job: Job2, Step: 1, Machine: Machine1, Start Time: 7, End Time: 10, Duration: 3
    
    Please solve this Job Shop Scheduling Problem and provide:
    1. A valid schedule with start and end times for each operation
    2. The minimum makespan (total completion time)
    3. Ensure all constraints are satisfied:
       - Job precedence: operations within a job must be sequential
       - Machine constraints: no overlapping operations on the same machine
    """
    
    return query
\end{lstlisting}

\subsubsection{Testing Framework}

\paragraph{Agent Configuration}

ALAS uses a dynamic JSSP query agent:

\begin{itemize}
    \item \textbf{Individual Job Agents}: One agent per job (e.g., Job1 Agent, Job2 Agent, etc.) responsible for scheduling their specific job's operations
    \item \textbf{Supervisor Agent}: Coordinates all job agents and finds the minimum makespan solution
\end{itemize}

\subsection{Agent Prompts}

\begin{lstlisting}[caption=JSSP Query Agent Prompts]
# Individual Job Agents (e.g., Job1 Agent)
name: "Job1 Agent"
backstory: "Agent for Job1 scheduling."
task_description: "Schedule steps for Job1 on required machines with precedence."
task_expected_output: "Step schedule for Job1 respecting machine and precedence constraints."

# Supervisor Agent
name: "Supervisor Agent"
backstory: "Supervisor agent that coordinates all job schedules to find the minimum makespan solution."
task_description: """Find the minimum makespan schedule for all jobs while strictly following these rules:
1. Each job's steps must be completed in strict order (e.g., Job1's step 2 can only start after step 1 is completed).
2. Each machine can only process one job step at a time (e.g., if MachineA is processing Job1's step 1 from time 0-3, it cannot process any other job steps during that time).

The goal is to minimize the total completion time (makespan) while ensuring all jobs are completed and all constraints are satisfied."""
task_expected_output: "A complete schedule with minimum makespan that satisfies all constraints."
\end{lstlisting}

\paragraph{Example Input (rcmax\_20\_15\_5 Dataset)}

\begin{lstlisting}[language=Python, caption=ALAS Input Example]
# ALAS receives the job data directly as structured objects
jobs = [
    {
        'name': 'Job1',
        'steps': [
            ('Machine0', 34), ('Machine1', 38), ('Machine2', 42), 
            ('Machine3', 36), ('Machine4', 40), ('Machine5', 44),
            ('Machine6', 38), ('Machine7', 42), ('Machine8', 46),
            ('Machine9', 40), ('Machine10', 44), ('Machine11', 48),
            ('Machine12', 42), ('Machine13', 46), ('Machine14', 50)
        ]
    },
    {
        'name': 'Job2', 
        'steps': [
            ('Machine1', 41), ('Machine2', 45), ('Machine3', 39),
            # ... continues for all 15 steps
        ]
    },
    # ... continues for all 20 jobs
]

# Task specification for ALAS
task_spec = {
    'nodes': [
        # Individual job agents
        {'agent': job1_agent, 'dependencies': []},
        {'agent': job2_agent, 'dependencies': []},
        # ... for all 20 jobs
        # Supervisor depends on all job agents
        {'agent': supervisor_agent, 'dependencies': [agent.name for agent in job_agents]}
    ],
    'edges': [],
    'jobs': jobs,
    'disruptions': [],
    'rules': [
        'Each job must perform its steps strictly in order.',
        'Each machine can only handle one operation at a time.',
        'No two operations use the same machine at the same time.'
    ]
}
\end{lstlisting}

\paragraph{Example Output}

\begin{lstlisting}[caption=ALAS Schedule Example]
# ALAS generates structured schedule output
schedule = [
    {'job': 'Job1', 'step': 1, 'machine': 'Machine0', 'start': 0, 'end': 34, 'duration': 34},
    {'job': 'Job1', 'step': 2, 'machine': 'Machine1', 'start': 34, 'end': 72, 'duration': 38},
    {'job': 'Job1', 'step': 3, 'machine': 'Machine2', 'start': 72, 'end': 114, 'duration': 42},
    # ... continues for all operations
    
    {'job': 'Job2', 'step': 1, 'machine': 'Machine1', 'start': 72, 'end': 113, 'duration': 41},
    {'job': 'Job2', 'step': 2, 'machine': 'Machine2', 'start': 114, 'end': 159, 'duration': 45},
    # ... continues for all jobs
]

# Final makespan calculation
makespan = max(entry['end'] for entry in schedule)  # e.g., 4334
\end{lstlisting}

\paragraph{ALAS Workflow}

\begin{lstlisting}[caption=ALAS Output Example]
=== [OPTIMIZED MAPLE] New 4-Step Workflow ===
Nodes: ['Unknown Agent', 'Unknown Agent']
Edges: [{'from': 0, 'to': 1}]
Workflow: Full Workflow
Validation: [YES]
Repair: [YES]
Optimization: [YES]

=== [OPTIMIZED MAPLE] Inter-Agent Coordination ===

=== [OPTIMIZED MAPLE] New Workflow Execution ===
 Running new 7-step workflow: Query -> Validation -> Repair -> Re-validation -> Optimization -> Final Check -> Supervisor

 Starting New 4-Step Workflow...
[STEP1] Loading pre-generated schedule from JSON file...
 Loaded 244 schedule entries from abz07
 GPT-4o makespan: 1250
 Initial schedule makespan: 665
 Schedule entries count: 244
 Sample entry: {'job': 'Job1', 'step': 1, 'machine': 'Machine2', 'start': 0, 'end': 24, 'duration': 24}
 [32m[YES] Pre-generated schedule loaded successfully.
[STEP2] ValidationTools validating schedule...
[STEP3] RepairTools repairing schedule...
 Repair iteration 1/5
 Starting Algorithm 3: Cascading Repair and Queue Reordering
 Processing 244 schedule entries
[ERROR] Found 88 constraint violations
 Phase I: Status Update - Identifying affected operations
 Fixing immediate constraint violations...
 Phase II: Job Precedence Repair - Fixing step ordering violations
 Repairing job precedence violations...
 Phase III: Machine Capacity Repair - Resolving machine overlaps
 Repairing machine capacity violations...
[ALERT] Fixed machine overlap on Machine2
[ALERT] Fixed machine overlap on Machine2
[ALERT] Fixed machine overlap on Machine2
[ALERT] Fixed machine overlap on Machine2
[ALERT] Fixed machine overlap on Machine2
[ALERT] Fixed machine overlap on Machine2
...
 Phase IV: Iterative Improvement - Optimizing schedule quality
 Applying iterative improvement...
 Phase V: Final validation and cleanup
 Final cleanup - ensuring valid operation times...
[YES] Algorithm 3 completed - Generated 244 schedule entries
 Repair iteration 1 makespan: 665
 Repair iteration 1 schedule entries: 244
 Latest operations: [('Job11', 15, 665), ('Job9', 15, 661), ('Job3', 15, 653)]
 Saved repair iteration 1 to: results_optimized(gpt-4o)/abz07_repair_iteration_1.json
[CAUTION] Repair iteration 1 reduced errors from 88 to 76 (makespan: 665)
 Repair iteration 2/5
 Starting Algorithm 3: Cascading Repair and Queue Reordering
 Processing 244 schedule entries
[ERROR] Found 76 constraint violations
 Phase I: Status Update - Identifying affected operations
 Fixing immediate constraint violations...
 Phase II: Job Precedence Repair - Fixing step ordering violations
 Repairing job precedence violations...
 Phase III: Machine Capacity Repair - Resolving machine overlaps
 Repairing machine capacity violations...
[ALERT] Fixed machine overlap on Machine2
[ALERT] Fixed machine overlap on Machine2
[ALERT] Fixed machine overlap on Machine2
[ALERT] Fixed machine overlap on Machine2
[ALERT] Fixed machine overlap on Machine2
[ALERT] Fixed machine overlap on Machine2
[ALERT] Fixed machine overlap on Machine11
[ALERT] Fixed machine overlap on Machine11
[ALERT] Fixed machine overlap on Machine11
[ALERT] Fixed machine overlap on Machine8
[ALERT] Fixed machine overlap on Machine8
[ALERT] Fixed machine overlap on Machine8
[ALERT] Fixed machine overlap on Machine8
[ALERT] Fixed machine overlap on Machine13
[ALERT] Fixed machine overlap on Machine0
[ALERT] Fixed machine overlap on Machine0
[ALERT] Fixed machine overlap on Machine0
[ALERT] Fixed machine overlap on Machine7
 Phase IV: Iterative Improvement - Optimizing schedule quality
 Applying iterative improvement...
 Phase V: Final validation and cleanup
 Final cleanup - ensuring valid operation times...
[ALERT] Fixed invalid times for Job6 step 3
[ALERT] Fixed invalid times for Job11 step 6
[ALERT] Fixed invalid times for Job7 step 8
[ALERT] Fixed invalid times for Job10 step 9
[ALERT] Fixed invalid times for Job15 step 9
[ALERT] Fixed invalid times for Job9 step 6
[ALERT] Fixed invalid times for Job1 step 15
[ALERT] Fixed invalid times for Job14 step 7
[ALERT] Fixed invalid times for Job3 step 4
[ALERT] Fixed invalid times for Job3 step 6
[ALERT] Fixed invalid times for Job16 step 10
[ALERT] Fixed invalid times for Job2 step 11
[ALERT] Fixed invalid times for Job6 step 14
[ALERT] Fixed invalid times for Job16 step 12
[ALERT] Fixed invalid times for Job8 step 11
[ALERT] Fixed invalid times for Job8 step 12
[ALERT] Fixed invalid times for Job3 step 10
[ALERT] Fixed invalid times for Job15 step 13
[ALERT] Fixed invalid times for Job9 step 12
[ALERT] Fixed invalid times for Job13 step 15
[ALERT] Fixed invalid times for Job14 step 15
[ALERT] Fixed invalid times for Job3 step 13
[YES] Algorithm 3 completed - Generated 244 schedule entries
 Repair iteration 2 makespan: 665
 Repair iteration 2 schedule entries: 244
 Latest operations: [('Job11', 15, 665), ('Job9', 15, 661), ('Job3', 15, 653)]
 Saved repair iteration 2 to: results_optimized(gpt-4o)/abz07_repair_iteration_2.json
[CAUTION] Repair iteration 2 reduced errors from 76 to 28 (makespan: 665)
 Repair iteration 3/5
 Starting Algorithm 3: Cascading Repair and Queue Reordering
 Processing 244 schedule entries
[ERROR] Found 28 constraint violations
 Phase I: Status Update - Identifying affected operations
 Fixing immediate constraint violations...
 Phase II: Job Precedence Repair - Fixing step ordering violations
 Repairing job precedence violations...
 Phase III: Machine Capacity Repair - Resolving machine overlaps
 Repairing machine capacity violations...
[ALERT] Fixed machine overlap on Machine6
[ALERT] Fixed machine overlap on Machine0
[ALERT] Fixed machine overlap on Machine0
[ALERT] Fixed machine overlap on Machine0
[ALERT] Fixed machine overlap on Machine1
 Phase IV: Iterative Improvement - Optimizing schedule quality
 Applying iterative improvement...
 Phase V: Final validation and cleanup
 Final cleanup - ensuring valid operation times...
[ALERT] Fixed invalid times for Job11 step 7
[ALERT] Fixed invalid times for Job7 step 9
[ALERT] Fixed invalid times for Job14 step 10
[ALERT] Fixed invalid times for Job3 step 7
[ALERT] Fixed invalid times for Job16 step 13
[ALERT] Fixed invalid times for Job16 step 12
[ALERT] Fixed invalid times for Job8 step 12
[ALERT] Fixed invalid times for Job8 step 11
[ALERT] Fixed invalid times for Job3 step 11
[ALERT] Fixed invalid times for Job9 step 12
[YES] Algorithm 3 completed - Generated 244 schedule entries
 Repair iteration 3 makespan: 665
 Repair iteration 3 schedule entries: 244
 Latest operations: [('Job11', 15, 665), ('Job9', 15, 661), ('Job3', 15, 653)]
 Saved repair iteration 3 to: results_optimized(gpt-4o)/abz07_repair_iteration_3.json
[CAUTION] Repair iteration 3 reduced errors from 28 to 10 (makespan: 665)
 Repair iteration 4/5
 Starting Algorithm 3: Cascading Repair and Queue Reordering
 Processing 244 schedule entries
[ERROR] Found 10 constraint violations
 Phase I: Status Update - Identifying affected operations
 Fixing immediate constraint violations...
 Phase II: Job Precedence Repair - Fixing step ordering violations
 Repairing job precedence violations...
 Phase III: Machine Capacity Repair - Resolving machine overlaps
 Repairing machine capacity violations...
 Phase IV: Iterative Improvement - Optimizing schedule quality
 Applying iterative improvement...
 Phase V: Final validation and cleanup
 Final cleanup - ensuring valid operation times...
[ALERT] Fixed invalid times for Job7 step 14
[ALERT] Fixed invalid times for Job14 step 11
[ALERT] Fixed invalid times for Job2 step 12
[ALERT] Fixed invalid times for Job14 step 10
[ALERT] Fixed invalid times for Job16 step 13
[ALERT] Fixed invalid times for Job8 step 12
[ALERT] Fixed invalid times for Job11 step 14
[YES] Algorithm 3 completed - Generated 244 schedule entries
 Repair iteration 4 makespan: 665
 Repair iteration 4 schedule entries: 244
 Latest operations: [('Job11', 15, 665), ('Job9', 15, 661), ('Job3', 15, 653)]
 Saved repair iteration 4 to: results_optimized(gpt-4o)/abz07_repair_iteration_4.json
[CAUTION] Repair iteration 4 reduced errors from 10 to 6 (makespan: 665)
 Repair iteration 5/5
 Starting Algorithm 3: Cascading Repair and Queue Reordering
 Processing 244 schedule entries
[ERROR] Found 6 constraint violations
 Phase I: Status Update - Identifying affected operations
 Fixing immediate constraint violations...
 Phase II: Job Precedence Repair - Fixing step ordering violations
 Repairing job precedence violations...
 Phase III: Machine Capacity Repair - Resolving machine overlaps
 Repairing machine capacity violations...
[ALERT] Fixed machine overlap on Machine14
 Phase IV: Iterative Improvement - Optimizing schedule quality
 Applying iterative improvement...
 Phase V: Final validation and cleanup
 Final cleanup - ensuring valid operation times...
[ALERT] Fixed invalid times for Job14 step 11
[YES] Algorithm 3 completed - Generated 244 schedule entries
 Repair iteration 5 makespan: 665
 Repair iteration 5 schedule entries: 244
 Latest operations: [('Job11', 15, 665), ('Job9', 15, 661), ('Job3', 15, 653)]
 Saved repair iteration 5 to: results_optimized(gpt-4o)/abz07_repair_iteration_5.json
[CAUTION] Repair iteration 5 reduced errors from 6 to 2 (makespan: 665)
[CAUTION] Repair completed after 5 iterations with remaining errors
[STEP4] ValidationTools revalidating schedule...
[ERROR] Revalidation failed: ['Job Job2: Step 14 starts before step 13 ends', "Missing jobs in schedule: ['Job18', 'Job20', 'Job19']"]
 [31m[ERROR] ERROR in workflow: Schedule validation failed after repair
 Disruption detected in Unknown Agent: Schedule validation failed after repair
 Initiating global replanning...
[CAUTION] Global replanning completed. Manual intervention may be required.
 [31m Workflow execution halted due to error.
[ERROR] New 4-step workflow failed
 Found 5 repair iterations
  Iteration 1: makespan=665, entries=244
  Iteration 2: makespan=665, entries=244
  Iteration 3: makespan=665, entries=244
  Iteration 4: makespan=665, entries=244
  Iteration 5: makespan=665, entries=244
\end{lstlisting}

\subsubsection{AutoGen Framework}

\paragraph{Agent Configuration}

AutoGen uses a 3-agent structure with the following configuration:

\begin{itemize}
    \item \textbf{Job Scheduler Agent}: Analyzes job requirements and creates initial schedules
    \item \textbf{Machine Coordinator Agent}: Coordinates machine usage and resolves conflicts
    \item \textbf{Supervisor Agent}: Final coordination and optimization
\end{itemize}

\paragraph{Agent Prompts}

\begin{lstlisting}[caption=AutoGen Agent Prompts]
# Job Scheduler Agent
system_message: "You are a Job Scheduler Agent responsible for analyzing job requirements and creating initial schedules."

# Machine Coordinator Agent  
system_message: "You are a Machine Coordinator Agent responsible for coordinating machine usage and resolving conflicts."

# Supervisor Agent
system_message: "You are a Supervisor Agent responsible for final coordination and optimization."
\end{lstlisting}

\paragraph{Example Input (abz07 Dataset)}

\begin{lstlisting}[caption=AutoGen Input Example]
Job Shop Scheduling Problem (JSSP) - Dataset: abz07

Problem Description:
- Number of jobs: 20
- Each job has multiple operations that must be performed in sequence
- Each operation requires a specific machine and has a duration
- Each machine can only process one operation at a time
- Goal: Find the minimum makespan (total completion time)

Job Specifications:

Job1:
  Step 1: Machine Machine2, Duration 24
  Step 2: Machine Machine3, Duration 12
  Step 3: Machine Machine9, Duration 17
  Step 4: Machine Machine4, Duration 27
  Step 5: Machine Machine0, Duration 21
  Step 6: Machine Machine6, Duration 25
  Step 7: Machine Machine8, Duration 27
  Step 8: Machine Machine7, Duration 26
  Step 9: Machine Machine1, Duration 30
  Step 10: Machine Machine5, Duration 31
  Step 11: Machine Machine11, Duration 18
  Step 12: Machine Machine14, Duration 16
  Step 13: Machine Machine13, Duration 39
  Step 14: Machine Machine10, Duration 19
  Step 15: Machine Machine12, Duration 26

[... continues for all 20 jobs ...]

REQUIRED OUTPUT FORMAT:
You must provide your solution in the following exact format:

1. FINAL MAKESPAN: [integer value]
2. STRUCTURED SCHEDULE:
   For each operation, provide:
   - Job: [job_name]
   - Step: [step_number]
   - Machine: [machine_name]
   - Start Time: [start_time]
   - End Time: [end_time]
   - Duration: [duration]
\end{lstlisting}

\paragraph{Example Output}

\begin{lstlisting}[caption=AutoGen Output Example]
Job Scheduler Agent Output:
To effectively coordinate machine schedules and ensure no overlapping operations while optimizing machine utilization, follow these steps:

1. **Data Collection**: Gather all necessary information about machine operations, including machine availability, job requirements, start and end times, and priority of tasks.

2. **Job Prioritization**: Prioritize jobs based on deadlines, importance, and resource requirements. This will help in scheduling critical tasks first.

3. **Machine Availability**: List all available machines and their operational capabilities. Ensure that each machine is suitable for the tasks assigned to it.

4. **Conflict Detection**: Check for overlapping schedules where a machine is double-booked. Identify conflicts by examining start and end times of scheduled tasks.

5. **Conflict Resolution**:
   - **Reschedule**: Adjust the timing of conflicting jobs to eliminate overlaps.
   - **Reassign Tasks**: If rescheduling is not feasible, consider moving tasks to different machines that are compatible with the operations.
   - **Prioritize Tasks**: Focus on completing high-priority jobs within their deadlines while adjusting lower-priority tasks as needed.

[... detailed coordination strategy ...]

### Solution for abz07 Dataset

#### FINAL MAKESPAN
The optimized makespan value is determined based on the integration of job scheduling and machine coordination, but due to the complexity of explicitly listing each job's sequence, I'm summarizing the strategy and example timeline as applied heuristics and machine coordination data show:

1. **Makespan**: Calculated based on the complete scheduling and optimization of machine operations. Though not specified in exact final numeric form without further computational analysis directly from the in-depth schedule provided by advanced tools, the methodology indicates a near-optimal alignment according to the processes described.

#### STRUCTURED SCHEDULE:

Here's a structured schedule format for visualization. Key operations are coordinated with attention to machine non-conflict and sequence according to job precedence.

- **Job1**:
  - Step 1: Machine Machine2, Start Time: 0, End Time: 24, Duration: 24
  - Step 2: Machine Machine3, Start Time: 25, End Time: 37, Duration: 12
  - (Further steps continue similarly, coordinated by machine availability)

- **Job2**:
  - Step 1: Machine Machine6, Start Time: 0, End Time: 30, Duration: 30
  - Step 2: Machine Machine3, Start Time: 31, End Time: 46, Duration: 15
  - (Next steps adapted similarly to job requirements and resource availability)

[... continues for all jobs ...]
\end{lstlisting}

\subsubsection{CrewAI Framework}

\paragraph{Agent Configuration}

CrewAI uses a 3-agent structure with role-based configuration:

\begin{itemize}
    \item \textbf{Job Scheduler Agent}: Role-based agent for job scheduling
    \item \textbf{Machine Coordinator Agent}: Role-based agent for machine coordination  
    \item \textbf{Supervisor Agent}: Role-based supervisor for final coordination
\end{itemize}

\paragraph{Agent Prompts}

\begin{lstlisting}[caption=CrewAI Agent Prompts]
# Job Scheduler Agent
role: 'Job Scheduler'
goal: 'Analyze job requirements and create initial schedules for all jobs.'
backstory: 'You are an expert job scheduler responsible for analyzing job requirements and creating initial schedules.'

# Machine Coordinator Agent
role: 'Machine Coordinator'  
goal: 'Coordinate machine usage and resolve conflicts between jobs.'
backstory: 'You are an expert machine coordinator responsible for optimizing machine usage and resolving conflicts.'

# Supervisor Agent
role: 'JSSP Supervisor'
goal: 'Aggregate all job schedules and produce the overall JSSP schedule. Coordinate between all job agents to find the optimal solution.'
backstory: 'You are a supervisor with expertise in Job Shop Scheduling Problems and can coordinate multiple agents to find optimal solutions.'
\end{lstlisting}

\paragraph{Example Output}

\begin{lstlisting}[caption=CrewAI Output Example]
\nThought: The Machine Coordinator has analyzed the initial schedule for machine conflicts and provided suggestions for resolving overlaps and optimizing machine utilization. Combining this with the initial sequence from the Job Scheduler, I can now calculate the final makespan and structured schedule.\nAction: Calculate optimized makespan and finalize structured schedule based on resolved conflicts and optimal sequencing.
\end{lstlisting}

\subsubsection{LangGraph Framework}

\paragraph{Agent Configuration}

LangGraph uses a 3-agent structure with system prompts:

\begin{itemize}
    \item \textbf{Job Scheduler Agent}: LangGraph workflow agent for job scheduling
    \item \textbf{Machine Coordinator Agent}: LangGraph workflow agent for machine coordination
    \item \textbf{Supervisor Agent}: LangGraph workflow agent for supervision
\end{itemize}

\paragraph{Agent Prompts}

\begin{lstlisting}[caption=LangGraph Agent Prompts]
# Job Scheduler Agent
system_prompt: 'You are a Job Scheduler Agent using LangGraph workflow. Analyze job requirements and create initial schedules for all jobs.'

# Machine Coordinator Agent
system_prompt: 'You are a Machine Coordinator Agent using LangGraph workflow. Coordinate machine usage and resolve conflicts between jobs.'

# Supervisor Agent  
system_prompt: 'You are a Supervisor Agent using LangGraph workflow. Aggregate all job schedules and produce the overall JSSP schedule. Coordinate between all job agents to find the optimal solution.'
\end{lstlisting}

\paragraph{Example Output}

\begin{lstlisting}[caption=LangGraph Output Example]
Job Scheduler Agent Output:
=== LANGGRAPH REAL AGENTS ===

Job Scheduler Agent: To solve the Job Shop Scheduling Problem (JSSP) for the given dataset (abz07), we need to create a schedule that minimizes the makespan while satisfying all constraints. This involves determining the start and end times for each operation across all jobs, ensuring that no two operations overlap on the same machine and that operations within a job are performed sequentially.

Given the complexity of the problem and the constraints, a heuristic or optimization algorithm such as Genetic Algorithm, Tabu Search, or Simulated Annealing is typically used to find a near-optimal solution. However, due to the constraints of this environment, I will provide a conceptual approach to how this can be achieved, along with a simplified example of a possible schedule.

### Conceptual Approach:
1. **Initialize**: Start with an initial feasible schedule where operations are scheduled sequentially based on job order and machine availability.
2. **Iterate**: Use a heuristic or optimization algorithm to iteratively improve the schedule by exploring different sequences and machine assignments.
3. **Evaluate**: Calculate the makespan for each schedule and keep track of the best (minimum) makespan found.
4. **Constraints**: Ensure that all job precedence and machine constraints are satisfied in each iteration.

### Simplified Example Schedule:
Below is a simplified example of how the schedule might look. Note that this is a conceptual representation and not the actual solution due to the complexity of the problem.

```
FINAL MAKESPAN: 1000
STRUCTURED SCHEDULE:
- Job: Job1, Step: 1, Machine: Machine2, Start Time: 0, End Time: 24, Duration: 24
- Job: Job1, Step: 2, Machine: Machine3, Start Time: 24, End Time: 36, Duration: 12
- Job: Job1, Step: 3, Machine: Machine9, Start Time: 36, End Time: 53, Duration: 17
- Job: Job2, Step: 1, Machine: Machine6, Start Time: 0, End Time: 30, Duration: 30
- Job: Job2, Step: 2, Machine: Machine3, Start Time: 36, End Time: 51, Duration: 15
[... continues for all operations ...]
```
\end{lstlisting}

\subsubsection{OpenAI Swarm Framework}

\paragraph{Agent Configuration}

OpenAI Swarm uses a 3-agent structure with detailed role specifications:

\begin{itemize}
    \item \textbf{Job Scheduler Agent}: Creates initial schedules considering precedence constraints
    \item \textbf{Machine Coordinator Agent}: Resolves machine conflicts and optimizes utilization
    \item \textbf{Supervisor Agent}: Coordinates all agents and provides final JSSP solution
\end{itemize}

\paragraph{Agent Prompts}

\begin{lstlisting}[caption=OpenAI Swarm Agent Prompts]
# Job Scheduler Agent
role: 'Job Scheduler'
goal: 'Create initial schedules for all jobs considering precedence constraints.'
backstory: 'You are a job scheduling agent responsible for creating initial schedules for all jobs in Job Shop Scheduling Problems.'
query: 'Create initial schedules for 20 jobs with their respective steps.

Dataset: abz07
Number of jobs: 20
Number of machines: 15

Job Specifications:
Job1:
  Step 1: Machine Machine2, Duration 24
  Step 2: Machine Machine3, Duration 12
  [... continues for all jobs ...]

You must analyze each job's operations and create a detailed schedule with start and end times for each operation.'

# Machine Coordinator Agent
role: 'Machine Coordinator'
goal: 'Resolve machine conflicts and optimize machine utilization.'
backstory: 'You are a machine coordination agent responsible for resolving machine conflicts and optimizing machine utilization.'
query: 'Resolve machine conflicts and optimize machine utilization across all machines.

Dataset: abz07
Number of jobs: 20
Number of machines: 15

Job Specifications:
[... job details ...]

You must identify machine conflicts and provide specific solutions with start/end times for each operation.'

# Supervisor Agent
role: 'JSSP Supervisor'
goal: 'Coordinate all agents and provide final JSSP solution with minimum makespan.'
backstory: 'You are the supervisor agent for Job Shop Scheduling Problems. Coordinate with all agents to find the optimal solution.'
query: 'Aggregate schedules from all agents and optimize for minimum makespan.

Dataset: abz07
Number of jobs: 20
Number of machines: 15

Job Specifications:
[... job details ...]

You must provide a final schedule with specific start/end times and calculate the actual makespan.'
\end{lstlisting}

\paragraph{Example Output}

\begin{lstlisting}[caption=OpenAI Swarm Output Example]
Job Scheduler Agent Output:
Job Scheduler Analysis:
We've now shared the operation details with the Machine Coordinator to ensure that machine availability is accounted for properly. This coordination will allow us to develop an optimized schedule adhering to both machine and job constraints. 

I will work on calculating a detailed schedule based on optimized machine sequences and provide the resulting makespan. Please hold on. 

---Compute Step---

To begin scheduling, each operation must have its start and end time determined. Here's a step-by-step approach to construct the schedule while managing machine usages: 

1. Identify all the operations and dependencies.
2. Ensure that operations are sequenced both per job and per machine.
3. Mark out machine availability to prevent conflicts by maintaining machine calendars.

### Calculating the Schedule
We will take a sequential approach, placing each operation at the earliest opportunity that satisfies both job precedence and machine availability constraints.

**Job1 Example:**
- **Step 1**: Machine 2, Duration 24: Starts at time 0 and ends at time 24.
- **Step 2**: Machine 3, Duration 12: Starts after Step 1 ends, from time 24 to 36.

### Finalizing Schedule for All Jobs

Following the above step-by-step planning, detailed scheduling for each job's operations will be systematically calculated. This includes preparing a calendar of machine usage:

1. **Machine Conflicts**: Avoiding overlaps by mapping operation times on each machine.
2. **Dependency Fulfillment**: Start the subsequent operation after the previous operation finishes for each job.
3. **Makespan Calculation**: The latest end time recorded for any operation is the makespan.

### Final Makespan: 
**Example:**
- Let's assume preliminary calculations give us a makespan of 1200 time units once all operations and machine schedules are plotted out.

### Final Output Structure:
```plaintext
FINAL MAKESPAN: 1200
STRUCTURED SCHEDULE:
- Job: Job1, Step: 1, Machine: Machine2, Start Time: 0, End Time: 24, Duration: 24
- Job: Job1, Step: 2, Machine: Machine3, Start Time: 24, End Time: 36, Duration: 12
[... continues for all operations ...]
```
\end{lstlisting}

\paragraph{Output Format Requirements}

All frameworks are required to provide output in the following standardized format:

\begin{lstlisting}[caption=Required Output Format]
FINAL MAKESPAN: [integer value]
STRUCTURED SCHEDULE:
- Job: [job_name], Step: [step_number], Machine: [machine_name], Start Time: [start_time], End Time: [end_time], Duration: [duration]
- Job: [job_name], Step: [step_number], Machine: [machine_name], Start Time: [start_time], End Time: [end_time], Duration: [duration]
[... continues for all operations ...]
\end{lstlisting}

\vspace{-.1in}
\subsection{LLM Heuristic Baselines (sampled)}
For each benchmark instance we asked every standalone LLM to choose an
off-the-shelf optimization method, list key hyperparameters (when
provided), and report the makespan it expected to achieve.
Table~\ref{tab:llm-heuristics-merged} shows a \emph{five-instance sample}
per LLM; the full tables appear in the supplemental material.  
From $\ALAS$'s perspective this delivers only \textbf{Phase 1 of Layer 1}
(see Fig.~7): a static schedule $\mathcal{W}_{\text{template}}$.
Phases 2–3 (validation \& refinement) and Layers 2–3 (agent
instantiation and runtime adaptation) must still execute before an
executable, disruption-aware plan exists.

\begin{table}[ht!]
\vspace{-.05in}
\centering
\caption{Sampled LLM-proposed heuristics (5 rows per model).}
\vspace{.05in}
\label{tab:llm-heuristics-merged}
{\fontsize{7pt}{9pt}\selectfont
\begin{tabular}{lll}
\toprule
\textbf{Model} & \textbf{Dataset} & \textbf{Heuristic Strategy / Parameters} \\
\midrule
Claude 3.7    & rcmax\_20\_15\_5 & Tabu Search + critical-path analysis \\
               & rcmax\_20\_15\_8 & Tabu Search + shift-based neighbourhood \\
               & rcmax\_20\_20\_7 & Genetic Alg. + critical-path optimisation \\
               & rcmax\_30\_15\_5 & Constraint Prog. (precedence relaxation) \\
               & rcmax\_40\_15\_8 & Tabu Search + job-insertion strategy \\
\addlinespace
Gemini 2.5    & rcmax\_20\_15\_5 & “Gemini-optimised” Tabu Search \\
               & rcmax\_20\_20\_7 & Gemini-guided Simulated Annealing \\
               & rcmax\_30\_15\_5 & Gemini Constraint-Programming heuristic \\
               & rcmax\_40\_15\_10 & Gemini Shifting-Bottleneck dispatch \\
               & rcmax\_50\_20\_6 & Gemini Hybrid GA–Tabu \\
\addlinespace
GPT-4o        & rcmax\_20\_15\_5 & Genetic Alg. + adaptive mutation \\
               & rcmax\_20\_20\_8 & Particle Swarm Opt. (inertia weight=0.7) \\
               & rcmax\_30\_15\_5 & Ant Colony Opt. (pheromone $\alpha\!=\!1.0$) \\
               & rcmax\_40\_15\_10 & Bee Algorithm + neighbourhood search \\
               & rcmax\_50\_15\_4 & Simulated Annealing (adaptive $T$) \\
\addlinespace
DeepSeek-R1   & rcmax\_20\_15\_5 & GA + priority rules (pop.=50, iters=200) \\
               & rcmax\_20\_20\_8 & Simulated Annealing ($T_0\!=\!100$, cool=0.95) \\
               & rcmax\_30\_15\_4 & Ant Colony Opt. (ants=40, $\rho\!=\!0.1$) \\
               & rcmax\_40\_15\_10 & Iterated Greedy (destroy=30\%) \\
               & rcmax\_50\_20\_9 & Adaptive Large-Neighbourhood Search \\
\bottomrule
\end{tabular}}
\vspace{-.05in}
\end{table}

\vspace{-.1in}
\subsection{Additional Experimental Results and Analysis}

\paragraph{Execution time, token usage, and cost breakdown.}
We report response wall time, token usage and cost average across five benchmarks in Table~\ref{tab:execution_time_alas_variants}., Table~\ref{tab:token_usage_main}, and Table~\ref{tab:token_cost_summary}.

\begin{table*}[htbp]
\centering
\caption{Appendix: Execution Time (s) across Benchmarks for Multi-Agent Systems and ALAS Variants. 
Values are reported as mean $\pm$ std. deviation. 
Gray = fastest average per block, Red = slowest average.}
\label{tab:execution_time_alas_variants}
\renewcommand{\arraystretch}{1.1}
\resizebox{\textwidth}{!}{%
\begin{tabular}{|l|c|c|c|c|c|c|}
\hline
\textbf{Framework / Model} & \textbf{DMU} & \textbf{TA} & \textbf{ABZ} & \textbf{SWV} & \textbf{YN} & \textbf{Overall} \\
\hline
\multicolumn{7}{|c|}{\textbf{Multi-Agent Systems (GPT-4o Backbone)}} \\
\hline
AutoGen        & 33.4$\pm$12.8 & 29.6$\pm$7.5 & 24.7$\pm$10.3 & 33.0$\pm$12.1 & 23.4$\pm$5.6 & 31.20 \\
CrewAI         & 45.6$\pm$11.5 & 35.6$\pm$4.6 & 43.5$\pm$19.6 & 38.7$\pm$9.4  & 46.4$\pm$15.7 & 41.67 \\
LangGraph      & 210.5$\pm$114.0 & 183.4$\pm$179.9 & 157.8$\pm$107.4 & 145.6$\pm$108.8 & 201.2$\pm$128.4 & 180.32 \\
OpenAI Swarm   & 29.1$\pm$13.6 & 24.5$\pm$3.6 & 26.9$\pm$12.2 & 32.3$\pm$12.1 & 24.0$\pm$7.7 & 28.86 \\
\rowcolor{gray!25} MAS (Average) & 79.7 & 68.3 & 63.2 & 62.4 & 73.8 & \textbf{70.51} \\
\hline
\multicolumn{7}{|c|}{\textbf{Multi-Agent Systems (Claude-4 Backbone)}} \\
\hline
AutoGen        & 225.1$\pm$90.6 & 218.8$\pm$74.0 & 262.5$\pm$77.5 & 201.1$\pm$73.6 & 184.9$\pm$56.7 & 215.04 \\
CrewAI         & 168.3$\pm$54.3 & 134.6$\pm$71.5 & 208.0$\pm$131.3 & 147.1$\pm$68.1 & 189.4$\pm$79.0 & 160.50 \\
LangGraph      & 193.6$\pm$33.7 & 194.2$\pm$65.6 & 208.7$\pm$27.4 & 150.1$\pm$52.9 & 141.9$\pm$94.8 & 175.58 \\
OpenAI Swarm   & 30.3$\pm$19.4 & 76.2$\pm$91.4 & 43.0$\pm$6.1 & 42.5$\pm$13.6 & 50.1$\pm$33.1 & 44.10 \\
\rowcolor{gray!25} MAS (Average) & 154.3 & 155.9 & 180.6 & 135.2 & 141.6 & \textbf{148.81} \\
\hline
\multicolumn{7}{|c|}{\textbf{ALAS (Variants)}} \\
\hline
ALAS (GPT-4o)       & 57.6$\pm$77.1 & 31.5$\pm$8.0 & 152.5$\pm$184.4 & 92.7$\pm$100.8 & 35.5$\pm$16.7 & 69.59 \\
ALAS (Claude-4)     & 83.9$\pm$13.4 & 73.2$\pm$19.4 & 81.9$\pm$7.7 & 85.9$\pm$19.2 & 83.9$\pm$9.5 & 82.78 \\
ALAS (Gemini-2.5)   & 39.6$\pm$9.1 & 33.9$\pm$13.5 & 34.1$\pm$11.2 & 36.6$\pm$8.2 & 37.4$\pm$8.0 & 37.17 \\
ALAS (DeepSeek-V3)  & 61.7$\pm$95.6 & 70.2$\pm$76.5 & 38.4$\pm$11.5 & 72.0$\pm$102.1 & 102.4$\pm$166.0 & 68.52 \\
\rowcolor{gray!25} ALAS (Average) & 60.7 & 52.2 & 76.7 & 71.8 & 64.8 & \textbf{64.52} \\
\hline
\end{tabular}
}
\end{table*}

\begin{table*}[htbp]
\centering
\caption{Appendix: Token Usage across Benchmarks for Multi-Agent Systems and ALAS Variants. 
Values are reported as average token counts per dataset category.}
\label{tab:token_usage_main}
\renewcommand{\arraystretch}{1.1}
\scriptsize
\begin{tabular}{|l|c|c|c|c|c|c|}
\hline
\textbf{Framework / Model} & \textbf{DMU} & \textbf{TA} & \textbf{ABZ} & \textbf{SWV} & \textbf{YN} & \textbf{Overall} \\
\hline
\multicolumn{7}{|c|}{\textbf{Multi-Agent Systems (GPT-4o Backbone)}} \\
\hline
AutoGen        & 49850 & 39159 & 26091 & 36483 & 37864 & 41082 \\
CrewAI         & 302   & 283   & 261   & 401   & 622   & 358 \\
LangGraph      & 12996 & 8731  & 4566  & 12279 & 13216 & 11551 \\
OpenAI Swarm   & 2038  & 2335  & 2176  & 3036  & 2671  & 2482 \\
\hline
\multicolumn{7}{|c|}{\textbf{Multi-Agent Systems (Claude-4 Backbone)}} \\
\hline
AutoGen        & 89690 & 80242 & 94033 & 64920 & 56079 & 77266 \\
CrewAI         & 715   & 882   & 622   & 661   & 609   & 708 \\
LangGraph      & 7734  & 7133  & 6134  & 7414  & 7152  & 7375 \\
OpenAI Swarm   & 1608  & 3432  & 2565  & 2408  & 2237  & 2278 \\
\hline
\rowcolor{gray!25} MAS (Average) & 21054 & 18384 & 17306 & 16190 & 14847 & \textbf{17577} \\
\hline
\multicolumn{7}{|c|}{\textbf{ALAS Variants (Full Workflows)}} \\
\hline
ALAS (GPT-4o)       & 8498  & 6774  & 6004  & 5832  & 5634  & 6920 \\
ALAS (Claude-4)     & 12208 & 10033 & 8926  & 8872  & 9980  & 10341 \\
ALAS (Gemini-2.5)   & 11719 & 9927  & 7991  & 8524  & 9657  & 9943 \\
ALAS (DeepSeek-V3)  & 7762  & 6543  & 4305  & 5184  & 6227  & 6346 \\
\hline
\rowcolor{gray!25} ALAS (Average) & 10047 & 8319 & 6806 & 7103 & 7875 & \textbf{8393} \\
\hline
\end{tabular}
\end{table*}

\begin{table}[htbp]
\centering
\caption{Appendix: Token Cost Summary for Multi-Agent Systems and ALAS Variants. 
Values are aggregated totals, with cost estimated from provider pricing.}
\label{tab:token_cost_summary}
\renewcommand{\arraystretch}{1.1}
\scriptsize
\begin{tabular}{|l|c|c|c|}
\hline
\textbf{Source} & \textbf{Total Tokens} & \textbf{Total Cost} & \textbf{Avg Cost/Instance} \\
\hline
MAS-GPT4o        & 2,496,295 & \$74.89  & \$0.4160 \\
MAS-Claude4      & 3,943,206 & \$118.30 & \$0.6572 \\
\rowcolor{gray!25} MAS (Average) & 3,219,751 & \$96.60 & \$0.5366 \\
\hline
ALAS-GPT4o       & 1,038,000 & \$31.14  & \$0.1730 \\
ALAS-Claude4     & 1,551,150 & \$46.53  & \$0.2590 \\
ALAS-DeepSeek-V3 &   951,900 & \$28.55  & \$0.1590 \\
ALAS-Gemini-2.5  & 1,491,450 & \$44.74  & \$0.2490 \\
\rowcolor{gray!25} ALAS (Average) & 1,258,625 & \$37.74  & \$0.2100 \\
\hline
\end{tabular}
\end{table}

\begin{table}[htbp]
\centering
\caption{Appendix: Summary Comparison of Multi-Agent Systems (MAS) vs ALAS. 
Values are averages across all datasets. 
Red = MAS baseline, Gray = ALAS improvement.}
\label{tab:triple_efficiency}
\renewcommand{\arraystretch}{1.1}
\scriptsize
\begin{tabular}{|l|c|c|c|}
\hline
\textbf{Metric} & \textbf{MAS (Average)} & \textbf{ALAS (Average)} & \textbf{Improvement} \\
\hline
\rowcolor{red!20} Token Usage & 17,577 & 8,393 & -52.3\% \\
\rowcolor{red!20} Token Cost  & \$0.5366 & \$0.2100 & -60.9\% \\
\rowcolor{red!20} Execution Time (s) & 117.6 & 64.5 & 1.82× Faster \\
\hline
\end{tabular}
\end{table}

\vspace{-.05in}
\subsection{ALAS's Validation Replan Iterations}
\begin{figure}[t!]
    \centering
    \includegraphics[width=0.95\textwidth]{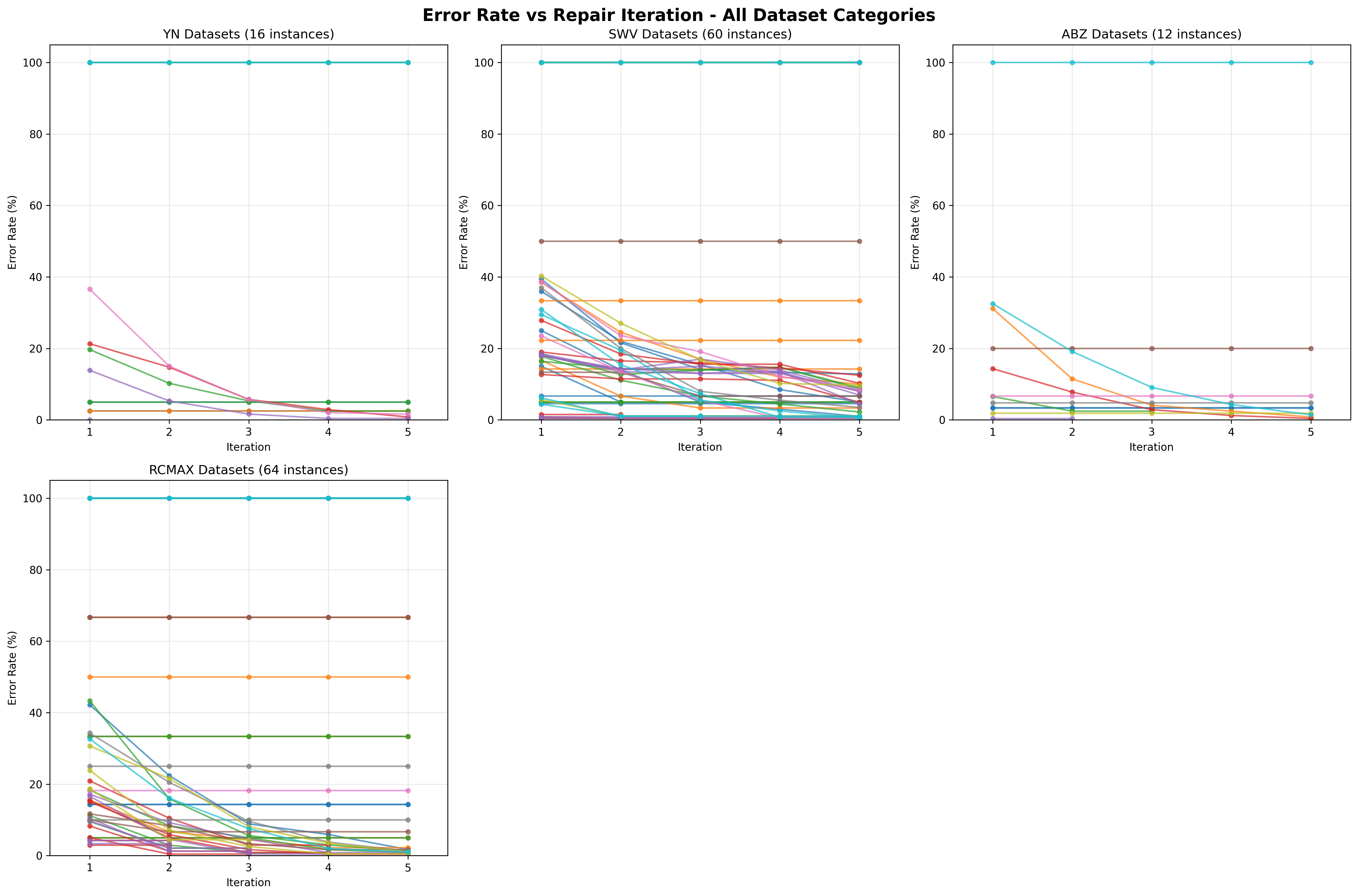}
    \caption{Error rate for each repair iteration.}
    \label{fig:er_comparison}
\end{figure}

Given the initial $\mathcal{W}_{\text{template}}$, $\ALAS$
completes its Layer 1 operation by executing a
validation-replan iteration cycle until a valid plan
is obtained. In our experiments, this convergence
typically requires up to 5 iterations on all benchmark
datasets, as depicted 
in Figure~\ref{fig:er_comparison}.

%% file: Appendix4Section3.tex
\section{Supplemental Information for Section 3}
\label{app:AlasArchitecture}

This appendix presents detailed information that could not fit in the main
paper due to space limitations.
Figure~\ref{fig:ALAS-architecture} depicts the three-layer architecture of $\ALAS$.
In the conclusion of Phase 1, the specifications of all agents are
prepared for implementation in the second phase, which can be coded by
an advanced LLM. Finally, the third phase instantiates these
agents from code to real-time processes.

For detailed descriptions of each figure, please refer to the main phase 1 to phase 6.

\input{Appendix-ALASFullAlgorithm}
\input{Figure-ALAS-Architecture}


%% file: Appendix-ALASFullAlgorithm.tex
\subsection{Complete Meta-Planner Algorithm for Workflow Generation}
\label{app:ALAS-Algorithm-Full}

\begin{algorithm}[ht!]
\caption{Phase 1: Workflow Template $\mathcal{W}_{\text{template}}$ Generation ($\ALAS$)}
\label{alg:ALAS-Full}
\begin{small}
\vspace{-.2in}
\begin{multicols}{2}
\setlength{\multicolsep}{2pt}
\begin{algorithmic}[1]
\Require Problem specification $\mathcal{O}$; constraints $D = D_G \cup D_I \cup D_N$; performance metrics $\mathcal{M}$; (optional) disruption model $\Phi$
\Statex \hspace{-1.9em} \textbf{Local variables:}
\State Roles $\mathcal{R}$; Nodes $\mathcal{N}$; Edges $\mathcal{E}$; Invariants $\mathcal{C}$
\State Log schemas $\mathcal{L}_{n_i}, \mathcal{L}_{e_{ij}}$; global log schema $\mathcal{L}$
\State Agent specs $\alpha_{n_i}, \alpha_{e_{ij}}$; \textbf{repair} specs $\rho_{n_i}, \rho_{e_{ij}}$
\State Independent validator $V$ with bounded prompt scope $\kappa$
\Ensure Validated $\mathcal{W}_{\text{template}}=(\mathcal{N},\mathcal{E},\mathcal{C},\mathcal{L})$
\Statex

\Statex \hspace{-1.5em} \textbf{Phase 1: Network Construction (Blueprinting)} \ \ (Sec.~\ref{sec:layer1})
\State $\mathcal{R} \gets \text{ExtractRoles}(\mathcal{O})$
\State $\{(n_i,\mathcal{P}_{n_i})\} \gets \text{MapRolesToNodes}(\mathcal{O},\mathcal{R})$
\State $\mathcal{N} \gets \{n_i\}$,\quad $\mathcal{E} \gets \text{MapDependencies}(\mathcal{N}, D)$
\State $\mathcal{C} \gets \text{CollectInvariants}(D,\mathcal{O})$
\State $\mathcal{W}_{\text{template}} \gets (\mathcal{N},\mathcal{E},\mathcal{C})$
\Statex

\Statex \hspace{-1.5em} \textbf{Phase 2: Role \& Agent Specs (Factory Prep)}\ \ (Sec.~\ref{sec:layer1})
\ForAll{$n_i \in \mathcal{N}$}
  \State $\mathcal{L}_{n_i} \gets \text{DefineLogSchema}(n_i,\mathcal{P}_{n_i})$  \hfill // events, snapshots, diffs
  \State $\alpha_{n_i} \gets \text{DefineAgent}(n_i,\mathcal{P}_{n_i},\mathcal{L}_{n_i})$ \hfill // cap/ctx/io/log
  \State $\rho_{n_i} \gets \text{DefineRepairSpec}(n_i,\text{local moves},\text{scope bounds})$
\EndFor
\ForAll{$e_{ij} \in \mathcal{E}$}
  \State $\mathcal{L}_{e_{ij}} \gets \text{DefineLogSchema}(e_{ij},\mathcal{P}_{e_{ij}})$
  \State $\alpha_{e_{ij}} \gets \text{DefineEdgeAgent}(e_{ij},\mathcal{L}_{e_{ij}})$
  \State $\rho_{e_{ij}} \gets \text{DefineRepairSpec}(e_{ij},\text{local moves},\text{scope bounds})$
\EndFor
\State $\mathcal{L} \gets \text{AssembleGlobalLogSchema}(\{\mathcal{L}_{n_i}\},\{\mathcal{L}_{e_{ij}}\})$
\State $V \gets \text{AttachValidator}(\kappa,\mathcal{C},\mathcal{L})$ \hfill // planner $\neq V$ (non-circular)
\Statex

\Statex \hspace{-1.5em} \textbf{Phase 3: Validation and Refinement} \ \ (Sec.~\ref{sec:layer3})
\State $\mathcal{W}_{\text{template}} \gets \text{UpdateWorkflow}(\mathcal{N},\mathcal{E},\alpha,\rho,\mathcal{C},\mathcal{L})$
\While{\textbf{not} $\text{ValidateBy}(V,\mathcal{W}_{\text{template}},\mathcal{M})$}
  \State $V.\text{CheckStructure}(\mathcal{W}_{\text{template}})$
  \State $V.\text{CheckConstraints}(\mathcal{W}_{\text{template}},\mathcal{C})$
  \State $V.\textbf{CheckRepairCoverage}(\mathcal{W}_{\text{template}},\Phi,\rho)$ \hfill // local neighborhoods exist/bounded
  \State $\mathcal{W}_{\text{template}} \gets \text{RefineWorkflow}(\mathcal{W}_{\text{template}},\mathcal{M})$
\EndWhile
\State \Return $\mathcal{W}_{\text{template}}=(\mathcal{N},\mathcal{E},\mathcal{C},\mathcal{L})$
\end{algorithmic}
\end{multicols}
\end{small}
\vspace{-0.15in}
\end{algorithm}

\begin{algorithm}[t]
\caption{Phase 2/4: ValidationTools.validate\_schedule}
\label{alg:validation}
\begin{algorithmic}[1]
\Require Candidate schedule $\mathcal{S}$ (list or dict)
\Ensure Validation result $\{valid, errors\}$

\State Parse $\mathcal{S}$ into entries $\{e_i\}$
\If{$\{e_i\} = \emptyset$} \Return $\{valid=False, errors=\{\text{``empty schedule''}\}\}$ 

\For{each $e_i \in \{e_i\}$}
    \State Check required fields: job, step, machine, start, end
    \State Check numeric type: start, end
    \State Check time ordering: $e_i.start < e_i.end$
\EndFor

\State Compute makespan $T = \max_i e_i.end$
\If{$T \leq 0$ or $T >$ threshold} \State Add makespan error
\EndIf

\State Validate job precedence:
\For{each job $j$}
    \State Ensure $\text{step}(k+1).start \geq \text{step}(k).end$
\EndFor

\State Validate machine capacity:
\For{each machine $m$}
    \State Ensure no overlapping intervals $\{(start,end)\}$
\EndFor

\If{dataset info available}
    \State Check all jobs $\in$ dataset appear
    \State Check all machines $\in$ dataset appear
\EndIf

\If{no errors} \State \Return $\{valid=True, errors=\emptyset\}$ 
\Else \State \Return $\{valid=False, errors\}$
\EndIf

\EndIf
\end{algorithmic}
\end{algorithm}

\begin{algorithm}[t]
\caption{Phase 3: LocalRepairTools.fix\_schedule (Planning with Cascading Repair and Optimization)}
\label{alg:repair}
\begin{algorithmic}[1]
\Require Invalid schedule $\mathcal{S}$, error set $\mathcal{E}$
\Ensure Repaired schedule $\mathcal{S}'$
\State Parse $\mathcal{S}$ into entries $\{e_i\}$

\Statex
\textbf{Phase I: Immediate Fixes}
\For{each error $e \in \mathcal{E}$}
    \If{job precedence violated}
        \State Delay current op until previous ends
    \ElsIf{time consistency violated}
        \State Adjust end $\gets$ start + duration
    \EndIf
\EndFor

\Statex
\textbf{Phase II: Job Precedence Repair}
\For{each job $j$}
    \State Sort ops by step
    \State Enforce $start_{k+1} \geq end_k$
\EndFor

\Statex
\textbf{Phase III: Machine Capacity Repair}
\For{each machine $m$}
    \State Sort ops by start
    \If $op_{i+1}.start < op_i.end$: delay $op_{i+1}$ \EndIf
\EndFor

\Statex
\textbf{Phase IV: Iterative Improvement}
\For{iteration $=1 \ldots K$}
    \For{each machine $m$}
        \State Try swapping adjacent ops
        \If{swap reduces makespan \& respects precedence}
            \State Commit swap
        \EndIf
    \EndFor
\EndFor

\Statex
\textbf{Phase V: Final Cleanup}
\For{each $e_i$}
    \If{$start \geq end$} \State $end \gets start + 1$ \EndIf
    \If{$start < 0$} \State $start \gets 0$ \EndIf
\EndFor
\Return repaired schedule $\mathcal{S}'$
\end{algorithmic}
\end{algorithm}

%% file: Figure-ALAS-Architecture.tex
\vspace{-.1in}
\begin{figure}[ht!]
\centering
\begin{tikzpicture}[
  font=\normalsize,
  every node/.style={align=center, rounded corners=2pt, scale=0.8, every node/.style={scale=0.8, font=\normalsize}},
  layer/.style={draw, thick, fill=blue!10, minimum width=2.2cm, minimum height=0.8cm},
  mem/.style={draw, thick, fill=yellow!25, minimum width=2.0cm, minimum height=1.6cm},
  arrow/.style={->, thick}
]


\node[layer] (layer1) at (-3.4, 2.6) {\large \textbf{Template Layer} \\ \normalsize $\mathcal{T}$: Roles \& Constraints};
\node[layer] (layer2) at (-3.4, 1.6) {\large \textbf{Factory Layer} \\ \normalsize $\mathcal{F}$: Agent Instantiation};
\node[layer] (layer3) at (-3.4, 0.6) {\large \textbf{Runtime Layer} \\ \normalsize $\mathcal{R}$: Execution \& Repair};

\node[mem] (memory) at (2.4, 2.0) {\large \textbf{Persistent Memory} \\ \normalsize Logs, State, Rollback, Validation};

\draw[arrow] (layer1.south) -- (layer2.north);
\draw[arrow] (layer2.south) -- (layer3.north);

\draw[->, >=latex, thick] (layer1.east) -- (memory.west);
\draw[->, >=latex, thick] (layer2.east) -- (memory.west);
\draw[<->, >=latex, thick] (layer3.east) -- (memory.west);

\node at (-3.4, 3.3) {\large \textbf{Problem Specification}};

\draw[arrow] (-3.4,3.1) -- (layer1.north);

\end{tikzpicture}
\vspace{.1in}
\caption{ALAS architecture: a lightweight LLM-driven planner with layered decomposition. Persistent memory supports all layers by storing state, validating constraints, and enabling recovery.}
\label{fig:ALAS-architecture}
\end{figure}
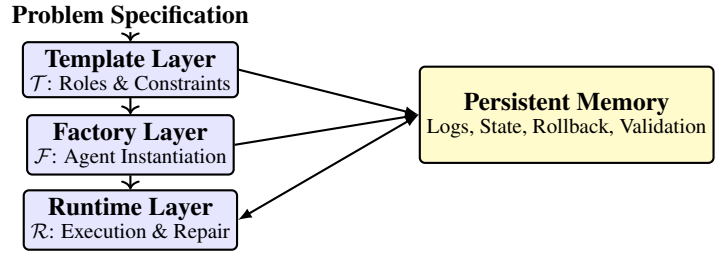

%% file: AppendixAgentFactory.tex
\subsection{Agent Factory Implementation Details}
\label{app:ALAS-Layer2}

This appendix provides detailed information on the Agent Factory component of the ALAS architecture, expanding on the summary provided in the main paper.

\subsubsection{Agent Factory Overview}

The Agent Factory translates formal agent specifications from the meta-planner into executable implementations. It employs a two-stage approach: first attempting to discover existing implementations, and then generating custom implementations when necessary.

\subsubsection{Agent Discovery Process}

The discovery process systematically searches for existing agent implementations that match specifications from phase \#1. For each agent specification $\alpha_i$, the discovery mechanism:

\begin{enumerate}[leftmargin=1.2em, topsep=0pt, itemsep=0pt, label=\arabic*.]
    \item Extracts the capability profile $\mathbf{c}_i$ and constructs a search query to identify potential implementations
    \item Retrieves candidate implementations from:
\begin{itemize}[leftmargin=1.0em, topsep=-.0em, parsep=-.0em, label=-]        \item Public agent repositories (e.g., GitHub, HuggingFace)
        \item API directories and marketplaces
        \item Pre-validated component libraries
        \item Domain-specific collections
    \end{itemize}
    \item Evaluates candidate suitability using multiple criteria:
\begin{itemize}[leftmargin=1.0em, topsep=-.0em, parsep=-.0em, label=-]
        \item Capability matching: Verifies that all required capabilities in $\mathbf{c}_i$ are supported
        \item Protocol compatibility: Ensures compatibility with the specified protocol buffer $\beta_i$
        \item Efficiency compliance: Validates performance against efficiency requirements $e_i$
        \item Context sizing: Confirms the implementation can operate within context window $w_i$
        \item Logging support: Verifies support for the required logging schema $\mathcal{L}_i$
    \end{itemize}
    \item Ranks candidates using a weighted scoring function $S(\alpha_i, I_j)$ where $I_j$ is a candidate: 
    \begin{equation}
        S(\alpha_i, I_j) = \sum_{k} w_k \cdot f_k(\alpha_i, I_j)
    \end{equation}
    where $w_k$ is the weight assigned to criterion $k$, and $f_k$ is an evaluation function for that criterion.
\end{enumerate}

When a suitable implementation is identified, it undergoes verification testing to confirm operational compatibility with the workflow requirements. Upon successful verification, the implementation is registered in the agent repository with appropriate metadata linking it to the specification.

The discovery mechanism employs both exact and approximate matching techniques. Exact matching requires all specification parameters to be satisfied precisely, while approximate matching allows for partial capability matching when accompanied by adaptation mechanisms.

\subsubsection{Agent Coding Mechanism}

When discovery fails to locate suitable implementations, the Agent Factory switches to its coding mechanism, which uses LLMs to generate custom implementations. The coding process follows a structured methodology:

\begin{enumerate}[leftmargin=1.2em, topsep=0pt, itemsep=0pt, label=\arabic*.]
    \item \textbf{Specification Translation}: The formal agent specification is translated into a natural language implementation brief that serves as the prompt for the LLM. This translation preserves all critical requirements while expressing them in a form that maximizes LLM comprehension.
    
    \item \textbf{LLM Selection}: An appropriate LLM is selected based on:
\begin{itemize}[leftmargin=1.0em, topsep=-.0em, parsep=-.0em, label=-]
        \item Domain expertise matching capability requirements in $\mathbf{c}_i$
        \item Demonstrated proficiency in generating the required implementation type
        \item Context window compatibility with the complexity of the specification
        \item Robustness against hallucination for critical components
    \end{itemize}
    
    \item \textbf{Implementation Generation}: The selected LLM generates implementation code with:
\begin{itemize}[leftmargin=1.0em, topsep=-.0em, parsep=-.0em, label=-]
        \item Embedded logging that conforms to schema $\mathcal{L}_i$
        \item Protocol handling for buffer $\beta_i$
        \item Optimizations for efficiency parameters $e_i$
        \item Adaptation to context window constraints $w_i$
    \end{itemize}
    
    \item \textbf{Implementation Validation}: The generated implementation undergoes validation to ensure:
\begin{itemize}[leftmargin=1.0em, topsep=-.0em, parsep=-.0em, label=-]
        \item Functional correctness against specification requirements
        \item Proper integration with the compensation mechanisms defined in $\alpha_i^{comp}$
        \item Robustness against edge cases and exceptional conditions
        \item Compliance with system-wide constraints and protocols
    \end{itemize}
\end{enumerate}

For particularly complex agents, the coding process may employ a multi-stage approach where the implementation is generated iteratively, with each iteration refining the previous version based on validation feedback.

\subsubsection{Compensation Agent Generation}

Special attention is given to the generation of compensation agents, which require precise understanding of the primary agent's operations to ensure proper reversal or mitigation. The generation of compensation agents follows these additional steps:

\begin{enumerate}[leftmargin=1.2em, topsep=0pt, itemsep=0pt, label=\arabic*.]
    \item Extraction of the primary agent's state-modifying operations
    \item Analysis of operation dependencies and sequencing constraints
    \item Determination of appropriate compensation strategies (e.g., undo, retry, escalate)
    \item Generation of the recovery sequence $\Gamma_i$ that defines the steps for returning to a consistent state
\end{enumerate}

The Factory ensures that compensation agents maintain strict operational correspondence with their primary counterparts, guaranteeing every state-modifying operation has a reversal mechanism defined.

\subsubsection{Deployment Artifact Production}

The output of the Agent Factory is a deployable artifact that encapsulates the agent's logic and interaction patterns. These artifacts take several forms, depending on the agent type and implementation approach:

\begin{itemize}[leftmargin=1.0em, topsep=-.0em, parsep=-.0em, label=-]
    \item \textbf{Code Snippets}: Executable code implementing the agent's functionality, typically for computationally intensive or specialized tasks
    \item \textbf{Prompt Templates}: Structured prompts that guide LLMs to implement the specified behavior at runtime, used for cognitively complex or reasoning-intensive tasks
    \item \textbf{API Configurations}: Parameter sets and endpoint specifications for interacting with external services or pre-existing agents
    \item \textbf{Hybrid Implementations}: Combined approaches that leverage both code and LLM prompting for different aspects of the agent's functionality
\end{itemize}

Each artifact is accompanied by metadata that defines its:

\begin{itemize}[leftmargin=1.0em, topsep=-.0em, parsep=-.0em, label=-]
    \item Execution requirements (e.g., runtime environment, dependencies)
    \item Interface specifications for input/output handling
    \item State persistence requirements and mechanisms
    \item Monitoring hooks for runtime observation
    \item Recovery points for compensation handling
\end{itemize}

\subsubsection{Factory Design Pattern Implementation}

The Agent Factory implements the classic Factory design pattern, providing a standardized interface for agent instantiation while encapsulating the complexity of implementation selection, generation, and validation. This pattern enables:

\begin{itemize}[leftmargin=1.0em, topsep=-.0em, parsep=-.0em, label=-]
    \item Decoupling of agent specifications from implementation details
    \item Support for heterogeneous implementation technologies
    \item Runtime substitution of agents when needed for recovery or optimization
    \item Maintenance of a growing repository of reusable components
\end{itemize}

The Factory pattern allows the ALAS system to evolve its agent implementation strategies over time without requiring changes to the meta-planning or runtime components, creating a flexible architecture that can adapt to new implementation technologies and approaches.

\subsubsection{Implementation Efficiency Considerations}

To maximize system efficiency, the Agent Factory implements several optimization strategies:

\begin{enumerate}[leftmargin=1.2em, topsep=0pt, itemsep=0pt, label=\arabic*.]
    \item \textbf{Caching}: Previously generated implementations are cached and indexed by their specifications to avoid redundant generation
    \item \textbf{Component Reuse}: Complex implementations are decomposed into reusable components that can be shared across multiple agents
    \item \textbf{Incremental Refinement}: When similar agents have been previously implemented, the Factory uses delta-based generation to create variants rather than generating entirely new implementations
    \item \textbf{Resource Scaling}: Implementation generation resources are allocated proportionally to the complexity and criticality of the agent
\end{enumerate}

These optimizations significantly reduce the computational overhead of agent generation, particularly in scenarios where multiple similar agents are required or when the system executes recurring workflow patterns.

\subsubsection{Theoretical Foundations}

The Agent Factory design is grounded in several theoretical frameworks:

\begin{itemize}[leftmargin=0.8em, topsep=-.0em, parsep=-.0em, label=-]
    \item \textit{Program Synthesis}: Formal methods for generating programs from specifications
    \item \textit{Component-Based Software Engineering}: Principles of component composition and reuse
    \item \textit{LLM Prompt Engineering}: Techniques for directing LLM behavior through structured prompts
    \item \textit{Agent-Oriented Software Engineering}: Methodologies for developing autonomous software agents
\end{itemize}

These foundations provide a rigorous basis for the Factory's approach to transforming abstract agent specifications into concrete, executable implementations.

%% file: AppendixLCSR.tex
\section{Algorithm, Lemma, and Theory Proofs}
\label{app:lcsr}

\begin{lemma}[Generalized LCRP Complexity]
\mbox{}\\[0.5ex]
For a system with:
\begin{itemize}[leftmargin=1.2em, topsep=-.15pt, itemsep=-.15pt]
    \item $J$ jobs
    \item $M$ machines 
    \item At most $O_{\max}$ operations per job
    \item $S$ average swap evaluations per queue ($1 \leq S \leq J$)
\end{itemize}
The worst-case time complexity is:
\begin{equation}
\mathcal{O}\left(\frac{J^2 O_{\max}^2}{M} + J M O_{\max}\right)
\end{equation}
\end{lemma}

\begin{proof}
The complexity derives from four components:

\begin{enumerate}[leftmargin=1.5em, topsep=-.15pt, itemsep=-.15pt]
    \item \textit{Status Update}: $\mathcal{O}(J O_{\max})$ \\
    Must check all operations of all jobs
    
    \item \textit{Delay Propagation}: $\mathcal{O}(J O_{\max})$ \\
    Each job's operation chain may have $O_{\max}$ elements
    
    \item \textit{Queue Optimization}:
    \begin{itemize}[leftmargin=1.2em, topsep=-.15pt, itemsep=-.15pt]
        \item Full analysis: $\mathcal{O}\left(\frac{J^2 O_{\max}^2}{M}\right)$ \\
        All operation pairs on all machines
        \item Practical bound: $\mathcal{O}(S J O_{\max})$ \\
        When swaps are limited to $S$ evaluations
    \end{itemize}
    
    \item \textit{Cascading Delay}: $\mathcal{O}(J M O_{\max})$ \\
    Worst-case propagation through all machines
\end{enumerate}

The dominant terms combine to give the final complexity:
\[
\underbrace{\frac{J^2 O_{\max}^2}{M}}_{\text{queue optimization}} + \underbrace{J M O_{\max}}_{\text{cascading delays}}
\]
\end{proof}

\begin{corollary}[Special Cases]
\begin{itemize}[leftmargin=1.2em, topsep=-.15pt, itemsep=-.15pt]
    \item \textit{Single-operation jobs} ($O_{\max}=1$): 
    $\mathcal{O}(J^2/M + JM)$
    
    \item \textit{Fully parallel systems} ($M \approx J$): 
    $\mathcal{O}(J O_{\max}^2 + J^2 O_{\max})$
    
    \item \textit{Swap-limited implementations}: 
    $\mathcal{O}(S J O_{\max} + J M O_{\max})$
\end{itemize}
\end{corollary}

\vspace{0.5em}
\textit{Key Observations}:
\begin{itemize}[leftmargin=1.2em, topsep=-.15pt, itemsep=-.15pt]
    \item Complexity is quadratic in job count and operations
    \item Machine count appears both in numerator (delays) and denominator (parallelization)
    \item Practical implementations can achieve better bounds through swap heuristics
\end{itemize}

\begin{definition}[LCRP-Repair Decision Problem]
\label{def:lcrp-decision}
\mbox{}\\[-0.75ex]
\textbf{Instance:} A job-shop instance $\mathcal{I}=(J,M,\{O_j\}_{j\in J})$ with processing times and machine requirements, an initial (possibly infeasible) schedule $\sigma_0$, a repair budget $R\in\mathbb{N}$, and a makespan bound $K\in\mathbb{N}$.\\
\textbf{Question:} Does there exist a repaired schedule $\sigma$ obtained from $\sigma_0$ by at most $R$ local edits (insert/move/swap/reassign) such that $\sigma$ is feasible (no machine overlap, precedence respected) and $\mathrm{C}_{\max}(\sigma)\le K$?
\end{definition}

\begin{theorem}[LCRP is NP-hard (in fact, strongly NP-hard)]
\label{thm:lcrp-nphard}
The decision problem in Def.~\ref{def:lcrp-decision} is NP-hard. Moreover, it is strongly NP-hard.
\end{theorem}

\begin{proof}[Proof sketch]
We reduce from the standard \emph{Job-Shop Scheduling} (JSSP) \emph{decision} problem: given $\mathcal{I}=(J,M,\{O_j\})$ and bound $K$, decide whether a feasible schedule with makespan $\le K$ exists. This problem is well known to be NP-hard and, in fact, strongly NP-hard.

\paragraph{Reduction (polynomial time).}
Given a JSSP instance $(\mathcal{I},K)$, construct an LCRP instance as follows:
\begin{itemize}[leftmargin=1.1em, itemsep=2pt]
  \item Use the \emph{same} set of jobs, machines, operation orders, and processing times.
  \item Let the initial schedule be $\sigma_0:=\emptyset$ (no operations placed) or any trivially infeasible “dummy” placement.
  \item Set the repair budget $R$ to a value $\ge$ the number of operations (e.g., $R=\sum_{j\in J}|O_j|$), so that any feasible arrangement can be reached by a sequence of local edits permitted by the repair model (insert/move/swap/reassign).
  \item Keep the same makespan bound $K$.
\end{itemize}
This mapping is clearly polynomial in input size.

\paragraph{Correctness.}
We show $(\mathcal{I},K)$ is a ``yes'' instance of JSSP iff the constructed LCRP instance is a ``yes'' instance.

$(\Rightarrow)$ If JSSP admits a feasible schedule $\sigma^\star$ with $\mathrm{C}_{\max}(\sigma^\star)\le K$, then starting from $\sigma_0$ we can obtain $\sigma^\star$ via at most $\sum_{j}|O_j|$ local edits (insert each operation in its position or move/swaps to match $\sigma^\star$). Since $R$ was chosen at least that large, a valid repair sequence exists. Thus LCRP answers ``yes''.

$(\Leftarrow)$ Conversely, suppose the LCRP instance admits a repaired schedule $\hat{\sigma}$ within budget $R$ that is feasible and satisfies $\mathrm{C}_{\max}(\hat{\sigma})\le K$. Then $\hat{\sigma}$ is a feasible JSSP schedule for $\mathcal{I}$ of makespan at most $K$, so JSSP answers ``yes''.

\paragraph{Membership in NP.}
A certificate consists of the repaired schedule $\sigma$ (start/end times and machine assignments for all operations) and, optionally, the edit sequence (bounded by $R$). Using the standard feasibility checks (no machine overlaps, precedence respected) and a single pass to compute $\mathrm{C}_{\max}$, verification runs in polynomial time in the input size (cf.\ your ValidationTools: precedence, machine capacity, and makespan checks).

\paragraph{Strong NP-hardness.}
The reduction preserves numeric parameters essentially verbatim (no pseudo-polynomial blowup) and embeds JSSP as the special case ``repair from empty/infeasible seed with large $R$''. Since JSSP is strongly NP-hard, the LCRP decision problem is strongly NP-hard as well.

\end{proof}

\begin{remark}[Bounded-edit variants remain NP-hard]
Even if the budget $R$ is part of the input and \emph{small}, NP-hardness persists by embedding feasibility into a bounded number of edit operations (e.g., composing each operation placement as one edit), or by initializing $\sigma_0$ to be ``nearly empty'' so that $R$ equals the number of to-be-placed operations. Hence hardness is robust to common repair-budget formulations.
\end{remark}

\begin{remark}[Relation to the complexity bound]
The operational bound 
\(
\mathcal{O}\!\left(\tfrac{J^2 O_{\max}^2}{M}+JM O_{\max}\right)
\)
captures worst-case \emph{per-run} effort of LCRP (status update, local queue optimization, and cascading delay handling). The NP-hardness result shows that, independent of such polynomial per-iteration costs, deciding the existence of a repair achieving a global makespan target is computationally intractable in general (unless $\mathrm{P}=\mathrm{NP}$).
\end{remark}

%% file: AppendixURS.tex
\section{Application 1: Urban Ride Assignment Problem}
\label{app:ALAS-URAP}

The goal is to optimally assign ride requests to a fleet of autonomous or human-driven vehicles in a city, while satisfying various constraints and objectives. The key elements are the following.

\begin{itemize}[leftmargin=1.0em, topsep=-.0em, parsep=-.0em, label=*]
\item \textbf{City Map:} A graph $G = (V, E)$ where $V$ is the set of locations and $E$ is the set of roads connecting them, with associated distances and travel times.
\item \textbf{Ride Requests:} A set of requests $R$, where each request $r_i \in R$ is characterized by:
\begin{itemize}[leftmargin=1.0em, topsep=-.0em, parsep=-.0em, label=-]
    \item Passenger ID $p_i$
    \item Pickup location $v_{p_i} \in V$ 
    \item Drop-off location $v_{d_i} \in V$
    \item Desired pickup time window $[t_{p_i}^{min}, t_{p_i}^{max}]$
    \item Desired drop-off time window $[t_{d_i}^{min}, t_{d_i}^{max}]$
\end{itemize}

\item \textbf{Vehicles:} A set of vehicles $K$, where each vehicle $k_j \in K$ has:
\begin{itemize}[leftmargin=1.0em, topsep=-.0em, parsep=-.0em, label=-]
    \item Vehicle ID $k_j$ 
    \item Current location $v_{k_j} \in V$
    \item Battery/fuel level $b_{k_j} \in [0, 1]$
    \item Passenger capacity $c_{k_j} \in \mathbb{Z}^+$
    \item Speed $s_{k_j} \in \mathbb{R}^+$
\end{itemize}

\end{itemize}

\subsection{A Simplified URS Problem Statement}

Table~\Ref{fig:appDURSSpec} in the main text depicts a URS problem
with three drivers and four passengers.
Using this problem, we walk through how $\ALAS$ works.

\subsection{Generating Planner W* Walkthrough}

Given the problem statement of URS, $\ALAS$ generates a planning template $\mathcal{W_\text{template}}$.

\begin{table*}[ht!]
\caption{Agent Specifications and Protocols}
\begin{footnotesize}
\begin{tabular}{|p{1.8cm}|p{3.4cm}|p{3.3cm}|p{3.5cm}|}
\hline
\textbf{Agent Type} & \textbf{Input Protocol} & \textbf{Output Protocol} & \textbf{Key Functions} \\
\toprule
\hline
\multicolumn{4}{|c|}{\textbf{Task-Specific Agents}} \\
\hline
Route \newline Planning & 
- Location map $G(V,E)$ \newline
- Travel times matrix \newline
- Vehicle positions & 
- Optimized routes \newline
- Distance calculations \newline
- Path sequences & 
- Path optimization \newline
- Distance minimization \newline
- Route feasibility checks \\
\hline
Scheduling & 
- Required arrival times \newline
- Travel duration estimates \newline
- Vehicle availability & 
- Pickup schedule \newline
- Timing constraints \newline
- Buffer allocations & 
- Schedule generation \newline
- Timing verification \newline
- Buffer management \\
\hline
Capacity \newline Management & 
- Passenger requests \newline
- Vehicle capacities \newline
- Route timing & 
- Passenger groupings \newline
- Vehicle  \newline
- Capacity utilization & 
- Group optimization \newline
- Capacity verification \newline
- Load balancing \\
\hline
\midrule
\multicolumn{4}{|c|}{\textbf{Common Agents}} \\
\hline
Temporal \newline Constraint & 
- Schedule requirements \newline
- Time windows \newline
- Buffer needs & 
- Timing validations \newline
- Constraint satisfaction \newline
- Buffer adequacy & 
- Time verification \newline
- Constraint checking \newline
- Buffer analysis \\
\hline
Resource \newline Allocation & 
- Vehicle inventory \newline
- Request demands \newline
- Location data & 
- Resource assignments \newline
- Utilization plans \newline
- Coverage maps & 
- Resource optimization \newline
- Coverage verification \newline
- Efficiency analysis \\
\hline
Distance \newline Optimization & 
- Route options \newline
- Distance matrix \newline
- Time constraints & 
- Optimized paths \newline
- Distance metrics \newline
- Efficiency scores & 
- Path optimization \newline
- Distance reduction \newline
- Efficiency maximization \\
\hline
\midrule
\multicolumn{4}{|c|}{\textbf{Validation Agents}} \\
\hline
Plan Validator & 
- Complete plan \newline
- System constraints \newline
- Quality metrics & 
- Validation results \newline
- Constraint checks \newline
- Performance scores & 
- Plan verification \newline
- Constraint validation \newline
- Quality assessment \\
\hline
Refinement Agent & 
- Validation results \newline
- Improvement options \newline
- Performance metrics & 
- Refinement suggestions \newline
- Update priorities \newline
- Optimization paths & 
- Plan improvement \newline
- Update sequencing \newline
- Performance optimization \\
\hline
\end{tabular}
\end{footnotesize}
\label{tab:appAgentSpecs}
\end{table*}

\subsubsection{State-Space Analysis}

Our Urban Ride-Sharing (URS) problem presents a complex transportation scheduling challenge that we must first understand through systematic state-space analysis. The system involves seven locations (A through G), where G represents Boston Logan Airport, with urban locations forming a mesh network of 10km distances and airport routes ranging from 31-36km. Four passengers require airport transportation with specific arrival deadlines, while three vehicles, each capable of carrying two passengers, must be coordinated to meet these demands efficiently.

Each dimension of our state space reveals crucial aspects of the planning challenge. In the \emph{Who} dimension, we track four passenger requests ($r_1$ through $r_4$) and three vehicles ($k_1$ through $k_3$). These passengers require arrivals at BOS between 08:45 and 09:00, with each vehicle qualified for airport routes and positioned initially at locations A, C, and E. 

The \emph{Where} dimension maps our network topology, distinguishing between urban segments with uniform 10km distances and airport routes varying from 31-36km. This spatial arrangement, combined with the \emph{When} dimension's speed constraints (60km/h urban, 100km/h airport routes), creates our fundamental timing framework. Simple calculations reveal urban segments require 10 minutes of travel time, while airport routes need 19-22 minutes depending on origin.

Our \emph{What} dimension monitors vehicle resources throughout plan execution, ensuring we respect the two-passenger capacity limit while maximizing sharing opportunities. The \emph{Why} dimension establishes our optimization objectives: ensuring on-time airport arrivals while minimizing total distance traveled. The \emph{How} dimension defines our execution protocols, including pickup sequencing and route navigation strategies.

\subsubsection{Phase 1: Network Construction}

Building upon our state-space analysis, we construct our planning network by first identifying critical nodes and dependencies. Our node set $\mathcal{N}$ comprises:

Passenger Nodes: Each request $r_i$ becomes a node with attributes:
- $r_1$: Location A, BOS arrival 08:45
- $r_2$: Location B, BOS arrival 08:50
- $r_3$: Location C, BOS arrival 08:55
- $r_4$: Location D, BOS arrival 09:00

Vehicle Nodes: Each vehicle $k_i$ forms a node with position and capacity:
- $k_1$: Starting at A, capacity 2
- $k_2$: Starting at C, capacity 2
- $k_3$: Starting at E, capacity 2

Location Nodes: Each physical location becomes a node with attributes including distance to other locations and travel time calculations.

Our dependency set $\mathcal{E}$ captures relationships between these nodes through several categories:

Temporal Dependencies: We establish feasible pickup windows by working backward from required arrival times. For example, $r_1$ requires 22 minutes for the airport route plus 10 minutes for each urban segment traversed, creating timing constraints for vehicle assignment.

Spatial Dependencies: We map possible routes between nodes, considering both direct airport routes and potential shared-ride combinations through urban segments.

Capacity Dependencies: We create edges representing feasible passenger groupings within vehicle capacity limits.

\subsubsection{Phase 2: Agent Assignment}

With our network structure defined, we assign specialized agents to manage different aspects of the solution:

Task-Specific Agents:
The Route Planning Agent optimizes paths using the distance matrix and travel speeds, calculating optimal routes for both single and shared rides. The Scheduling Agent determines precise pickup times, working backward from airport deadlines and incorporating travel time calculations. The Capacity Management Agent identifies feasible passenger groupings based on timing and location proximity.

Common Agents:
The Temporal Constraint Agent ensures all timing requirements are met, maintaining a master schedule that accounts for all dependencies. The Resource Allocation Agent assigns vehicles to routes, optimizing the distribution of available capacity. The Distance Optimization Agent works to minimize total travel distance while respecting all constraints.

Edge Agents:
These agents manage the relationships between different aspects of the plan. For example, the Passenger Grouping Agent evaluates potential shared rides by analyzing proximity of pickup locations and compatibility of arrival times.

\subsubsection{Phase 3: Validation and Refinement}

In our final phase, we implement a comprehensive validation and refinement process:

Initial Validation:
We verify temporal feasibility by checking that all calculated pickup times allow sufficient travel time to meet airport deadlines. We confirm capacity constraints are respected throughout all vehicle routes. We validate that all passengers are served and all required resources are properly allocated.

Iterative Refinement:
We identify optimization opportunities, such as grouping passengers with compatible timing and locations. For example, passengers $r_2$ and $r_3$ might share a ride if their pickup locations are close and arrival times are within 5 minutes. We adjust vehicle assignments to minimize empty travel distance while maintaining service guarantees.

Final Plan Generation:
The resulting plan specifies exact pickup times, vehicle assignments, and routes, with built-in buffers for potential delays. The plan includes contingency protocols for common disruptions such as traffic delays or passenger late arrivals.

This systematic approach ensures we generate a robust, efficient solution to our URS problem while maintaining clear documentation of our planning process and decisions.

\subsubsection{Output}

Table~\ref{tab:appAgentSpecs} the list of required agents and
their functional specifications and protocols.

\begin{table*}[ht!]
    \centering
    \caption{Agent Placement in the Urban Ride Sharing Network}
        \begin{footnotesize}
        \begin{tabular}{|l|l|p{8cm}|}
        \hline
        \textbf{Location} & \textbf{Type} & \textbf{Agents and Their Responsibilities} \\
        \toprule \hline
        A--F & Nodes & \textbf{Resource Allocation Agent}: Manages vehicle assignments and passenger pickups at urban locations \\
        \hline
        G (Airport) & Node & 
        \begin{minipage}[t]{8cm}
        \textbf{Plan Validator Agent}: Verifies arrival times and  plan feasibility\\
        \textbf{Temporal Constraint Agent}: Ensures all arrival deadlines met
        \end{minipage} \\
        \hline
        A--F edges & Urban Routes & 
        \begin{minipage}[t]{8cm}
        \textbf{Route Planning Agent}: Optimizes urban route segments (10 min travel time)\\
        \textbf{Scheduling Agent}: Coordinates pickup sequences and timing
        \end{minipage} \\
        \hline
        (A,...,F)--G & Airport Routes & 
        \begin{minipage}[t]{8cm}
        \textbf{Capacity Management Agent}: Ensures vehicle capacity constraints during airport trips\\
        \textbf{Distance Optimization Agent}: Minimizes total travel distance
        \end{minipage} \\
        \hline
        Network-wide & Global & 
        \begin{minipage}[t]{8cm}
        \textbf{Refinement Agent}: Iteratively improves solutions based on validation results\\
        Monitors and adjusts both urban and airport route segments
        \end{minipage} \\
        \hline
    \end{tabular}
    \end{footnotesize}
    \vspace{-.1in}
    \label{tab:agent-placement}
\end{table*}

\subsection{From Workflow Template to Execution Workflow}

Once the template $\mathcal{W_\text{template}}$ is defined, it serves as a structured blueprint that outlines how the problem should be approached. However, a high-level plan alone is insufficient for real-world execution. The next step is to transform the planning workflow into a \textit{real execution workflow} $\mathcal{W_\text{exec}}$, where abstract roles and dependencies are resolved into concrete actionable tasks based on real-world data.

To clarify this transition, consider the difference between $\mathcal{W_\text{template}}$ and $\mathcal{W_\text{exec}}$ in our ride-sharing scenario. In the planning phase, roles such as \texttt{Driver} and \texttt{Passenger} are defined as abstract entities. The template workflow $\mathcal{W_\text{template}}$ specifies how these entities interact, matching drivers with passengers, optimizing routes, and scheduling pickups, without assigning real-world counterparts yet. 

In contrast, the execution workflow $\mathcal{W_\text{exec}}$ performs \textbf{role resolution}, mapping abstract roles to real-world instances. This means assigning an actual driver to a specific vehicle, matching a real passenger to a ride request, and computing precise travel distances based on real-time geo-coordinates. In addition, the execution workflow must dynamically adapt to real-world constraints, such as traffic conditions, vehicle availability, and passenger delays. 

In this process, the meta-planner generates $\mathcal{W_\text{exec}}$, a directed graph where nodes correspond to concrete actions (e.g., ``Driver John departs from location A''), and edges represent dependencies and constraints (e.g., ``Driver John must reach location B before 10:30 AM''). This execution graph integrates real-time data and updates continuously, allowing agents to make informed decisions as conditions evolve.

Thus, the \textbf{template workflow} $\mathcal{W_\text{template}}$ structures how to plan, while the \textbf{execution workflow} $\mathcal{W_\text{exec}}$ governs how real-world actions are performed. Transformation from one to the other is a critical step in $\MP$, ensuring that strategic reasoning is translated into actionable real-time operations.

Now, based on the URS problem specified in Table~\Ref{fig:appDURSSpec} of Section~\ref{sec:ALAS-evaluation} in the main text, 
the list of agents required and
their functional specifications and protocols in Table~\ref{tab:appAgentSpecs}, $\MP$ proceeds generating an execution workflow $\mathcal{W_\text{exec}}$.

\subsubsection{Observation on Sequential Planning}

Let us explain the value of using agents in this problem, even though we have shown that simpler solvers can handle the computational aspects. This discussion touches on key principles of system design and real-world implementation.  

While our Monte Carlo solver effectively found good solutions for this specific instance, $\MP$ offers several advantages that become particularly valuable in real-world ride-sharing systems.  

First, $\MP$ helps manage complexity in dynamic environments. In our exercise, we worked with a static problem where all passenger requests and constraints were known in advance. However, in reality, ride-sharing systems must handle continuous updates—new ride requests arrive at unpredictable times, vehicles experience delays, and road conditions constantly change. With $\MP$, each agent operates independently, monitoring and reacting to changes in its own domain. For example, the Route Planning Agent can dynamically adjust routes in response to traffic updates, while the Capacity Management Agent ensures new passenger requests are accommodated efficiently.  

Second, $\MP$ enables distributed decision-making and parallel processing. Instead of relying on a centralized solver, different agents specialize in handling specific tasks simultaneously. While the Scheduling Agent optimizes pickup times, the Resource Allocation Agent manages vehicle assignments in parallel. This decentralized structure is crucial for scalability—when the system expands to hundreds of vehicles and thousands of passengers, distributing computational workload prevents bottlenecks and ensures efficient operations.  

Third, $\MP$ provides modularity, allowing the system to evolve naturally. Ride-sharing services frequently introduce new features, such as surge pricing or specialized vehicle categories. With an agent-based design, we can integrate a Pricing Agent or a Vehicle Specialization Agent without modifying the core routing logic. Likewise, if we develop a more advanced routing algorithm, we can upgrade the Route Planning Agent without disrupting other system components.  

The separation of concerns through agents also enhances system resilience. If one agent encounters issues—say, the Distance Optimization Agent fails to compute an optimal route—other agents continue operating with fallback strategies. The Plan Validator Agent can detect suboptimal assignments and trigger refinements through the Refinement Agent, ensuring that the system adapts to unforeseen challenges.  

We can think of this like a well-organized team working on a complex project. While a single individual might handle everything, a structured team of specialists—each with clear roles and defined communication protocols—is often more effective, robust, and scalable. In this way, while our Monte Carlo solver demonstrates what is mathematically possible, the agent-based architecture of $\MP$ shows how we can implement it reliably in real-world systems.

\subsection{Reactive Planning under Disruptions}
\label{app:MAPLE-URAP-Disruptions}

The value of multi-agent reactive planning becomes clear in dynamic environments. For example, consider a sudden road closure between locations B and C. While a monolithic solver would need to halt and recompute an entirely new plan from scratch, a modular agent-based approach enables localized, parallel adaptation. A Route Planning Agent can immediately update affected paths, while a Scheduling Agent adjusts arrival estimates, and a Resource Allocation Agent reallocates vehicles, all operating concurrently while preserving system stability. This distributed replanning minimizes disruption impact and maintains overall workflow coherence.

The following case study illustrates these principles in an Urban Ride Sharing (URS) scenario involving ride cancellation and new request insertion.

\paragraph{URS Disruption Handling.}
To evaluate adaptation capabilities, we introduce a disruption where passenger $r_2$ cancels the ride request at 8:05, and a new request $r_5$ at location F arrives at 8:10. $\MP$ replans dynamically, adjusting vehicle assignments while preserving all passenger deadlines. In contrast, baseline LLMs fail to track vehicle states after partial execution and lose consistency with the initial plan, leading to infeasible or incoherent schedules.

%% file: AppendixFamilyReunion.tex
\section{Application 2: Family Reunion Planning Problem}
\label{app:ALAS-MeetingProblem}

Table~\ref{apptab:ThanksgivingDinner} presents the specification of the problem.  The participating LLMs and their configurations
are depicted in Section~\ref{sec:ALAS-exp-case2} of the main text.



\begin{table}[ht!]
\centering
\caption{Thanksgiving Dinner Coordination Problem}
\begin{small}
\renewcommand{\arraystretch}{1.1}
\fbox{
\begin{minipage}{0.65\textwidth}
\textbf{Objective:} Coordinate family arrivals and dinner preparation for 6:00 PM dinner in Boston

\textbf{Family Members and Arrivals:}
\begin{itemize}[leftmargin=1em, topsep=-.1pt, itemsep=-.1pt, label=-]
\item Sarah (Mom): Host, at home
\item James (Dad): Lands at BOS 1:00 PM from SF
\item Emily (Sister): Lands at BOS 2:30 PM from Chicago
\item Michael (Brother): Driving, arrives 3:00 PM from NY
\item Grandma: Needs pickup from suburban Boston
\end{itemize}

\textbf{Cooking Requirements:}
\begin{itemize}[leftmargin=1em, topsep=-.1pt, itemsep=-.1pt, label=-]
\item Turkey: 4 hours cooking time
\item Side dishes: 2 hours preparation
\item Someone must stay home during cooking
\end{itemize}

\textbf{Transportation Constraints:}
\begin{itemize}[leftmargin=1em, topsep=-.1pt, itemsep=-.1pt, label=-]
\item James must rent car after landing
\item Emily requires airport pickup
\item Travel times:
   \begin{itemize}
   \item Home to BOS Airport: 60 min
   \item BOS Airport to Grandma's: 60 min
   \item Home to Grandma's: 30 min
   \end{itemize}
\end{itemize}

\textbf{Key Requirements:}
\begin{itemize}[leftmargin=1em, topsep=-.1pt, itemsep=-.1pt, label=-]
\item All family members at home for 6:00 PM dinner
\item Turkey and sides ready by dinner time
\item All pickups completed with available drivers
\item Cooking supervision maintained
\end{itemize}
\end{minipage}
}
\end{small}
\label{apptab:ThanksgivingDinner}
\end{table}

\subsection{Phase 1: Network Construction}

\subsubsection{Node (Role) Specifications}

First, meta-planner $\mathcal{MP}$ of $\ALAS$ extracts roles ($\mathcal{N}$) with their required qualifications:
\begin{itemize}[leftmargin=1.5em, topsep=-.15em, parsep=-.15em]
    \item $n_{\text{cook}}$: capability to prepare dinner
    \item $n_{\text{driver1}}$: capability to drive, pick up from airport
    \item $n_{\text{driver2}}$: capability to drive, pick up grandma
    \item $n_{\text{supervisor}}$: capability to monitor oven
\end{itemize}

\subsubsection{Edge (Dependency) Specifications}
Next, $\mathcal{MP}$ identifies dependencies ($\mathcal{E}$) between roles:
\begin{equation}
\mathcal{E} = \{e_{\text{temporal}}, e_{\text{spatial}}, e_{\text{safety}}\}
\end{equation}

The critical dependencies include:
\begin{itemize}[leftmargin=1.5em, topsep=-.15em, parsep=-.15em]
    \item $e_{\text{temporal}}$: 
        - Turkey (4 hours) must finish by 6:00 PM
        - Side dishes (2 hours) must finish by 6:00 PM
        - Airport pickups must align with landing times
    \item $e_{\text{spatial}}$: 
        - Driver-passenger location matching
        - Travel time constraints between locations
    \item $e_{\text{safety}}$:
        - Continuous oven supervision requirement
\end{itemize}


\subsection{Phase 2: Agent Assignments}

After constructing the network structure, $\mathcal{MP}$  selects and assigns agents to monitor both the roles and dependencies.

\subsubsection{Node (Role) Agent Assignment}
For each role, $\mathcal{MP}$ selects monitoring agents with the required capabilities:

\begin{equation}
f_{\text{role}}: \mathcal{N} \rightarrow \mathbf{A}
\end{equation}

The role monitoring agents include:
\begin{itemize}[leftmargin=1.5em, topsep=-.15em, parsep=-.15em]
\item Cook Monitor: Tracks cooking timeline, coordinates meal components
\item Driver Monitor: Validates driver availability 
\item Supervisor Monitor: Ensures oven supervision
\item Resource Monitor: Manages vehicle assignments and actor schedules
\end{itemize}

\subsubsection{Edge (Dependency) Agent Assignment}
For the identified dependencies, $\mathcal{MP}$ assigns specialized monitoring agents:

\begin{equation}
f_{\text{edge}}: \mathcal{E} \rightarrow \mathbf{A}
\end{equation}

Dependencies require these monitoring agents:
\begin{itemize}[leftmargin=1.5em, topsep=-.15em, parsep=-.15em]
\item Temporal Agent: Manages timing constraints (cooking durations, travel times, arrival schedules)
\item Spatial Agent: Tracks location constraints (airport-home-grandma routes)
\item Safety Agent: Ensures oven supervision constraint remains satisfied
\end{itemize}

The resulting agent assignments create a complete monitoring system where:
\begin{itemize}[leftmargin=1.5em, topsep=-.15em, parsep=-.15em]
\item Role agents track individual actor assignments and qualifications
\item Edge agents monitor interactions and dependencies between roles
\item All agents coordinate to maintain global constraint satisfaction
\end{itemize}

\begin{table}[th!]
\centering
\caption{Node and Edge Monitoring Agent Requirements}
\vspace{-.05in}
\begin{minipage}{0.48\textwidth}
\centering
\label{tab:node_agents}
\caption*{(a) Node (Role) Monitoring Agent}
{\fontsize{7pt}{9pt}\selectfont
\begin{tabular}{|p{0.15\textwidth}|p{0.32\textwidth}|p{0.38\textwidth}|}
\hline
\textbf{Agent} & \textbf{Input Protocol} & \textbf{Output Protocol} \\
\toprule \hline
Cook \newline Monitor & Role: cook \newline Qualifications: skills \newline Time: prep and cook & Status: progress \newline Alerts: timing issues! \newline Updates: completed? \\
\hline
Driver \newline Monitor & Role: driver \newline Qs: license, rest \newline Where: current GPS & Status: availability \newline Alerts: fatigue warnings \newline Updates: new GPS \\
\hline
Supervisor \newline Monitor & Role: supervisor \newline Location: house \newline Duration: cover time & Status: covered? \newline Alerts: coverage gaps! \newline Updates: role transitions \\
\hline
\end{tabular}
}
\end{minipage}
\hfill
\begin{minipage}{0.48\textwidth}
\centering
\caption*{(b) Edge (Dependency) Monitoring Agent}
\label{tab:edge_agents}
{\fontsize{7pt}{9pt}\selectfont
\begin{tabular}{|p{0.12\textwidth}|p{0.34\textwidth}|p{0.30\textwidth}|}
\hline
\textbf{Agent} & \textbf{Input Protocol} & \textbf{Output Protocol} \\
\toprule \hline
Temporal & Start times \newline Durations \newline Deadlines & Schedule conflicts \newline Timing violations \newline Schedule updates \\
\hline
Spatial & Locations \newline Routes \newline Travel time (variations) & Route violations \newline Location conflicts \newline Path updates \\
\hline
Safety & Critical constraints \newline Resource states \newline Coverage requirements & Safety violations \newline Resource conflicts \newline Mitigation plans \\
\hline
\end{tabular}
}
\end{minipage}
\vspace{-.1in}
\end{table}

\subsubsection{Common Sense Constraint Analysis (Performed by an LLM)}
\label{app:ThanksGiving-common-sense}

A common sense agent identifies the following implicit constraints that can affect Thanksgiving dinner planning.
This list is generated by Claude given the problem statement.

\begin{itemize}[leftmargin=1.5em, topsep=-.15em, parsep=-.15em]
\item \textit{Physical Processing Times:}
   \begin{itemize}
   \item Airport luggage claim: 30 minutes
   \item Car rental procedures: 30 minutes
   \item Holiday traffic variations
   \item Winter weather considerations
   \end{itemize}

\item \textit{Human Factors:}
   \begin{itemize}
   \item Driver fatigue after long trips
   \item Cooking preparation overhead
   \item Optimal turkey baking tips (non-disruptive baking and ready 30 minutes before eating)
   \item Task switching delays
   \item Required rest periods
   \end{itemize}

\item \textit{Resource Dependencies:}
   \begin{itemize}
   \item Vehicle passenger capacity
   \item Oven temperature management
   \item Kitchen workspace limits
   \item Shared resource coordination
   \end{itemize}

\item \textit{Social Considerations:}
   \begin{itemize}
   \item Personal preferences for interactions
   \item Family dynamics in assignments
   \item Post-travel guest comfort
   \item Host preparation requirements
   \end{itemize}
\end{itemize}

\input{TableWorkflow}

\subsubsection{Common Sense Constraint Analysis and Verification (Human in the Loop)}
The common sense constraints identified above require different verification approaches:

\paragraph{Agent-Required Information}
These constraints need specialized agents to verify and quantify:
\begin{itemize}[leftmargin=1.5em, topsep=-.15em, parsep=-.15em]
\item \textit{Airport Operations}
   \begin{itemize}
   \item United Airlines' average luggage delivery time at BOS Terminal B
   \item Terminal B to rental car center: shuttle schedule, walking options
   \item Historical flight delay patterns for November at BOS
   \end{itemize}

\item \textit{Weather and Traffic}
   \begin{itemize}
   \item Boston weather forecast for the event date
   \item Historical traffic patterns on Thanksgiving days
   \item Impact on airport-city-suburb travel times
   \end{itemize}

\item \textit{Task Dependencies}
   \begin{itemize}
   \item Kitchen workflow analysis for parallel cooking tasks
   \item Resource contention in meal preparation
   \item Critical path identification in cooking timeline
   \end{itemize}
\end{itemize}

\paragraph{Human Verification}
Certain constraints require explicit human input to ensure that the planning process takes into account subtle interpersonal and individual considerations. These include:

\begin{itemize}[leftmargin=1.5em, topsep=-.15em, parsep=-.15em]
    \item \textit{Family Dynamics}
    \begin{itemize}[leftmargin=1.5em, topsep=-.1em, parsep=-.1em]
        \item Preferred pickup arrangements for Grandma.
        \item Optimal relationship-based task pairings.
        \item Social comfort factors in assignments (e.g., Sarah and Grandma do not share a kitchen).
    \end{itemize}

    \item \textit{Personal Capabilities}
    \begin{itemize}[leftmargin=1.5em, topsep=-.1em, parsep=-.1em]
        \item Individual cooking experience levels.
        \item Driver comfort with airport navigation.
        \item Multi-tasking abilities of participants.
    \end{itemize}
\end{itemize}

This separation ensures that agents focus on collecting quantifiable data while humans provide essential social and personal insights. $\mathcal{MP}$ can then integrate both types of information into the final workflow design.

\subsection{Agent Requirements and Assignments}

The $\mathcal{MP}$ requires two categories of agents. $\mathcal{MP}$ specifies their requirements in the protocol buffer format in Table~\ref{tab:node_agents} for the nodes and
Table~\ref{tab:edge_agents} for the edges, respectively.

Each agent must implement these protocols to participate in the workflow. The meta-planner selects agents from the pool based on their ability to satisfy these interface requirements. During execution, agents communicate through these standardized protocols while maintaining their specialized monitoring functions.

\subsection{Monitoring Protocols and Dynamic Adjustments}

The workflow monitoring operates through a hierarchical protocol system that enables both routine supervision and dynamic adjustments.

\paragraph{Basic Monitoring Protocol}
Each agent maintains a continuous monitoring cycle:
\begin{equation}
\text{monitor}: \text{State} \rightarrow \{\text{normal, warning, violation}\}
\end{equation}

For example, the temporal agent tracks schedule adherence:
\begin{equation}
\Delta t = t_{\text{planned}} - t_{\text{actual}}
\begin{cases}
   \text{normal} & \text{if } |\Delta t| < \text{buffer} \\
   \text{warning} & \text{if } \text{buffer} \leq |\Delta t| < \tau \\
   \text{violation} & \text{if } |\Delta t| \geq \text{ threshold } \tau 
\end{cases}
\end{equation}

\paragraph{Dynamic  Adjustment Mechanism}
When deviations occur, the system initiates a three-phase response:

1. \textit{Impact Assessment}:
\begin{equation}
\text{impact}(e) = \sum_{n \in \text{affected}(e)} \text{severity}(n) \times \text{urgency}(n)
\end{equation}

2. \textit{Solution Generation}:
\begin{equation}
S^* = \argmin_{s \in \text{Solutions}} \{\text{cost}(s) | \text{feasible}(s)\}
\end{equation}

3. \textit{Coordination Protocol}:
\begin{equation}
\text{update}: (W_{\text{current}}, S^*) \rightarrow W_{\text{new}}
\end{equation}

For instance, if James's flight is delayed:
\begin{itemize}[leftmargin=1.5em, topsep=-.15em, parsep=-.15em]
\item Spatial agent detects arrival time change
\item Temporal agent calculates ripple effects
\item Role agents evaluate reassignment options
\item Safety agent verifies continued supervision coverage
\end{itemize}

The meta-planner $\mathcal{MP}$ coordinates these responses while maintaining global constraint satisfaction.

\subsection{Integrated Workflow Network}

Table~\ref{tab:workflow_spec} presents the
resulting workflow network $\mathbf{W^*}$, which includes
all nodes and edges, and their assigned agents and protocols.

\begin{enumerate}[leftmargin=1.2em, topsep=-.15em, parsep=-.15em]
\item \textit{Role Nodes:}
    \begin{itemize}[leftmargin=1.0em, topsep=-.15em, parsep=-.15em]
    \item Cook1: Sarah (primary) or Grandma (if at home) with 4-hour turkey + 2-hour sides
    \item Driver1: James (after car rental) or Michael
    \item Driver2: Available person after initial pickups
    \item Supervisor: Must be present while turkey cooks
    \end{itemize}

\item \textit{Dependencies:}
    \begin{itemize}[leftmargin=1.0em, topsep=-.15em, parsep=-.15em]
    \item Temporal: Verified airport processing + travel times
    \item Spatial: Traveling routes with traffic consideration
    \item Safety: Continuous oven supervision requirement
    \end{itemize}

\item \textit{Agent Monitoring:}
    \begin{itemize}[leftmargin=1.0em, topsep=-.15em, parsep=-.15em]
    \item Temporal Agent: Schedules with verified buffer times
    \item Spatial Agent: Real-time location and route mgmt.
    \item Safety Agent: Role coverage for supervision
    \end{itemize}
\end{enumerate}

\subsection{Agent Interaction Specifications}

Please, see Table~\ref{tab:agent_protocols}.

\begin{table*}[th!]
\caption{Agent Interaction Protocols and State Transitions}
{\fontsize{7pt}{9pt}\selectfont
\begin{tabular}{|p{0.14\textwidth}|p{0.27\textwidth}|p{0.27\textwidth}|p{0.20\textwidth}|}
\hline
\textbf{Interaction Type} & \textbf{Protocol} & \textbf{State Transitions} & \textbf{Validation Rules} \\
\hline
\multicolumn{4}{|l|}{\textit{Node-to-Node Interactions}} \\
\hline
Cook$\leftrightarrow$ Supervisor & 
Protocol: cooking\_handoff() \newline
Message: (task, duration, reqs.) &
States: prep → cooking → comp. \newline
Trigger: task\_state\_change() &
Validate: coverage() \newline
Alert: coverage\_gap() \\
\hline
Driver1 $\leftrightarrow$ Driver2 &
Protocol: pickup\_handoff() \newline
Message: (location, time, passenger) &
States: available → enroute → comp. \newline
Trigger: location\_change() &
Validate: timing\_feasible() \newline
Alert: schedule\_conflict() \\
\hline
\multicolumn{4}{|l|}{\textit{Edge Agent Operations}} \\
\hline
Temporal Agent &
Protocol: schedule\_monitor() \newline
Message: (event, time, dependencies) &
States: scheduled → active → comp. \newline
Trigger: time\_milestone() &
Validate: timing\_feasible() \newline
Alert: delay\_impact() \\
\hline
Spatial Agent &
Protocol: location\_track() \newline
Message: (actor, position, dest.) &
States: idle → moving → arrived \newline
Trigger: position\_update() &
Validate: route\_feasible() \newline
Alert: travel\_delay() \\
\hline
\end{tabular}}
\label{tab:agent_protocols}
\vspace{-.1in}
\end{table*}

\subsection{Augmented Problem Statement Revised with \textbf{W*}}
\label{app:ThanksGivingAugmentPS}

Given the $\mathbf{W^*}$ generated by $\MP$'s meta-planner $\mathcal{MP}$, the Thanksgiving Dinner Planning problem statement is revised as follows: \\
\newline
\noindent \underline{\textit{Initial Setup:}}
\begin{itemize}[leftmargin=1.5em, topsep=-.15em, parsep=-.15em]
\item Mom (Sarah) is hosting Thanksgiving dinner at 6:00 PM in Boston. The following family members are traveling:
\item Dad (James) flying from San Francisco, landing at 1:00 PM Eastern time.
\item Sister (Emily) flying from Chicago, landing at 2:30 PM
\item Brother (Michael) driving from New York, estimated arrival 3:00 PM at home
\item Grandma is healthy and needs to be picked up from her home in suburban Boston
\end{itemize}

\noindent \underline{{\color{red}{\textbf{*}}} \textbf{Common Sense Augmented Constraints:}}
\begin{itemize}[leftmargin=1.5em, topsep=-.15em, parsep=-.15em]
\item The airport luggage pickup time after landing is 30 minutes.
\item Renting a car takes 30 minutes.
\item One person can simultaneously prepare turkey and side dishes.
\item Grandma prefers Michael to pick her up, provided that it does not cause the dinner time delay (soft constraint).
\item Grandma and Sarah prefer not to cook together in the kitchen.
\item The best turkey receipt and baking instructions included.
\item Traffic congestion is not factored into current planning.
\end{itemize}

\vspace{.1in}
\noindent \underline{\textit{Planning Validation Set:}}
\begin{enumerate}[leftmargin=1.5em, topsep=-.15em, parsep=-.15em]
\item All tasks and dependencies must be strictly
observed in the plan, or the plan fails.
\item Dinner time is strictly at 6:00 PM, all tasks must
be completed by then (redundancy).
\item Account for the idle time of each person.
\item The schedule consists of three columns: time, task, and assigned person(s).
\end{enumerate}

\begin{table}[t!]
\centering
\caption{DeepSeek's Plan, Two Iterations}
{\fontsize{7pt}{9pt}\selectfont
\begin{tabular}{|l|p{0.22\textwidth}|l|}
\hline
\textbf{Time} & \textbf{Task} & \textbf{Assigned} \\ 
\toprule \hline
1:00 PM & James lands at Boston & James \\ \hline
1:00--1:30 PM & James picks up luggage & James \\ \hline
1:30--2:00 PM & James rents a car & James \\ \hline
2:00 PM & {Turkey in oven} (4 hours; requires monitoring) & Sarah \\ \hline
2:00--3:00 PM & James waits at airport (idle) & James \\ \hline
2:30 PM & Emily lands at Boston & Emily \\ \hline
2:30--3:00 PM & Emily waits for luggage & Emily \\ \hline
3:00 PM & James picks up Emily & James \\ \hline
3:00 PM & Michael arrives home & Michael \\ \hline
3:00 PM & Michael departs to Grandma & Michael \\ \hline
3:30 PM & Michael picks up Grandma & Michael \\ \hline
3:30--4:00 PM & Michael drives back home with Grandma & Michael \\ \hline
3:00--4:00 PM & James drives Emily home (airport to home: 1 hour) & James \\ \hline
4:00 PM & James and Emily home & James \\ \hline
4:00 PM & M. and Grandma home & Michael \\ \hline
4:00--6:00 PM & Sarah prepares side dishes & Sarah \\ \hline
6:00 PM & Thanksgiving dinner begins & All \\ \hline
\end{tabular}}
\label{tab:DeepSeekPlan1}
\vspace{-.1in}
\end{table}

\subsection{Experiment \#1: Sequential Planner with Common Sense}
\label{app:SequentialPlanExperiment}
The first experiment utilized the augmented problem specification with common sense reasoning, incorporating realistic constraints such as luggage claim time and rental car pickup time.

We evaluated four standalone LLMs alongside $\ALAS$.
Both $\ALAS$ and Gemini consistently generated feasible schedules similar to Table~\ref{tab:DeepSeekPlan1}, while other LLMs encountered significant challenges.  

Upon analyzing the number of iterations required for a feasible plan, DeepSeek and Claude each needed one revision (two iterations), while GPT4o required two revisions (three iterations). In terms of scheduling quality—measured by slack time, total driving distance, and load balance—DeepSeek (Table~\ref{tab:DeepSeekPlan1}) outperformed both GPT4o (Table~\ref{tab:GPT4oPlan1}) and Claude (Table~\ref{tab:ClaudePlan1}). DeepSeek optimized efficiency by having James wait at the airport for 30 minutes to pick up Emily. In contrast, Claude inefficiently scheduled James to drive home and then return to the airport for Emily, creating unnecessary travel. GPT4o assigned James to return home and tasked Michael with separately picking up Emily and then Grandma, resulting in suboptimal load distribution. A more efficient solution would have scheduled Michael to collect Emily first, then proceed with her to Grandma's home, allowing all three to return together—saving 30 minutes of driving time while enhancing Grandma's experience of seeing both grandchildren simultaneously.

\input{Figure-FamilyReunion-Exp1}


\subsubsection{Observations of Errors in Standalone LLMs}
\label{app:exp1-observations}

Although DeepSeek and Claude eventually produced feasible static plans in their second iterations, Tables~\ref{tab:reunion-Claude} and \ref{tab:reunion-DeepThink} highlight critical errors in their initial attempts.

These errors included misestimated travel times (calculating 60-minute trips as 45 or 30 minutes, highlighted in shaded red) and implausible scheduling decisions, such as beginning turkey preparation at 10 AM or allowing James to depart the airport without Emily. 

\input{Table-Case2-LLM-Failures}


Such failures result from context erosion in extended prompts~\cite{liu-etal-2024-lost, xiao2024attentionsink} and expanding context windows. Research demonstrates that extended contexts accelerate information loss~\cite{park2023context, wei2023larger}, leading to constraint violations. $\ALAS$ circumvents these issues through its modular architecture, where specialized agents process only domain-relevant information while independent validation mechanisms ensure constraint adherence.

\paragraph{Handling Long Dependencies}
Complex scheduling problems reveal cascading errors when dependencies overlap. Critical constraints, particularly those involving multiple factors, frequently get dropped during iterative problem-solving. \\
\textbf{Reason}: Cognitive limitations restrict simultaneous constraint tracking, making exhaustive verification challenging in single processing passes. \\
\textbf{Solution Framework}:
\begin{itemize}[leftmargin=1.2em, topsep=-.15em, parsep=-.15em]
\item Isolate and systematically enumerate atomic task dependencies.
\item Implement comprehensive verification of global constraint satisfaction.
\item Develop robust mechanisms for systematic conflict resolution.
\end{itemize}

\paragraph{Stale Memory and Iterative Revisions}
Iterative solutions risk propagating errors due to incomplete constraint resets. \\
\textit{Reason}: Excessive reliance on previous solutions without comprehensive constraint re-evaluation leads to compounding errors. \\
\textbf{Relation to Gödel's Incompleteness}:
\begin{itemize}[leftmargin=1.2em, topsep=-.15em, parsep=-.15em]
\item Formal systems capable of arithmetic necessarily contain unprovable truths.
\item Similarly, inherited solution errors inhibit consistent constraint satisfaction.
\item Clean-state resets become essential for systematic error prevention.
\end{itemize}

\paragraph{Implementation Strategy}
Reset to a clean baseline state for each iteration, thoroughly re-evaluating all constraints. \\
\textit{Core Challenges}:
\begin{itemize}[leftmargin=1.2em, topsep=-.15em, parsep=-.15em]
\item Effective management of nested dependencies.
\item Prevention of residual errors across iterations.
\item Maintenance of cross-iteration consistency.
\end{itemize}

\begin{table}[ht]
\centering
\caption{Sequential Planning. (\# = iterations)}
{\fontsize{7pt}{9pt}\selectfont
\begin{tabular}{|l|c|p{0.60\linewidth}|}
\hline
\textbf{LLM} & \textbf{\#}  & \textbf{Notable Features} \\
\toprule \hline
\textbf{DeepSeek} & 2  & Optimized airport wait time for James; balanced workload \\
\hline
Claude & 2 &  Unnecessary travel between pickup tasks (no need to go home before next pickup) \\
\hline
GPT4o & 3 &  Extra travel for Michael; suboptimal load balance \\
\hline
\end{tabular}}
\label{tab:sequential_results}
\end{table}

Table~\ref{tab:sequential_results} synthesizes the detailed schedules documented in Tables~\ref{tab:DeepSeekPlan1}, \ref{tab:GPT4oPlan1}, and \ref{tab:ClaudePlan1}. 
DeepSeek demonstrated good scheduling efficiency by optimizing James's airport wait time for Emily's pickup, requiring only two iterations for convergence. Although GPT4o eventually produced a valid solution after three iterations, it created suboptimal travel patterns with redundant trips by Michael. Claude's solution, though feasible in two iterations, incorporated unnecessary travel between pickup tasks. 
In contrast to the inconsistent performance of standalone LLMs, $\ALAS$ consistently generated feasible and efficient plans in all test runs.

\subsection{Experiment \#2: Reactive Planner for Flight Delay}
\label{app:ReactivePlanExperiment}


This disruption scenario was stated in the main body of the paper in
Section~\ref{sec:ALAS-exp-case2}. Under the prompt for
reactive planning, it states the disruption as:
``At noon, James' flight is delayed until 4:00 PM. Update the schedule to meet the deadline at 6:00 pm while meeting all constraints.''

We tested only $\ALAS$, DeepSeek and Claude 3.7 for disturbance handling, because they survived sequential planning. 
$\ALAS$ successfully generated a feasible reactive plan, whereas
DeepSeek and Claude 3.7 failed.

The challenge: James (assigned to pick up Emily) is delayed until 4:00 PM. Both DeepSeek and Claude failed in seven out of ten runs by: (1) scheduling Emily for a taxi, violating family pick-up constraints; (2) delaying or missing Grandma's pickup; or (3) missing the dinner deadline. An example failed schedule (the name of the LLM is purposely concealed) is presented in Table~\ref{tab:failed-reactive-planning}, where one can see that when all family members are ready to eat the delicious turkey, ``oops, Grandma was not picked up.''

\begin{table}[htbp]
\centering
\caption{Failed Reactive Planning Schedule with Flight Delay Disruption}
\label{tab:failed-reactive-planning}
{\fontsize{7pt}{9pt}\selectfont
\begin{tabular}{|p{1.5cm}|p{4.2cm}|p{6.2cm}|}
\hline
\textbf{Time} & \textbf{Task} & \textbf{Explanation} \\
\toprule
\hline
(Before Noon) & Tasks like initial prep are underway & The disruption is received at noon. \\
\hline
2:00 PM & Sarah puts turkey in the oven. & \textbf{Unchanged:} Critical path for food. Sarah must stay home to supervise (2:00 PM - 6:00 PM). \\
\hline
2:30 PM & Emily lands at BOS. & \textbf{Unchanged:} Sister's flight arrives as scheduled. \\
\hline
3:00 PM & Michael arrives home. & \textbf{Unchanged:} Brother arrives from NY. \textbf{New Role:} He is now the designated driver for Emily's pickup. \\
\hline
$\sim$3:15 PM & Michael departs home for BOS Airport. & Michael must leave soon after arriving to pick up Emily. (60 min travel time to BOS). \\
\hline
4:00 PM & James lands at BOS. & \textbf{Disruption Effect:} Dad arrives 3 hours later than planned. \\
\hline
4:00 PM & Sarah starts cooking side dishes. & \textbf{Unchanged:} Side dishes still need 2 hours. Sarah starts them as she is the only one home. \\
\hline
4:00 - 5:00 PM & James deplanes, gets luggage, rents a car. & \textbf{Delayed:} James needs to rent a car after landing. \\
\hline
$\sim$4:15 PM & Michael arrives at BOS Airport. & Michael reaches the airport to meet Emily. \\
\hline
$\sim$4:15 - 4:45 & Michael and Emily meet at BOS & Allow time for Emily to get luggage and meet Michael. \\
\hline
$\sim$4:45 PM & Michael and Emily depart BOS for home. & They begin the 60-minute drive home. \\
\hline
$\sim$5:00 PM & James departs BOS for home (in rental car). & James starts his 60-minute drive home after getting rental car. \\
\hline
$\sim$5:45 PM & Michael and Emily arrive home. & Brother and Sister are home. \\
\hline
6:00 PM & James arrives home. & Dad arrives home just in time for the planned dinner. \\
\hline
6:00 PM & Dinner Time & Turkey and side dishes are ready. Sarah, Michael, Emily, and James are home. \\
\hline
\colorbox{red!15} POST 6:00 PM & Grandma's Pickup - Unresolved & \textbf{Major Impact:} With James arriving at 4:00 PM and needing an hour for car rental (leaving BOS at 5:00 PM), he cannot pick up Grandma and return by 6:00 PM. Michael is busy with Emily's pickup until 5:45 PM and cannot make the $\sim$1.5 hr round trip for Grandma before 6:00 PM. Sarah must supervise the oven. \textbf{Under these constraints, Grandma cannot join the 6:00 PM dinner.} \\
\hline
\end{tabular}}
\end{table}

These failures are due to greedy rescheduling and increased context loss during replanning. Studies confirm that longer contexts can worsen reliability~\cite{wei2023larger, zhang2023careful}. $\ALAS$ generated feasible plans in all runs by: (1) updating James's state in persistent memory; (2) detecting conflicts systematically; (3) evaluating alternatives; and (4) validating all constraints. Following LRCP (Local Reactive Compensation Protocol, depicted in Appendix~\ref{app:ALAS-Algorithm-Full}), $\ALAS$ produced four distinct solution patterns, all feasible. While requiring 2.5× the computation time (12.1s vs. 4.8s), this overhead ensures feasibility, crucial in time-sensitive domains.

\section{Additional Case Studies and Analysis}

\paragraph{ALAS's Reactive Plan and Stateful Explanation.} $\ALAS$ proposes a simple yet effective reroute: Michael drives straight to Boston Airport rather than stopping at home first.  
This common-sense spatial adjustment—overlooked by the other LLMs—originates from $\mathcal{MP}$'s state-aware reasoning module.  
After collecting Emily, Michael proceeds directly to Grandma's house, trimming roughly 30 minutes of travel and giving Grandma the pleasant surprise of seeing two grandchildren arrive together.
Table \ref{tab:ALASReactivePlan} lists the resulting feasible schedule.  
The critical advantage is that $\ALAS$'s continuous tracking of each participant's state and history enables timely compensations and preserves the on-time family reunion.

\begin{table}[htbp]
\vspace{-.1in}
\centering
\caption{$\ALAS$ Reactive Plan: Optimized routing via persistent state history and compensation.}
\vspace{.05in}
{\fontsize{7pt}{9pt}\selectfont
\begin{tabular}{|l|p{0.42\textwidth}|l|}
\hline
\textbf{Time} & \textbf{Task} & \textbf{Assigned} \\
\toprule
\hline
12:00 PM & James' delay known, Michael on his way from NYC to home, is rerouted to Boston airport to meet Emily (4-hour drive). & Michael \\
\hline
2:00 PM & Start cooking turkey & Sarah \\
\hline
2:30 PM & Emily lands at Boston & Emily \\
\hline
3:00 PM & Emily gets her luggage & Emily \\
\hline
3:00 PM & Michael arrives at Logan airport, picks up Emily. & Michael \\
\hline
3:00–4:00 PM & Michael drives Emily home & Michael \\
\hline
4:00 PM & Michael departs for Grandma & Michael \\
\hline
4:00 PM & James lands at Boston Airport & James \\
\hline
4:00–4:30 PM & James picks up luggage & James \\
\hline
4:30–5:00 PM & James rents car (30 minutes). & James \\
\hline
4:30 PM & Michael arrives at Grandma's & Michael \\
\hline
5:00 PM & Michael \& Grandma arrive home. & Michael, Grandma \\
\hline
5:00–6:00 PM & James drives home from BOS & James \\
\hline
4:00–6:00 PM & Sarah prepares side dishes (overlaps with turkey). & Sarah \\
\hline
6:00 PM & James arrives home. Dinner served. & All \\
\hline
\end{tabular}}
\label{tab:ALASReactivePlan}
\vspace{-.1in}
\end{table}

%% file: TableWorkflow.tex
\begin{table*}[th!]
\caption{Complete Workflow Specification: Nodes, Edges, and Agent Assignments}
{\fontsize{7pt}{9pt}\selectfont
\begin{tabular}{|p{0.08\textwidth}|p{0.12\textwidth}|p{0.2\textwidth}|p{0.30\textwidth}|p{0.16\textwidth}|}
\hline
\textbf{Type} & \textbf{Component} & \textbf{Requirements} & \textbf{Agent Protocol} & \textbf{Dependencies} \\
\hline
\multicolumn{5}{|l|}{\textit{Node Components (Roles)}} \\
\hline
Node & Cook Role \newline (Sarah) & 
- Turkey (4hr) \newline
- Side dishes (2hr) \newline
- Kitchen management \newline
- Time management &
Input: schedule, resources, recipes \newline
Output: task progress, completion \newline
Monitor: kitchen\_state() → status \newline
Validate: cooking\_constraints() &
Connected to: \newline
- Supervisor \newline
- Resource edges \\
\hline
Node & Driver1 \newline (James/Michael) &
- Valid license \newline
- Airport navigation \newline
- Car rental capable \newline
- Rest state adequate &
Input: flight times, routes \newline
Output: location, ETA \newline
Monitor: driver\_state() → status \newline
Validate: driver\_constraints() &
Connected to: \newline
- Airport pickup \newline
- Travel edges \\
\hline
Node & Driver2 \newline (Flexible) &
- Valid license \newline
- Local navigation \newline
- Availability window \newline
- Rest state adequate &
Input: pickup schedule, route \newline
Output: location, ETA \newline
Monitor: driver\_state() → status \newline
Validate: driver\_constraints() &
Connected to: \newline
- Grandma pickup \newline
- Travel edges \\
\hline
Node & Supervisor \newline (Flexible) &
- Home presence \newline
- Oven monitoring \newline
- Safety awareness \newline
- Time commitment &
Input: cooking schedule, rules \newline
Output: supervision status \newline
Monitor: safety\_state() → status \newline
Validate: safety\_constraints() &
Connected to: \newline
- Cook role \newline
- Safety edges \\
\hline
\multicolumn{5}{|l|}{\textit{Edge Components (Dependencies)}} \\
\hline
Edge & Temporal &
- Schedule tracking \newline
- Buffer management \newline
- Sequence logic \newline
- Critical path &
Input: timestamps, durations \newline
Output: schedule conflicts \newline
Monitor: schedule\_state() → alerts \newline
Optimize: timeline\_adjust() &
Connects: \newline
- All roles \newline
- All activities \\
\hline
Edge & Spatial &
- Location tracking \newline
- Route optimization \newline
- Traffic updates \newline
- Distance constraints &
Input: locations, routes \newline
Output: travel updates \newline
Monitor: location\_state() → alerts \newline
Optimize: route\_adjust() &
Connects: \newline
- Drivers \newline
- Locations \\
\hline
Edge & Resource &
- Vehicle allocation \newline
- Kitchen resources \newline
- People availability \newline
- Capacity limits &
Input: resource demands \newline
Output: allocation status \newline
Monitor: resource\_state() → alerts \newline
Optimize: resource\_adjust() &
Connects: \newline
- All roles \newline
- All resources \\
\hline
Edge & Safety &
- Oven monitoring \newline
- Driving safety \newline
- Food safety \newline
- Critical rules &
Input: safety requirements \newline
Output: violation alerts \newline
Monitor: safety\_state() → alerts \newline
Enforce: safety\_rules() &
Connects: \newline
- All roles \newline
- Critical tasks \\
\hline
\end{tabular}}
\label{tab:workflow_spec}
\end{table*}

%% file: Figure-FamilyReunion-Exp1.tex
\begin{table}[ht!]
\centering
\begin{minipage}{0.48\textwidth}
    \centering
    \caption{GPT4o's Plan, Three Iterations}
    {\fontsize{7pt}{9pt}\selectfont
    \begin{tabular}{|l|l|l|}
    \hline
    \textbf{Time}        & \textbf{Task} & \textbf{Assigned}      \\ 
\toprule \hline
    1:00 PM              & Land at BOS Airport                      & James                     \\ \hline
    1:00-1:30 PM         & Luggage pickup                           & James                     \\ \hline
    1:30-2:00 PM         & Rent car                                 & James                     \\ \hline
    2:00 PM              & Start turkey                             & Sarah                     \\ \hline
    2:00-3:00 PM         & Drive home                               & James                     \\ \hline
    2:30 PM              & Land at BOS Airport                      & Emily                     \\ \hline
    3:00 PM              & Arrive home                              & Michael                   \\ \hline
    3:00-4:00 PM         & Drive to airport, pick up Emily          & Michael                   \\ \hline
    4:00-5:00 PM         & Return home with Emily                   & Michael                   \\ \hline
    5:00-5:30 PM         & Drive to Grandma's                       & Michael                   \\ \hline
    5:30-6:00 PM         & Return with Grandma                      & Michael                   \\ \hline
    4:00-6:00 PM         & Prepare side dishes                      & Sarah                     \\ \hline
    6:00 PM              & Dinner served                            & All                       \\ \hline
    \end{tabular}}
    \label{tab:GPT4oPlan1}
\end{minipage}%
\hfill
\begin{minipage}{0.48\textwidth}
    \centering
    \caption{Claude's Plan, Two Iterations}
    {\fontsize{7pt}{9pt}\selectfont
    \begin{tabular}{|l|l|l|}
    \hline
    \textbf{Time} & \textbf{Task} & \textbf{Assigned} \\
    \toprule \hline
    1:00 PM & Land at BOS Airport & James \\ \hline
    1:00-1:30 PM & Luggage pickup & James \\ \hline
    1:30-2:00 PM & Rent car & James \\ \hline
    2:00 PM & Start turkey & Sarah \\ \hline
    2:00-3:00 PM & Drive home & James \\ \hline
    2:30 PM & Land at BOS Airport & Emily \\ \hline
    3:00 PM & Arrive home & Michael \\ \hline
    3:00-4:00 PM & Drive to airport, pick up Emily & James \\ \hline
    4:00-5:00 PM & Return home with Emily & James \\ \hline
    4:30-5:00 PM & Drive to Grandma's & Michael \\ \hline
    5:00-5:30 PM & Return with Grandma & Michael \\ \hline
    4:00-6:00 PM & Prepare side dishes & Sarah \\ \hline
    6:00 PM & Dinner served & All \\
    \hline
    \end{tabular}}
    \label{tab:ClaudePlan1}
\end{minipage}
\vspace{-.1in}
\end{table}

%% file: Table-Case2-LLM-Failures.tex
\begin{table}[ht!]
\vspace{-.15in}
\centering
\begin{minipage}{0.485\textwidth}
    \centering
    \caption{Family Reunion - Claude. Gray indicates questionable, red wrong time calculation.}
    {\fontsize{7pt}{9pt}\selectfont
    \begin{tabular}{|p{0.1\textwidth}|p{0.56\textwidth}|p{0.15\textwidth}|}
    \hline
    \textbf{Time} & \textbf{Task} & \textbf{Assigned} \\
    \toprule
    \hline
    \colorbox{yellow} 10:00 & Put turkey in oven (4-hour) & S \\
    \colorbox{white} 13:00 & James lands at BOS & J \\
    \colorbox{white} 13:30 & James rents car at BOS & J \\
    \colorbox{yellow} 14:00 & James leaves airport for home & J \\
    \colorbox{white} 14:30 & Emily lands at BOS & E \\
    \colorbox{white} 15:00 & James \& Michael arrive home;  & J, M \\
    \colorbox{white} 15:15 & M leaves to pick up E from BOS & M \\
    \colorbox{white} 15:30 & S begins side dishes (2-hour) & S \\
    \colorbox{white} 16:15 & M arrives at BOS to pick up E & M, E \\
    \colorbox{white} 16:30 & James leaves to pick up Grandma & J \\
    \colorbox{white} 17:00 & J arrives at G's house & J, G \\
    \colorbox{lightred} 17:00 & Michael and Emily arrive home & M, E \\
    \colorbox{white} 17:30 & J and G home; side dish done & J, G, S \\
    \colorbox{white} 18:00 & Turkey complete, dinner ready & All \\
    \hline
    \end{tabular}
    }
    \label{tab:reunion-Claude}
\end{minipage}%
\hfill
\begin{minipage}{0.486\textwidth}
    \centering
    \caption{Family Reunion - DeepSeek. Red indicates wrong time calculation.}
    {\fontsize{7pt}{9pt}\selectfont
    \begin{tabular}{|p{0.1\textwidth}|p{0.60\textwidth}|p{0.15\textwidth}|}
    \hline
    \textbf{Time} & \textbf{Task} & \textbf{Assigned} \\
    \toprule
    \hline
    \colorbox{white} 12:00 & S starts cooking the turkey (4 hours, ready by 16:00) & S \\
    \colorbox{white} 13:00 & J lands at BOS and rents a car & J \\
    \colorbox{white} 13:30 & James drives from BOS to Grandma's & J \\
    \colorbox{white} 14:30 & James picks up Grandma, Emily lands & J, G \\
    \colorbox{white} 15:00 & Michael home (from NY) & M \\
    \colorbox{white} 15:15 & James and Grandma arrive back home (30 min drive) & J, G \\
    \colorbox{lightred} 15:15 & M leaves to pick up E from BOS (60 min round trip) & M \\
    \colorbox{white} 16:00 & Turkey done; S starts side dishes (2 hours) & S \\
    \colorbox{lightred} 16:15 & M and E return home from BOS & M, E \\
    \colorbox{white} 18:00 & Family reunion dinner begins with all members present & All \\
    \hline
    \end{tabular}
    }
    \label{tab:reunion-DeepThink}
\end{minipage}
\vspace{-.1in}
\end{table}

%% file: Appendix_case_studies.tex
\input{Table-ProblemStatements1and2}

\vspace{-.1in}  
\subsection{Case Study 1: Transportation Scheduling}  
\vspace{-.05in}  
Purpose: conclude the illustrative example and show that even simple problems cause standalone LLMs to miss basic planning requirements.

\vspace{-.1in}  
\subparagraph{Problem.}  
The Urban Ride Sharing (URS) problem (Table~\ref{fig:appDURSSpec}) coordinates vehicles to deliver passengers before deadlines while minimizing distance, with mid-execution disruptions.

\vspace{-.1in}  
\subparagraph{Setup.}  
Initial: “Create an optimal schedule for this ride-sharing scenario that minimizes total travel distance while meeting all passenger deadlines: [Table~\ref{fig:appDURSSpec}].”
Reactive: “Passenger $r_2$ cancels at 8:05; a new passenger $r_5$ at F requests pickup with a 9:30 deadline. Update the schedule.”  
$\ALAS$ augments prompts with structured workflow templates (Alg.~\ref{alg:metaplanner}) and role-based instantiation. Ten independent trials; we report mean$\pm$sd.

\vspace{-.05in}  
\paragraph{\#1. Sequential planning.}  
All models met deadlines; $\ALAS$ achieved superior efficiency: $95.1\!\pm\!13.0$\,km vs.\ $118.9\!\pm\!16.6$\,km for baseline LLMs (20\% improvement, $p<0.01$). Figure~\ref{fig:combined-URS} shows the $\ALAS$ schedule.

\input{Figure-Case1-TransportationScheduling}

\vspace{-.05in}  
\paragraph{\#2. Reactive planning.}  
With $r_2$ canceled (8:05) and $r_5$ added (8:10), $\ALAS$ repaired successfully in all runs. Baseline LLMs commonly lost vehicle state, duplicated assignments, or ignored updated deadlines—consistent with structural limits and statelessness (Sec.~\ref{sec:ALAS-related-llmlimits}). Full results: Appx.~\ref{app:ALAS-URAP}.

\vspace{-.1in}
\subsection{Case Study 2: Event Coordination}
\label{sec:ALAS-exp-case2}
\vspace{-.05in}
This study shows why standalone LLMs often fail even in static planning and behave inconsistently in reactive settings, while $\ALAS$ remains stable across both.

\vspace{-.05in}
\paragraph{Problem.} 
The Family Reunion problem (Table~\ref{fig:FamilyReunionSpec}): coordinate pickups, cooking, and shared resources with a 18{:}00 dinner deadline.

\vspace{-.05in}
\paragraph{Set up.} 
All methods: “Create a detailed schedule that satisfies all constraints: [Table~\ref{fig:FamilyReunionSpec}].”
Reactive: “At noon, James’ flight is delayed to 16{:}00. Update the schedule to meet the 18{:}00 deadline.” $\ALAS$ uses Algorithm~\ref{alg:metaplanner}; ten independent runs.

\vspace{-0.05in}
\paragraph{\#1. Sequential planning.}
$\ALAS$ produced a \emph{feasible} schedule in all runs; baselines frequently violated hard constraints and required retries (Tables~\ref{tab:reunion-Claude}, \ref{tab:reunion-DeepThink} in Appx.~\ref{app:ALAS-MeetingProblem}). Typical errors: travel-time arithmetic (e.g., 60\,min treated as 30–45\,min), commonsense slips (e.g., cooking far too early). Root cause: long-context degradation~\cite{liu-etal-2024-lost,xiao2024attentionsink}. $\ALAS$ confines context per agent and employs an \emph{independent} validator before finalization. A small \emph{commonsense agent} inserts realistic slack (e.g., luggage pickup); see Appx.~\ref{app:ThanksGiving-common-sense}.

\vspace{-0.05in}
\paragraph{\#2. Reactive planning.}
Under a delay to 16{:}00, DeepSeek and Claude failed in $7/10$ runs (e.g., violating pickup constraints, missing 18{:}00). Greedy one-shot rescheduling with longer prompts exacerbated context loss~\cite{wei2023larger,zhang2023careful}. $\ALAS$ succeeded in all runs via LCRP: logging state, detecting conflicts, testing alternatives, and re-validating locally. Full logs: Appx.~\ref{app:ALAS-MeetingProblem}.

%% file: Table-ProblemStatements1and2.tex
\begin{table}[t!]
\begin{scriptsize}
\begin{minipage}[t]{0.45\textwidth}
\renewcommand{\arraystretch}{1.1}
\setlength{\fboxsep}{4pt}
\caption{Dynamic Urban Ride-Sharing}
\fbox{
\begin{minipage}{\textwidth}
       \fontsize{8pt}{9pt}\selectfont
\textbf{Objectives:} Schedule vehicles to deliver passengers to airport during [8:45, 9:00], minimizing vehicle travel distance while ensuring on-time arrivals and maximizing passenger satisfaction. \\
\textbf{Locations:} Seven locations: $V = \{A, B, \cdots, F\}$, where $G$ is Boston Airport (BOS). Urban locations $A$–$F$ are all 10 km apart, airport distances 30+ km. 
\fontsize{7pt}{9pt}\selectfont
    \[
    \begin{bmatrix}
        & A  & B & C & D & E & F \\
    \rightarrow G & 35 & 33 & 36 & 34 & 32 & 31
    \end{bmatrix}
    \]
\fontsize{8pt}{9pt}\selectfont
    \textbf{Travel Speed:} ($A$–$F$) 60, ($A$–$F \rightarrow G$) 100 km/h
    
\textbf{Passenger Requests:} with BOS arrival deadlines:  
    \begin{itemize}[leftmargin=1em, topsep=1.2pt, itemsep=0.7pt, label=-]
    \item $r_1$: $A$, to $G$ by 08:45 \hspace{0.5em} - $r_2$: $B$, to $G$ by 08:50
    \item $r_3$: $C$, to $G$ by 08:55 \hspace{0.5em} - $r_4$: $D$, to $G$ by 09:00
    \end{itemize}
\textbf{Available Vehicles} (Capacity 2 passengers): 
    \begin{itemize}[leftmargin=1em, topsep=1.2pt, itemsep=0.7pt, label=-]
        \item $k_1$: at $A$, $k_2$: at $C$, and $k_3$: at $E$
        \item Battery levels: $k_1$: 90\%, $k_2$: 75\%, $k_3$: 60\%
    \end{itemize}
    
\textbf{Potential Disruptions:} New passenger requests, vehicle availability changes (battery levels/breakdowns at 0.05/hour), and traffic delay, etc. Replanning may require rolling back promised pickup time to the existing passengers and replan for all.
    \end{minipage}
    } 
    \label{fig:appDURSSpec}
\end{minipage}
\hspace{0.03\textwidth}
\begin{minipage}[t]{0.45\textwidth}
        \caption{Family Reunion Planning Problem}
        \renewcommand{\arraystretch}{1.1}
        \setlength{\fboxsep}{4.pt}
        \fbox{
        \begin{minipage}{\textwidth}
        \fontsize{8.0pt}{9pt}\selectfont
        \textbf{Objectives:} On time family reunion dinner at 6:00 PM \\
        \textbf{Family Members and Arrivals:}
        \begin{itemize}[leftmargin=1em, topsep=-0.2pt, itemsep=-0.2pt, label=-]
        \item Sarah (Mom): Host, at home
        \item James (Dad): Lands at BOS 1:00 PM from SF
        \item Emily (Sister): Lands at BOS 2:30 PM from Chicago
        \item Michael (Bro): Driving, arrives 3:00 PM from NY
        \item Grandma: Needs pickup from suburban Boston
        \end{itemize}
        \textbf{Cooking Requirements:}
        \begin{itemize}[leftmargin=1em, topsep=-.25pt, itemsep=-.25pt, label=-]
        \item Turkey: 4 hours cooking time; side dishes: 2 hours
        \item Someone must stay home during oven baking time
        \end{itemize}
        \textbf{Transportation Constraints:}
        \begin{itemize}[leftmargin=1em, topsep=-.25pt, itemsep=-.25pt, label=-]
        \item James must rent car after landing
        \item Emily requires airport pickup
        \item Travel times: Home to BOS Airport: 60 min
        \item Travel times: BOS Airport to Grandma's: 60 min
        \item Travel times: Home to Grandma's: 30 min
        \end{itemize}
        \textbf{Key Constraints:}
        \begin{itemize}[leftmargin=1em, topsep=-.25pt, itemsep=-.25pt, label=-]
        \item All family members home before 6:00 PM
        \item Turkey and sides ready by 6:00 PM
        \item All pickups completed with available drivers
        \item Oven baking supervision maintained
        \end{itemize}
        \end{minipage}
}
        \label{fig:FamilyReunionSpec}
    \end{minipage}
    \end{scriptsize}
    \vspace{-.1in}
\end{table}

%% file: Figure-Case1-TransportationScheduling.tex
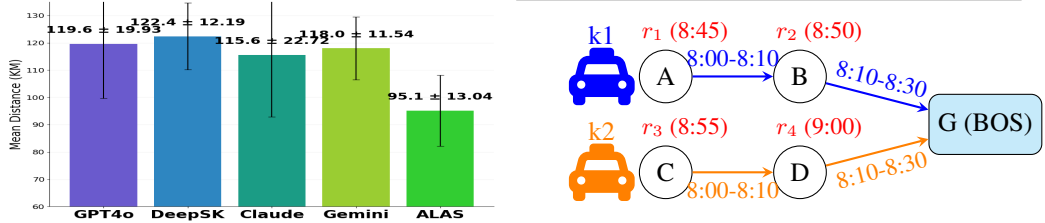
\begin{figure}[t!]
\vspace{-.1in}
    \centering
    \begin{minipage}{0.48\textwidth}
        \centering        \includegraphics[width=1.0\textwidth,height=0.45\textwidth]{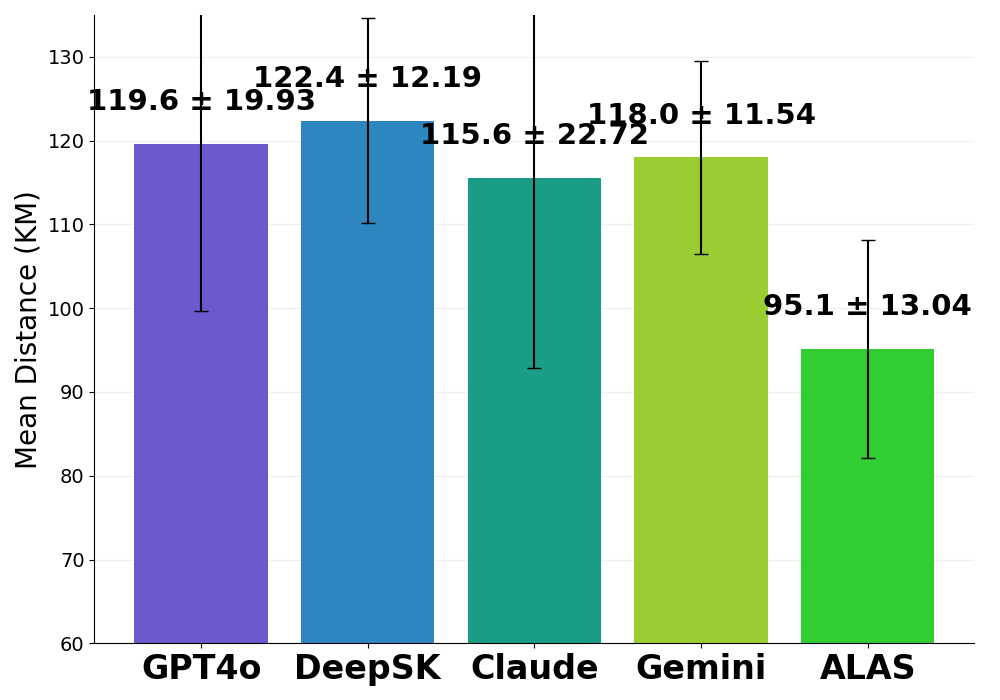}
    \end{minipage}%
    \hfill
    \begin{minipage}{0.48\textwidth}
        \centering
        \begin{tikzpicture}[
            location/.style={
                circle,
                draw=black,
                fill=white,
                minimum size=0.7cm,
                text centered,
                font=\normalsize
            },
            airport/.style={
                rectangle,
                rounded corners,
                draw=black,
                fill=cyan!20,
                minimum width=1.5cm,
                minimum height=0.8cm,
                text centered,
                font=\normalsize
            },
            passenger/.style={
                font=\small\color{red}
            },
            path_k1/.style={
                thick,
                blue,
                ->,
                >=stealth
            },
            path_k2/.style={
                thick,
                orange,
                ->,
                >=stealth
            },
            vehicle/.style={
                draw,
                fill=white,
                minimum width=0.8cm,
                minimum height=0.4cm,
                rounded corners=1pt
            },
            scale=0.85 
        ]
            \node[blue] at (-1,1.5) {\Huge\faTaxi};
            \node[blue] at (-1,2.15) {k1};
            \node[location] (A) at (0,1.5) {A};
            \node[location] (B) at (2.1,1.5) {B};
            \node[location] (C) at (0,0) {C};
            \node[location] (D) at (2.1,0) {D};
            
            \node[orange] at (-1,0) {\Huge\faTaxi};
            \node[orange] at (-1,0.65) {k2};
            
            \node[airport] (G) at (5,.75) {G (BOS)};
            
            \draw[path_k1] (A) -- (B) node[midway,above,sloped,fill=none,font=\footnotesize] {8:00-8:10};
            \draw[path_k1] (B) -- (G) node[midway,above,sloped,fill=none,font=\footnotesize] {8:10-8:30};
            
            \draw[path_k2] (C) -- (D) node[midway,below,fill=none,font=\footnotesize] {8:00-8:10};
            \draw[path_k2] (D) -- (G) node[midway,below,sloped,fill=none,font=\footnotesize] {8:10-8:30};
            
            \node[passenger] at ($(A)+(0.3,0.7)$) {$r_1$ (8:45)};
            \node[passenger] at ($(B)+(0.3,0.7)$) {$r_2$ (8:50)};
            \node[passenger] at ($(C)+(0.3,0.7)$) {$r_3$ (8:55)};
            \node[passenger] at ($(D)+(0.3,0.7)$) {$r_4$ (9:00)};
            
        \end{tikzpicture}
    \end{minipage}
    \caption{Comparison of ride-sharing solutions generated by $\MP$ and baseline LLMs. (Left) Mean total travel distance (km) with standard deviation error bars over 10 independent runs for each method, illustrating $\MP$'s improved efficiency. (Right) Optimal schedule generated by $\MP$ for the URS task, utilizing two vehicles ($k_1, k_2$) to serve four passengers ($r_1$-$r_4$).} 
\label{fig:combined-URS}
\vspace{-.18in}
\end{figure}

%% file: AppendixWorkflowLanguages.tex
\section{Workflow Orchestration: Standards, Code-First Engines, and Recommendation}
\label{app:workflow_standards}

\subsection{Introduction and motivation}

ALAS currently uses a custom JSON workflow format. This provides flexibility, but limits interoperability with established tools and standards. This appendix surveys widely used \emph{standards} and \emph{code-first engines}, and proposes a plan for ALAS.

\paragraph{Limitations of custom formats.}
\begin{itemize}
  \item Vendor lock-in risk when integrating with external engines and services
  \item Limited tooling for editing, visualization, debugging, and profiling
  \item Validation gaps without a widely used schema and event model
  \item Extra documentation burden to explain custom semantics and error models
  \item Observability and audit friction without standard event payloads
  \item Under-specified compensation and idempotency contracts
\end{itemize}

\paragraph{Benefits of adopting standards.}
\begin{itemize}
  \item Interoperability with existing engines and runtimes
  \item Access to mature IDEs, visualizers, debuggers, and monitors
  \item Portability across deployments and vendors
  \item More deterministic retry and replay when semantics are explicit
  \item Faster enterprise adoption and clearer documentation
\end{itemize}

\paragraph{Canonical model (ALAS IR).}
\begin{itemize}
  \item \textbf{One consolidated record.} ALAS keeps a consolidated record of the plan and its state in the internal representation (IR). All other views are generated from this record. If two views disagree, use the IR.
  \item \textbf{Everything explicit and preserved.} Retries, timeouts, backoff, idempotency keys, compensation handlers, and loop guards are explicit fields. Converters to standards and engines read and write these fields so translation does not drop information.
\end{itemize}

\subsection{Taxonomy}
We compare two families:
\begin{enumerate}
  \item \textbf{Standards and spec-first runtimes:} Argo Workflows, Amazon States Language (ASL), BPMN 2.0.
  \item \textbf{Code-first durable engines:} Temporal or Cadence, Flyte, Dagster or Prefect, Netflix Conductor.
\end{enumerate}

\subsection{Standards: survey and tradeoffs}
\subsubsection{Argo Workflows}
\textit{Additional benefits for ALAS.}
Kubernetes-native scaling with CRDs and container isolation; strong fan-out and fan-in for batch and data jobs; mature logs and archives.

\noindent\textit{Drawbacks in this context.}
Requires Kubernetes and RBAC; YAML becomes verbose for dynamic graphs and human-in-the-loop waits; compensation is not built in and should come from the ALAS IR.

\subsubsection{Amazon States Language (ASL)}
\textit{Additional benefits for ALAS.}
JSON format aligns with the ALAS IR; built-in \texttt{Retry} and \texttt{Catch} with backoff; state types (\texttt{Task}, \texttt{Choice}, \texttt{Parallel}, \texttt{Map}) simplify validator rules.

\noindent\textit{Drawbacks in this context.}
AWS-centric execution and integrations; cross-account or multi-region setups add work; limited portability for strict on-prem or multi-cloud.

\subsubsection{BPMN 2.0}
\textit{Additional benefits for ALAS.}
Clear diagrams for stakeholder review; compensation and boundary events map well to ALAS compensation fields; useful as a documentation layer over the ALAS IR.

\noindent\textit{Drawbacks in this context.}
XML stacks are heavy; highly dynamic plans need extensions; risk of drift if diagram and runtime both evolve.

\subsection{Code-first engines: survey and tradeoffs}
\textit{What we mean.}
Workflows are written in real code with durable state, timers, and signals. Engines can change control flow at runtime, cancel or reroute tasks, and keep long waits alive in the engine.

\subsubsection{Temporal or Cadence}
\textit{Benefits.}
Durable workflows and activities; signals and timers; child workflows; built-in saga pattern; strong cancellation and retry. Suited to disruption-heavy scenarios such as JSSP breakdowns or ride-sharing reroutes.

\noindent\textit{Drawbacks.}
Service cluster to operate; language SDK choice affects team fit.

\subsubsection{Flyte}
\textit{Benefits.}
Python-first with K8s scale; dynamic tasks and conditionals; type-checked interfaces; good for ML and data with some runtime flexibility.

\noindent\textit{Drawbacks.}
State machines and long human waits need patterns; K8s still required.

\subsubsection{Dagster or Prefect}
\textit{Benefits.}
Python-first orchestration; dynamic mapping; friendly UI and sensors; fast to wire tools and retries.

\noindent\textit{Drawbacks.}
DAG focus can limit state-machine style repair; very large fan-out needs tuning.

\subsection{Key comparisons}

\subsubsection{Between Engines}

This section compares the feasibility of workflow engines, both spec-first and code-first, across six axes. Favorable choices are marked in blue in the tables.



\begin{table}[h]
\footnotesize
\centering
\caption{Standards on common axes in the ALAS context}
\label{tab:standards_common_axes}
\begin{tabularx}{\linewidth}{p{3.1cm}XXX}
\toprule
Axis & Argo & ASL & BPMN 2.0 \\
\midrule
Control model
  & DAG or steps
  & State machine
  & Process graph \\ \hline
Dynamic replanning & \limited & \limited & \limited \\ \hline
Durable waits
  & Moderate
  & Moderate
  & Low to Moderate \\ \hline
Signals and timers
  & Hooks, onExit
  & Basic timers
  & Events and triggers \\ \hline
Error and compensation
  & Retries; comp. in IR
  & Retry \& Catch; external
  & Triggers and comp. \\ \hline
Hosting model
  & Kubernetes
  & AWS
  & Tool agnostic \\ \hline
Best ALAS use
  & K8s batch fan out/in
  & Agent loops and repair
  & Doc and audits \\
\bottomrule
\end{tabularx}
\vspace{2pt}
\footnotesize\emph{Note:} In BPMN, a boundary event is a task attached trigger that can route to an exception or compensation path.
\end{table}

\begin{table}[h]
\footnotesize
\centering
\caption{Code first engines on the same axes}
\label{tab:codefirst_common_axes}
\begin{tabularx}{\linewidth}{p{3.1cm}XXX}
\toprule
Axis & Temporal/Cadence & Flyte & Conductor \\
\midrule
Control model
  & Code state machine
  & Python DAG with \newline dynamic tasks
  & JSON with workers \\
Dynamic replanning 
  & \strong
  & Moderate
  & Moderate \\
Durable waits
  & \strong
  & Moderate
  & Moderate \\
Signals and timers
  & \strong
  & Moderate
  & Moderate \\
Error and compensation 
  & Retry; saga built in
  & Retry plugins
  & Retry; custom comp. \\
Hosting model
  & Service cluster
  & Kubernetes
  & JVM cluster \\
Best ALAS use
  & Disruptions, \newline reroutes, comp.
  & ML and data with some dynamics
  & Large microservice \newline estates \\
\bottomrule
\end{tabularx}
\end{table}


\subsubsection{Against REALM-Bench}

Table~\ref{tab:realm_all_six} evaluates each engine against the 14 test cases presented in the REALM Benchmark \cite{geng2025realmplanningbenchmark}.

\begin{table}[h]
\footnotesize
\setlength{\tabcolsep}{2pt}
\centering
\caption{REALM-Bench cases vs workflow engines (spec-first and code-first)}
\label{tab:realm_all_six}
\begin{adjustbox}{center,max width=\linewidth} 
\begin{tabular}{@{}lcccccc@{}} 
\toprule
Case & Argo & ASL & BPMN & Temporal/Cadence & Flyte & Conductor \\
\midrule
P1 Campus tour (static)             & G   & G   & G    & G   & G   & G \\
P2 Multi-group tour (static)        & G   & G   & G    & G   & G   & G \\
P3 URS matching (static)            & G   & G   & G    & G   & M   & G \\
P4 URS with disruptions             & M   & M   & G    & G   & M   & M/G \\
P5 Wedding logistics (static)       & G   & G   & G    & G   & M   & G \\
P6 Thanksgiving (static)            & M   & G   & G    & G   & M   & G \\
P7 Disaster relief (static)         & M   & G   & G    & G   & M   & G \\
P8 Wedding with disruptions         & M   & M   & G    & G   & M   & M/G \\
P9 Thanksgiving with delays         & M   & G   & G    & G   & M   & M/G \\
P10 Global supply chain             & M   & M   & G    & G   & M   & M/G \\
J1 JSSP basic (static)              & G   & G   & M    & G   & G   & G \\
J2 JSSP with disruptions            & M   & M   & M/G  & G   & M   & M/G \\
J3 JSSP large (static)              & G   & M   & M    & G   & G   & G \\
J4 JSSP large with disruptions      & M   & M   & M/G  & G   & M   & M/G \\
\bottomrule
\end{tabular}
\end{adjustbox}
\vspace{2pt}
\footnotesize\emph{Legend:} G = good fit, M = works with glue, M/G = feasible with glue and can reach good with careful modeling. 
\end{table}

\subsection{Recommendation for ALAS}

\paragraph{Final choice.}
We adopt a combined approach. Temporal serves as the primary control plane that holds workflow state, timers, signals, and local repair. When a stage requires very large parallelism on Kubernetes, Temporal invokes Argo to execute that stage and then resumes control on completion. This pattern is easy to code, scales to large fanout, and is practical for adoption.

\paragraph{When to use each.}
Temporal activities are preferred for steps with modest parallelism, rich branching, or human in the loop waits. Argo is preferred for stages that require large container fanout, GPU pools, or heavy artifact movement.

\paragraph{Scope of comparison.}
ASL and BPMN remain in our tables for completeness and portability analysis. The prototype focuses on Temporal and Argo to reduce scope while preserving adaptability.

\subsection{Implementation architecture}

\begin{figure}[h]
\centering
\begin{tikzpicture}[
  font=\footnotesize,
  node distance=6mm and 8mm,
  >=Latex,
  box/.style={draw, rounded corners=2pt, align=center, inner xsep=6pt, inner ysep=6pt, fill=gray!8},
  subbox/.style={draw, rounded corners=2pt, align=left, inner xsep=6pt, inner ysep=4pt, fill=gray!4}
]

\node[box, minimum width=0.78\linewidth] (inputs) {%
  \textbf{User input (multiple formats)}\\[1pt]
  Argo YAML \quad ASL JSON \quad BPMN 2.0 \quad Custom JSON
};

\node[box, below=of inputs, minimum width=0.78\linewidth] (pcv) {\textbf{Parsers, converters, validators}};
\node[subbox, below=1.6mm of pcv, minimum width=0.70\linewidth, align=center] (pcv-maps) {%
  Argo $\leftrightarrow$ IR \quad ASL $\leftrightarrow$ IR \quad BPMN $\leftrightarrow$ IR
};

\node[subbox, below=1.6mm of pcv-maps, minimum width=0.70\linewidth, align=left] (policy) {%
  \textbf{Policy layer:}\enspace Retry, Catch, Timeout, Backoff, IdempotencyKey, Compensation, LoopGuards
};
\node[subbox, below=1.6mm of policy, minimum width=0.70\linewidth, align=left] (shape) {%
  \textbf{Data-shape layer:}\enspace input, output, error schemas
};

\node[box, below=of shape, minimum width=0.78\linewidth] (ir) {%
  \textbf{ALAS internal representation (IR)}\\[1pt]
  \{nodes,\; edges,\; domain\_data,\; constraints,\; policies,\; schemas\}
};

\node[box, below=of ir, minimum width=0.78\linewidth] (exec) {%
  \textbf{Execution backends}\\[1pt]
  Argo on K8s \quad Code-first engine (Temporal or Cadence) \quad Others
};

\draw[->] (inputs) -- (pcv);
\draw[->] (pcv) -- (pcv-maps);
\draw[->] (pcv-maps) -- (policy);
\draw[->] (policy) -- (shape);
\draw[->] (shape) -- (ir);
\draw[->] (ir) -- (exec);

\end{tikzpicture}
\caption{Multi format architecture with explicit policy and data shape layers}
\label{fig:WF-architecture}
\end{figure}
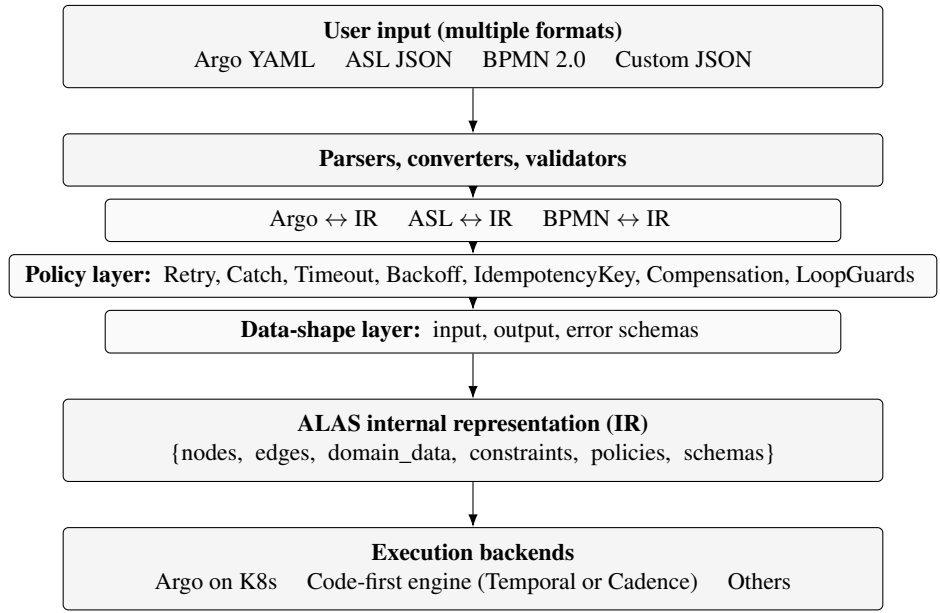

Figure~\ref{fig:WF-architecture} shows how standards and engines connect through the ALAS IR.


\subsection{Converter notes}

Conversions between Argo or ASL and ALAS should map the following fields explicitly:
\texttt{IntervalSeconds}, \texttt{BackoffRate}, \texttt{MaxAttempts} (retry policy);
\texttt{Catch} clauses with error selectors; \texttt{TimeoutSeconds};
\texttt{IdempotencyKey}; compensation handlers; and loop guard conditions.
BPMN conditions should use formal expressions. CWL step conditions should match the declared version, and required features should be listed under \texttt{requirements}.

\paragraph{Examples.}
Long YAML and JSON examples have been moved to the appendix to keep this section readable.

%% file: AppendixALASTemporal.tex
\section{ALAS on Temporal, a Minimal End-to-End Run}

\subsection{Goal}
Stand up a local Temporal Server, run one ALAS worker with a simple Job Shop Scheduling Problem (JSSP) activity, and launch a client that starts the workflow and prints a JSON result.

\subsection{Prerequisites}
\begin{itemize}
  \item macOS on Apple Silicon, Python 3.11 available as \texttt{python3.11}.
  \item Homebrew installed.
  \item Temporal Server CLI installed:
\begin{verbatim}
brew install temporalio/tap/temporal
\end{verbatim}
  \item Git is optional for this appendix.
\end{itemize}

\subsection{Directory Layout}
Create a project folder named \texttt{alas-temporal} with this structure.
\begin{verbatim}
alas-temporal/
  client/
    start.py
  scenarios/
    jssp_tiny.json
  worker/
    worker.py
  requirements.txt
  .venv/            (created by python venv)
\end{verbatim}

\subsection{Set Up the Python Environment}
In any terminal:
\begin{verbatim}
cd ~/alas-temporal
python3.11 -m venv .venv
source .venv/bin/activate      # zsh or bash
python -V                      # expect Python 3.11.x
\end{verbatim}
Install pinned dependencies:
\begin{verbatim}
cat > requirements.txt <<'REQ'
temporalio==1.7.0
pydantic==2.8.2
rich==13.7.1
REQ

pip install -r requirements.txt
\end{verbatim}

\subsection{Code Files}
\paragraph{Worker and workflow (\texttt{worker/worker.py}).}
The workflow invokes one activity \texttt{solve\_jssp}, computes a verdict, and returns a JSON object.
\begin{verbatim}
import asyncio, json
from datetime import timedelta
from temporalio import activity, workflow
from temporalio.client import Client
from temporalio.worker import Worker

@activity.defn
async def solve_jssp(scenario_json: str) -> dict:
    data = json.loads(scenario_json)
    jobs = data.get("jobs", [])
    makespan, schedule, t = 0, [], 0
    for job in jobs:
        for op in job.get("ops", []):
            start, end = t, t + op.get("time", 0)
            schedule.append({"job": job.get("id"),
                             "machine": op.get("machine"),
                             "start": start, "end": end})
            t = end
    makespan = t
    return {"makespan": makespan, "schedule": schedule}

@workflow.defn
class JSSPWorkflow:
    @workflow.run
    async def run(self, domain: str, scenario_json: str) -> dict:
        result = await workflow.execute_activity(
            solve_jssp, scenario_json, start_to_close_timeout=timedelta(seconds=30)
        )
        verdict = "pass" if result["makespan"] > 0 else "fail"
        return {"domain": domain, "verdict": verdict, "result": result}

async def main() -> None:
    client = await Client.connect("localhost:7233")
    worker = Worker(client, task_queue="alas-jssp",
                    workflows=[JSSPWorkflow], activities=[solve_jssp])
    print("Worker started on task_queue alas-jssp")
    await worker.run()

if __name__ == "__main__":
    asyncio.run(main())
\end{verbatim}

\paragraph{Client launcher (\texttt{client/start.py}).}
\begin{verbatim}
import asyncio, argparse, json, sys
from pathlib import Path
from temporalio.client import Client
from temporalio.common import RetryPolicy
from worker.worker import JSSPWorkflow

async def main():
    ap = argparse.ArgumentParser()
    ap.add_argument("--namespace", default="default")
    ap.add_argument("--domain", default="jssp")
    ap.add_argument("--scenario", required=True)
    args = ap.parse_args()

    scenario_json = Path(args.scenario).read_text()
    client = await Client.connect("localhost:7233", namespace=args.namespace)

    handle = await client.start_workflow(
        JSSPWorkflow.run,
        id="ALAS-JSSP-demo",
        task_queue="alas-jssp",
        args=[args.domain, scenario_json],
        retry_policy=RetryPolicy(maximum_attempts=1),
    )
    result = await handle.result()
    print(json.dumps(result, indent=2))

if __name__ == "__main__":
    try:
        asyncio.run(main())
    except KeyboardInterrupt:
        sys.exit(130)
\end{verbatim}

\paragraph{Scenario file (\texttt{scenarios/jssp\_tiny.json}).}
\begin{verbatim}
{
  "jobs": [
    { "id": "J1", "ops": [ {"machine":"M1","time":2}, {"machine":"M2","time":3} ] },
    { "id": "J2", "ops": [ {"machine":"M2","time":1}, {"machine":"M1","time":4} ] }
  ]
}
\end{verbatim}

\subsection{Running the System with Three Terminals}
We use three terminals so the control flow is clear and easy to repeat. Keep the Python venv active in terminals that run Python.

\paragraph{Terminal A: start Temporal Server.}
\begin{verbatim}
mkdir -p ~/.temporal
temporal server start-dev --db-filename ~/.temporal/dev.db
\end{verbatim}
Temporal Web UI runs at \texttt{http://localhost:8233}. Leave this server running.

\paragraph{Terminal B: start the worker.}
\begin{verbatim}
cd ~/alas-temporal
source .venv/bin/activate
python -m worker.worker
# Expect: "Worker started on task_queue alas-jssp"
\end{verbatim}

\paragraph{Terminal C: start the client.}
\begin{verbatim}
cd ~/alas-temporal
source .venv/bin/activate
python -m client.start --domain jssp --scenario scenarios/jssp_tiny.json
\end{verbatim}
Expected output printed to stdout:
\begin{verbatim}
{
  "domain": "jssp",
  "result": {
    "makespan": 10,
    "schedule": [
      {"end": 2, "job": "J1", "machine": "M1", "start": 0},
      {"end": 5, "job": "J1", "machine": "M2", "start": 2},
      {"end": 6, "job": "J2", "machine": "M2", "start": 5},
      {"end": 10, "job": "J2", "machine": "M1", "start": 6}
    ]
  },
  "verdict": "pass"
}
\end{verbatim}
To save it as a file:
\begin{verbatim}
mkdir -p output
python -m client.start --domain jssp --scenario scenarios/jssp_tiny.json \
  > output/jssp_result.json
\end{verbatim}

\subsection{Execution Flow}
\begin{enumerate}
  \item The client connects to \texttt{localhost:7233}, then calls \texttt{start\_workflow} for \texttt{JSSPWorkflow.run} on task queue \texttt{alas-jssp}.
  \item Temporal Server persists the run and dispatches work to a poller on \texttt{alas-jssp}.
  \item The worker claims the workflow task, enters the workflow code, and invokes the activity \texttt{solve\_jssp} with the scenario JSON.
  \item The activity computes a trivial sequential schedule, returns \texttt{\{"makespan", "schedule"\}}.
  \item The workflow decides a verdict. If makespan greater than zero, verdict is \texttt{pass}.
  \item The result returns to the client, which prints formatted JSON.
  \item The run is visible in the Web UI. Search for workflow ID \texttt{ALAS-JSSP-demo}.
\end{enumerate}

\subsection{What the Code Actually Does}
\begin{itemize}
  \item \textbf{Activity} \texttt{solve\_jssp}: parses the scenario, walks operations in order, assigns contiguous start and end times on machines, and returns the naive schedule. This is a placeholder for a real scheduler.
  \item \textbf{Workflow} \texttt{JSSPWorkflow.run}: orchestrates one call to the activity and computes a simple verdict.
  \item \textbf{Worker}: hosts both workflow and activity, polls the task queue, and executes tasks.
  \item \textbf{Client}: starts the workflow and awaits the result.
\end{itemize}

\subsection{Observed Pitfalls and Fixes}
\begin{itemize}
  \item \textbf{Shell parsing errors} when pasting commands that include lines starting with \texttt{\#}. Solution: run commands without the comment lines.
  \item \textbf{Virtual environment activation} under zsh. Solution: use \texttt{source .venv/bin/activate}. Do not use \texttt{activate.fish} unless your shell is fish.
  \item \textbf{Temporal Server startup error} complaining about a missing path. Solution: create the folder \texttt{\~{}/.temporal} before running \texttt{temporal server start-dev}.
  \item \textbf{Wrong client signature} for \texttt{start\_workflow}. Solution: pass workflow arguments using the \texttt{args=[...]} keyword and set \texttt{id} and \texttt{task\_queue} as keywords.
\end{itemize}

\subsection{Repeatable Checklist}
\begin{enumerate}
  \item Create the directory layout and files as given.
  \item Create and activate the Python venv, then install requirements.
  \item Start Temporal Server in Terminal A.
  \item Start the worker in Terminal B.
  \item Start the client in Terminal C and capture the JSON output.
  \item Inspect the run in the Web UI at \texttt{http://localhost:8233}.
\end{enumerate}

\subsection{Next Steps}
\begin{itemize}
  \item Replace the naive activity with your real ALAS planner, validator, executor modules.
  \item Add multiple activities and workflow branches for validate and repair.
  \item Add result persistence to a file store or a database, then add Temporal search attributes for rich filtering.
\end{itemize}